\documentstyle{article}
\newcommand{\blankline}{\vskip .3cm}
\newcommand{\f}{\begin{equation}}
\newcommand{\ff}{\end{equation}}
\begin{document}

\vskip  6 cm
\vfill
\centerline{\LARGE RECENT DEVELOPMENTS}
\blankline
\centerline{\LARGE  in}
\blankline
\centerline{\LARGE NONPERTURBATIVE QUANTUM GRAVITY}
\blankline
\blankline
\rm
\vskip1cm
\centerline{Lee Smolin}
\blankline
 \centerline{\it   Department of Physics,   Syracuse  University,
 Syracuse, New York, 13210 U.S.A.\footnote{bitnet address:
SMOLIN@SUHEP}}
 \vfill
\centerline{\today}
\vfill
\centerline{Abstract}
\noindent
 New results from the new variables/loop representation program of
nonperturbative quantum gravity are presented, with a focus on results
of Ashtekar, Rovelli and the author which greatly clarify the physical
interpretation of the quantum states in the loop representation.
These include:

1)  The construction of a class of states which approximate smooth
metrics for length measurements on scales, $L$, to order $l_{Planck}/L$.
2)  The discovery that any such state must have discrete structure at the
Planck length.  3)  The construction of operators for the area of arbitrary
surfaces and volumes of arbitrary regions and the
discovery that these operators are  finite.  4)  The diagonalization of
these operators and the demonstration that the spectra are discrete,
so that in quantum gravity areas and volumes are quantized in Planck
units.    5)  The construction of finite diffeomorphism invariant
operators that measure geometrical quantities such as the volume
of the universe and the areas of minimal surfaces.

These results are made possible by the use of new
techniques  for the regularization  of operator products that respect
diffeomorphism invariance.

Several new results in the classical theory are also reviewed including
the solution of the hamiltonian and diffeomorphism constraints
in closed form of Capovilla, Dell and Jacobson and a new form  of the
action that induces Chern-Simon theory on the boundaries of spacetime.
A new classical
discretization of the Einstein equations is also presented.

\vfill\eject

\tableofcontents
\vfill\eject

\section{Introduction}

 This review is devoted to the new variables/loop representation approach
to quantum gravity which has been under development for the last five
years.  It particularly focuses on several developments of 1991
which have expanded greatly our understanding of several related
issues:   What is the structure of space and time on the Planck
scale? How is the classical limit to be understood from a nonperturbative
point of view?    How are operators representing observables to be
regulated and constructed
nonperturbatively without breaking the diffeomorphism invariance
of the theory?

The central result I review here is the construction of a class of
nonperturbative quantum states which approximate classical metrics
on scales, $L$, large compared to the Planck length, but have discrete
structure when probed on the Planck
length\cite{weaveletter,weavepaper}.   This is closely related
to the discovery that several operators that characterize
the spatial geometry, such as the area of an
arbitrary surface and the volume of an arbitrary region, are finite
when constructed nonperturbatively.  Furthermore, their spectra are
quantized in multiples of the natural Planck units.

This review is organized into seven chapters.  In the remainder of this
introduction I attempt to set the scene with a general discussion of
the problem of quantum gravity.  The second chapter is devoted to
a description of quantum geometry that follows from the new
results; this is done with a minimum of formalities and no calculations.
The loop representation and Ashtekar variables, which underlie this
picture are introduced in chapter three and chapter four contains the
details of the calculations that support the picture given in chapter
two.  Chapter five reviews some recent developments in classical
relativity which complement and illuminate the new picture of
quantum geometry.  Three key open issues are discussed
 in chapter six, after which the review closes with a brief conclusion.

\subsection{Why is quantum gravity such a hard problem?}

As physicists, why should we be interested in the problem of
quantum
gravity?   It is a problem that is, if we are honest with ourselves,
far
removed for the frontiers of experimental and observational
science,
and it is also a problem that, quite conceivably, will not be solved
during any of our lifetimes.  Yet, in spite of a rate of progress
 slower than that of
even the search for a cure for cancer, more than a few of us
continue
to devote most of our energies to it.  Why
do we do it and what are we hoping to achieve?

The main reason, of course, is that the present situation in theoretical
physics is untenable.   We have two theories that work very
well in two different observational and
experimental regiemes\cite{problem}.
Somehow, in the world around us, general relativity and quantum
mechanics coexist; indeed, more than that, they must be different
aspects of a single theory.  The problem of quantum gravity
is to discover that theory.

The problem is that, as there are no obvious experimental clues
as to the shape of that theory\footnote{I say obvious, because it surely
will be the case that when we have the right theory we will realize
that there were clues all around us.}, it is anyone's guess which
postulates and formal structures from the two theories will
survive the unification and which should be discarded.
 As a result there has been, in the last twenty  years, a great
proliferation of approaches to the problem.  Here is a (partial)
list:  supergravity, higher derivative theories, nonpolynomial
lagrangians, large $N$ expansions, quantum field theory in curved
spacetimes, lattice quantum
gravity, induced gravity, twistor theory,  spacetime codes,
asymptotic safety, Kaluza-Klein theories, Euclidean quantum
cosmology,  nonperturbative hamiltonian methods,
nonperturbative Monto Carlo calculations,   wormholes, decoherence,
perturbative string theory, nonperturbative closed string theory,
matrix models,...  Certainly, we have learned something interesting
from most of these about how to go about trying to construct a quantum
theory of gravity.  I would not want to argue with anyone who
enthusiastically advocates any particular direction. I have my
own directions and I believe that the healthiest thing for the field
as a whole is to let a thousand
theories bloom so that we can all learn from
the success and failures of the different directions.  But it is rather
sobering, in my opinion, to realize
that only one of these has, so far, led to a firm prediction about nature.
This is the black hole evaporation of
Hawking\cite{bhHawking}, discovered in 1974.
It is further, at least to me, worrying, that since 1974 we have come
no closer to uncovering the connection between quantum mechanics,
relativity and thermodynamics, at  whose existence black hole
thermodynamics is  believed to be hinting\cite{bhthermo}.

In this situation, perhaps it is not inappropriate to reflect a little on
what we are trying to do when we work on an approach to quantum
gravity.   What follows are some remarks in this direction.

Thomas Khun, in his irritatingly influencial\footnote{At least in
English
speaking
academic circles.} "The Structure of
Scientific Revolutions \cite{structure}" draws
a distinction between what he calls "normal science" and
"revolutionary science".  Normal science is supposed to be what we
are
doing when we know and accept the basic principles and methods of
a subject (which make up what he calls the "paradigm");
revolutionary
science is what happens when those basic  foundations
change.  It is then commonplace
to say that the first third of this century-the period of the
development
of relativity and quantum theory-was a "revolutionary" period for
physics,
and that since then we have been mostly  engaged in "normal"
science.
As exciting as the rest of the twentieth century has been for
physics,  even in potentially revolutionary fields
such as particle physics and cosmology, over the last
fifty  years
almost
all work has accepted as given the basic principles of relativity
and quantum  mechanics.

Against this background quantum gravity is an anomaly; for this
field is nothing if it is not "revolutionary" science.   From this
point
of view one can, perhaps,  see a basis for the slow rate of progress:
to work on quantum
gravity now may be a mistimed attempt to do
"revolutionary science" during a  period of "normal science."

I believe, however, that this point of view results from too narrow
of
an historical focus.  To get some perspective, let's look at the last
great revolutionary period in science:  the transition from Aristotelian
to Newtonian science that is usually called, for short,
the Copernican revolution.  Two things are most impressive about
this
period, given the modern view of scientific revolutions:
first, how long it took and second, how clueless all of
the major participants were as to the shape of the theory and
world
view that were to be the outcomes of the revolution\cite{copernicanhis}.
The Copernican revolution was only begun by the publication of
{ \it  De Revolutionibus} in 1543,  in the early 1600's Kepler
and
Galileo made great contributions, but the synthesis of Newton
did not appear until 1687.   And,
as great as they
were,  neither Copernicus, Kepler nor Galileo believed any of the
following basic tenants of the Newtonian world view: that the
universe is infinite, that the sun is a star, that there are laws of
physics which govern both motion on Earth and in the heavens
and, moreover, that these laws are deterministic.  Indeed, Kepler
and
Galileo  did not, apparently,
feel themselves very much threatened when Bruno, who advocated
the first two of these ideas, was burned at the
stake in the Campo di Fiori in Rome in1600\cite{koestler-bruno}.

This, perhaps, is  the reason for the long period of transition.
The change from the Aristotelian
world view to the Newtonian world view was perhaps too much,
even for geniuses such as Copernicus, Galileo or Kepler.  At the
same time, some of the
basic features of the new theory could  only be glimpsed
by a
certifiable mystic such as Bruno.
Thus, the change had to
occur over several generations during which the groundwork for the
overthrow of Aristotle was laid by people who were,
in their outlooks and expectations,  mainly Aristotelian
themselves.  Only after that could Descartes
propose a radically new world view and only after that
could  a social outcast in a scientifically out of the
way place-self taught from his readings of Descartes-  carry out
the final steps of the revolution.

With this example in mind, let me suggest that,  rather than
think of ourselves as living in
a period following a great scientific revolution, we should think of
ourselves as living in the middle period of a revolution that is
taking
a long time for the same reason the Copernican revolution took
so long.  We are, perhaps, in the position
of Galileo and Kepler:  we are sure that quantum mechanics and
relativity are more right than Newtonian physics, but all of
our expectations about what physics is and how it works are
at heart Newtonian.  This is as it must be, because quantum
mechanics
and relativity have, neither separately or together, given us a
world
view or a view of science that is coherent and complete enough to
replace the Newtonian views.  For exactly this reason we are
convinced of the need to replace the current situation of having
two,
apparently incompatible, but allegedly fundamental, theories with
one synthesis.  This indeed, {\it is} the problem of quantum gravity.
But
we are having a hard time doing this and I, at least, suspect that
the
reason is that the final synthesis, when it comes, will be as far
from
our expectations as to what a physical theory should be as
Newtonian mechanics was to the expectations of Copernicus,
Kepler and
Galileo.

In this situation, what can we usefully do?  I think, first of all, that
we should lower a little our expectations.  We should stop trying
every five years to invent a candidate for the final theory of
everything.
Let me put forward the proposition that almost anything that we
can now invent, educated as we are mostly in a classical
framework, is
unlikely to be radical enough.  For what changes during a scientific
revolution is not only the answers to questions, but the questions
themselves.  Those brought up on Aristotle,
including Copernicus, Kepler and
Galileo,  were stuck on trying to answer the problem: what is the
shape  of the orbits of the planets.  It never occurred to them that
this
question was to become much less important and that the new
physics
would center around completely different questions: what are the
laws of motion and what are the forces.  Similarly, by
trying
to invent "The" lagrangian and "The" symmetry we are, perhaps,
acting out
of our Newtonian instincts; we are trying to answer the important
questions of the old science.  Mathematics will do us little good if
we have not yet stumbled upon the right new questions; had
Copernicus known Fourier analysis he could have made a much
better epicycle theory (indeed, he could have used it,
there were more epicycles in his
theory than in Ptolemy's) but he never would have hit on the idea
of a law of motion.

The main problem, then, is what to do while we are waiting to
stumble
upon the right questions.  One approach is to try to go out and
reinvent
physics.  This is, indeed, the most likely approach to succeed in the
long run and I believe that we should try to spend as much time
and energy as we can doing this.  But it's hard,
and, except for those rare individuals with strong imaginations and
no
fear of failure, it costs at least as much in terms of our own sense
of
self-confidence as it does in terms of our professional life.   But is
there
then anything else to do?

One thing that we can do is to take
the laws and principles of physics as we have them
and try as hard as we can to  make them
work in this new domain.  That is, let us take quantum mechanics,
as given by Heisenberg, Shroedinger and Dirac, and general
relativity,
as given by Einstein, and try to put them
together, making as few ad-hoc hypothesis
and approximations as we can get away with.
The aim is not to invent a new fundamental theory, but to try to
learn
as much as we can about what the problem with putting them
together
is.  By doing this we may gather clues that
could
help the eventual invention of a new theory.  We may also
learn how to speak about nature in a language which is
completely
consistent with both quantum mechanics and relativity.
By doing so we may find ourselves asking new questions that we
would not have asked had we not first rid ourselves of the
pervasive influence of notions which are clearly wrong.

This is, then, the philosophy of the work that I will be presenting
here.
The approach I will be describing   is
a nonperturbative approach that some
of us have been following for about the last five years.  By
nonperturbative I mean that no use is made of any classical
background metrics or connections, so that all the degrees of
freedom of the gravitational field are treated fully quantum
mechanically\footnote{There is good reason to believe that
any  description of quantum geometry
that is valid at the Planck scale we must be nonperturbative,
because, by the
dimensional character of the gravitational constant, the
conventional
perturbation theory blows up at the Planck scale.   More generally,
a general
conclusion we can draw from all of the work on quantum gravity
from the 1950's
to the present is that perturbative approaches simply don't
work, so that any successful approach to quantum gravity
must be nonperturbative.}.   Thus, this
work differs from some other approaches
to quantum gravity in which one  expands the fields around
a classical background and then quantizes, not the whole
field, but only the deviations from the background.

Indeed, one of the themes of this review will be that the idea that there
is a classical spatial or spacetime geometry is one of the
vestiges of the old Newtonian physics that we have to leave behind if
we are to understand how quantum theory can incorporate gravitation.
Classical metrics and connections may appear as approximate, coarse
grained, descriptions of some quantum states, when one is studying
phenomena on scales
much larger than the Planck scale.  But in a fundamental
theory these can have no place.

The question that any nonperturbative approach must then answer is
what description  of geometry is to replace  the differential
geometry of the classical theory.   The results that are
described in the following chapters address this question.    We find
that consistent application of quantum mechanics to  general
relativity reveals a rather simple picture of quantum geometry.
In this picture, quantum geometry is discrete  at
Planck scales and all physical quantities are topological and
combinatorial.  Local quantities, such as the metric at a point,
no longer have any meaning, but non-local quantities, such as the
areas and volumes of arbitrary surfaces and regions, are well
defined.  Indeed, these quantities are represented by finite operators
whose spectra are quantized in Planck units.  Classical metrics
do play a role, but only as approximate descriptions of the physics
of certain states, at scales much larger than the Planck scale.

These results have been found using
a new formulation of general relativity which is due to Ashtekar
and is called the new, or self-dual, variables\cite{abhay,book,poona}.
It also involves a
new representation for quantum field theories which is called
the loop representation\cite{gambini-loops,carlolee}.   The new
variables and the loop representation have led to a number of
interesting results, for example, large classes of
 exact solutions to all the
constraints of quantum gravity have been found.

The basics of the new variables and loop representation will
be presented in chapter three.  However,  because the new picture which
has emerged of quantum geometry
is so simple, I will begin in the next section to describe
it  without a lot of formalities and equations.
The rest of the review will then be devoted to explaining the details
of the calculations that give rise to this picture and to presenting
other results which illuminate it.

\section{A picture of quantum geometry}

In this chapter a new, and completely nonperturbative, description
of quantum geometry is given.  The calculations that support it
are described in chapter 4;  as we go along a series of footnotes
will tell the reader in which sections to find particular calculations.

\subsection{Preliminary remarks about nonperturbative quantum
geometry}

To confront quantum gravity nonperturbatively, we must first
forget a great
deal
that we have learned from conventional quantum field theory.  This
is because
all conventional quantum field theory depends on a fixed
background
metric.  Thus, the first thing we must do is to identify the  familiar
structures from conventional field theory that
will be absent in a nonperturbative formulation of quantum
gravity or any  diffeomorphism
invariant
quantum
field theory:

{\bf A)  There is no background metric, connection, or any other
structures given on space or spacetime
besides
(at most)  the topological and differentiable structure of a three
manifold.}  Further,
if diffeomorphism invariance is to be maintained no such classical
structures
may
be introduced by the quantization procedure.  This
poses a great challenge to
us as quantum field theorists, because literally all of the
regularization
and
renormalization procedures we have in our toolbox rely on the
presence of a
fixed
background metric.  We thus need to introduce new kinds of
regularization
procedures
in nonperturbative quantum gravity.  In practice, we so far make
use of a
kind of
a compromise: we introduce background metrics for the purposes of
constructing
regulated operators.  Than we insist that any dependence on them
disappears
when
we take the limits in which the regulators are removed.  We must
do this
to insure that  the final,
nonperturbative operators have no dependence on arbitrary
background
structures.

In chapter 4 will see several examples of regularization procedures in
which this
can be carried out.

{\bf B) There are no N-point functions.}
In general there are no observables
associated
with the values of fields at points of the manifold.  This is because
the
diffeomorphism
invariance means that points have no meaning.  For those
used to perturbation theory it is important to stress that in
a nonperturbative
context, diffeomorphism invariance has a rather different effect
than it does in the linearized or perturbative theory.  It has the
full force of its original meaning, which is that the manifold of
points
has no meaning.  The only meaningful objects are the equivalence
classes
of manifolds under all diffeomorphisms.   In the context of
the quantum theory is important
to stress that  diffeomorphisms
must
be thought of as active transformations.  Although their action is
related
to the action of general coordinate transformations, they are not
the same thing.    Under a general coordinate transformation the
points remain the same, but have different labels.  Under a
diffeomorphism points are taken to
different points\cite{hole}.

If a point, $p$, of a manifold has no meaning, then neither can a
field,
$\phi (p)$, evaluated at that point.  As a result, no local observables
are diffeomorphism invariant.  It is then a non-trivial problem
to
construct diffeomorphism invariant observables.  Indeed it is fair
to
say that most of the difficulties presently
faced by nonperturbative quantum
gravity are connected with this issue.

Nonperturbative diffeomorphism invariant quantum field theories
are then going to be very different from conventional quantum
field theories as, in the latter, essentially all observables we
use have to do with
 metrical relations among local observables.  Since there is no
background
metric and there are no local observables these will not exist
in a nonperturbative diffeomorphism invariant theory.
The question is then: what new kinds
of observables are to replace these? One answer, which has
emerged from the recent work on nonperturbative quantum
gravity, and which will be developed at
length below, is that
 all diffeomorphism invariant observables measure
topological relations of non-local observables.

In these remarks I have been refering, implicitly, to the effects
of spatial diffeomorphisms.  But there are also additional
problems associated with diffeomorphisms that change
the time coordinate.  This is because the effect on the fields of
a diffeomorphism that changes the time coordinate is
indistinguishable
from the effect of the Hamiltonian, that gives the change of the
fields under evolution in time.  Indeed, locally, they are generated
by the action of the same object: the Hamiltonian constraint.
Thus, in contrast to ordinary gauge
theories, the gauge symmetry is deeply interwoven with the
dynamics
and the problem of finding spacetime diffeomorphism invariant
observables is a dynamical problem.  This is the essence
of the problem of time in quantum gravity, a deep and central
issue about which a great deal has been
written\cite{problemoftime}.  I
will touch on this problem only in section 6.2.

It may be useful if I describe in a more formal way some of the
distinctions  I have been raising
informally in the last few paragraphs.
In the work that I will be  reviewing here, which is based on
what is called the Dirac quantization procedure\cite{diracproc},
it will be important
to distinguish three stages in the construction of a
nonperturbative quantum theory
of gravity.   The first stage is the
kinematical level,  where we construct the states as a
representation of
a certain algebra of observables which are defined in terms of
easily
accessible local quantities such as metrics and connections.  This
state
space, which will be denoted, ${\cal S}^{kin}$,
 is not physically meaningful, but in what is called the Dirac
approach to the quantization of a gauge theory it is the starting
point
for the construction of the physical states.

The second stage is achieved by
finding those states which can be constructed from the kinematical
states
which are spatially diffeomorphism invariant.  This is done by
constructing the operator that generates spatial diffeomorphisms
within
the kinematical state space and finding its kernel.   The resulting
space
of diffeomorphism invariant states will be denoted ${\cal
S}^{diff}$.

Finally, we must impose also the remaining part of the spacetime
diffeomorphism invariance-that associated with changing the
definition of the time coordinate.  We do this by defining,
in either ${\cal S}^{kin}$ or ${\cal S}^{diff}$, an operator
to represent the Hamiltonian  constraint  that
classically generates reparametrization of the time coordinate.
The
resulting simultaneous kernel of the Hamiltonian and
diffeomorphism
constraints is called the physical state space and is denoted
${\cal S}^{phys}$.

An important issue is how we choose the inner product of the state
spaces
at each of the three levels.  At present this is one of the issues
that is
not settled, indeed, this question is intimately related to the
problem of
time.  However,  in the last two years a definite point of view and
a program for resolving it has developed.  Because of the
importance of
this issue, I will deal with it  separately, in  section 6.2.

Finally, I want to emphasize that  I will be discussing primarily
the  Hamiltonian
approach to quantization.  Because of this there is not only not a
classical metric in the picture, there is not even in any sense a
four dimensional classical manifold.  The four dimensional
spacetime
manifold has gone to the same place in quantum gravity
as the trajectory of the electron in ordinary quantum mechanics:
classical physicists' heaven\cite{heaven}.
What
remains is  a spatial manifold, $\Sigma$, on which there is
given a fixed topological\footnote{Throughout the history of quantum
gravity there has
been speculation about quantum effects changing the topology of
spacetime.
{}From the present point of view, which is based on hamiltonian
quantization
this remains speculation.  The topological and differential structure
of
$\Sigma $ are fixed  because there are no variables that describe
them
in the classical phase space which is the starting point of the
quantization.  To use
the analogy of the electron again, the topology of
$\Sigma$ is fixed for the same reason that the topology of space is
fixed when we do one particle quantum mechanics.  Whether there
is
either a way or a good scientific reason to quantize the topology of
space remains an interesting open problem, which will not be
discussed here.}.
 and differentiable structure.
Associated
with each $\Sigma$ we then have the three  state spaces ${\cal
S}^{kin}_\Sigma$,
${\cal S}^{diff}_\Sigma$ and ${\cal S}^{phys}_\Sigma$

\subsection{Quantum states as functions of loops}

Having set the scene with the discussion of these preliminaries, I
can now return to my goal, which is to describe in simple terms
the state spaces of quantum gravity.   The description I will be
employing is based on a particular representation of the quantum
theory, which is called the loop representation.  I will be saying
much more about it later; for this first look it is enough to know
that in this representation the kinematical states,
${\cal S}^{kin}_\Sigma$, are described as functions over a certain
space of loops.  To describe this adequately  I need
 to say exactly what loops are involved in
these spaces.

Conventionally, the loop space of a manifold, $\Sigma$,  is taken
to be the space of maps, $\gamma $,  from $S^1$ into $\Sigma $ .
We will require that the loops be, in addition, piecewise smooth.
As has been discussed in many places, the space of such maps is
an infinite dimensional differentiable manifold.

However, when we are dealing with the loop
representation we are interested, not directly in these maps, but in
the equivalence classes of the maps under the following three
operations:

i)  Reparametrization invariance.

ii)  Equivalence under retracings:  if $\eta$ is a curve originating
on
a loop $\alpha$, then $\alpha \circ \eta \circ \eta^{-1} \approx
\alpha $,
where $\circ$ means composition of loops.

iii)Inversion:  $\alpha \equiv \alpha^{-1}$.

The space of loops defined modulo these relations will be called
the
space
of nonparametric loops.  It will be denoted\footnote{The {\cal H} stands
for holonomy.
Another way to define the space of loops is to say that two loops,
$\alpha$ and $\beta$ are equivalent whenever their $U(1)$ holonomies
$U_\alpha = exp \oint_\alpha A$ are equal for all connections $A_a$
on $\Sigma.$\cite{abhaychris}}
${\cal HL}_\Sigma$.
It is the quotient of a differentiable manifold
by the operations i) and ii) but it is not, in itself, a differentiable
manifold\cite{gambini-loops}.

It is natural when considering loops with the identifications
i) - iii) to extend the definition to include multi-loops.
Multi-loops are taken to be countable unordered sets of
loops.  We will generally use the word loop to refer either to a
single loop or to a multi-loop, but we will always mean loops
defined modulo the relations i) - iii).

Generally, a loop will be denoted by a lower case Greek letter,
such as $\alpha $, $\beta$ or $\gamma$.  When we need to
we will put on labels to distinguish the single loops (or
components)
of a multi-loop, as in $\gamma_i$, $i=1,...,N$.  I should stress that
the space of loops includes loops with arbitrary intersections,
self-intersections, multiple tracings and nondifferentiable points.
These will
be important for the physics, because certain important operators
act specially at such singular points.

However, in this introductory sketch, we will mostly restrict
attention to loops without intersections or multiple tracings.
These will be called {\it simple loops.}

In the loop representation, the
quantum states are taken to be functions on loops and
denoted $\Psi [\gamma ]$.  The meaning of this loop
representation will become clearer as we use it.

\subsection{Quantum geometry at the kinematical level}

Having  taken care of the preliminaries, I can now describe
the state spaces  of nonperturbative quantum gravity.   I will
begin at the kinematical level.

A good intuitive way to describe a quantum state space is to
give a basis that
consists of  eigenstates of a familiar set of observables.  Thus, one thing
that
one might like to do in quantum gravity at the kinematical level
is describe the basis of eigenstates of the three metric.  This is
exactly what I cannot do here, and for a good reason, which is that
in the representation I am describing there simply is no well
defined
operator which corresponds to the three metric at a point.   The
absence of any operator that measures anything about the metric
{\it at a point} is a key lesson and I will spend considerable time
below\footnote{This is the subject of section 4.1} to convince the
reader that this is the case\footnote{There are,
of course,  other representations, such
as
the metric representation, in which such an operator can be
constructed, at least at the formal level. }.

While  the metric makes no sense at a point, there are other
operators
which do measure metric information that can be defined in the
loop representation.  Moreover these operators, not only exist, they
are, as I will show in detail in chapter 4,
finite when they are defined through a
proper regularization procedure.  In this review I will be discussing
three such operators.

The first measures the area of any given two dimensional surface,
$\cal S$, in $\Sigma$.  It is denoted $\hat{\cal A}[{\cal S}]$.

The second measures the volume of any three dimensional region
$  \cal R$ in $\Sigma$.  It is denoted $\hat{\cal V}[{\cal R}]$.

The third measures the integrated norm of any one form $\omega$
on $\Sigma$.  It is a little funny looking, written in terms of the
classical three metric $q_{ab}$ it is
\f
Q[\omega ] = \int_\Sigma \sqrt{det(q)q^{ab}\omega_a \omega_b }  .
\ff
Note that the square root is a density and is thus integrable.

I will now describe a basis in ${\cal S}^{kin}_\Sigma$ that is
constructed
of simultaneous eigenstates of these three operators.  It is
composed
of characteristic states of loops.  That is, given any loop
$\alpha$,
we associate a state $\Psi_\alpha [\gamma ] $.
In the loop representation
this state has a simple representation when the loops
$\alpha$ and  $\gamma $ are both simple.   In that case,
\f
\Psi_\alpha [\gamma  ] = 1 \ \  {\rm if } \  \gamma = \alpha \ \ \ \
{\rm and  \ \ zero \ \ otherwise.}
\ff
The value of such a state on loops $\gamma $ that are not simple is
more complicated; this is discussed in sections 4.3 and 4.4.

The reader may object that such states seem very formal, we are
defining
a state to be a kronocker delta on the space of loops, which is a
continuous space.  Certainly there are norms we could put on the
state
space with respect to which this state, if it could be defined at all,
would have either vanishing or infinite norm.  Among these is the
Fock inner product, which can be defined in
the loop representation\cite{abhaycarlo-maxwell,selfdual,gravitons}.
Now, the Fock inner product is the appropriate norm to use if one is
defining the state space of Maxwell theory or of {\it free}
gravitons.
But its use makes absolutely no sense in nonperturbative quantum
gravity,
because it depends on a fixed background metric
and thus breaks diffeomorphism invariance.

To construct the non-perturbative theory we
would like to use a norm on the state space
that depends on no fixed background structure and is therefore
invariant under the action of the diffeomorphisms on the space of
loops.  To my knowledge,  there is  only one class of
inner products we can put on
the space of functions over loops that satisfies this requirement.
This
is the class which make use of the  discrete topology on the loop
space.
(The reader may recall that the discrete topology of a point set is
the
one in which every point constitutes an open set.  This
topology, and the corresponding
discrete measure, exists on every continuous topological space.)

We thus define a norm on ${\cal S}^{kin}_\Sigma$ which is given by
\f
\left |   \Phi \right |_{discrete}  =  \sum_\alpha | \Phi [\alpha ]|^2
\ff
where the sum is over all loops $\alpha$ on which the state
$\Phi$ has support\footnote{There are some complications in this
definition in the case that the loop is not simple.  These will
be discussed below in section 4.3 and 6.2.  Nothing we are saying here
will be changed by this.}

States which are normalizable under (3) can have support only on
a countable set of loops.  They can therefore be expressed in the form
\f
\Phi [\alpha ] = \sum_I c_I \Psi_{\gamma_I} [\alpha ]
\ff
where $\gamma_I$ are any countable set of loops, indexed by $I$
and
\f
\sum_I  | c_I|^2 < \infty  .
\ff

This may seem a weird sort of state space on which to base a
quantum
field theory.  However, one of the themes of this review is
that such a state space is adequate to serve as the Hilbert
space for nonperturbative quantum gravity  and for
diffeomorphism invariant quantum field theories in general,
 at the kinematical
level\footnote{This kind of representation has been discussed
earlier
in\cite{rayner,Gtozero} and has recently been studied rigorously by
Ashtekar and Isham\cite{abhaychris}.} .  Moreover,
it is not only adequate, but it has some definite advantages.
I will make,
as we go along, three kinds of arguments for its use.

 First of all,  to demonstrate its adequacy one can show that
it carries a faithful representation of an algebra of observables
that
completely coordinatize the phase space
of the classical theory\cite{abhaychris}.
Second,
it carries an unbroken representation of the spatial
diffeomorphism
group and is,  apparently, the only
kind of representation of the kinematical algebra of
classical observables that does so.

Third,  the space ${\cal S}^{kin}_\Sigma$
with the norm (3) contains within it  states that
are semiclassical, in the sense that all measurements
performed on them give, to a certain degree of approximation, the
same
result as we would obtain with the flat space metric of the
classical
theory.

We will then take the state space ${\cal S}^{kin}_\Sigma$
with the norm (3) as a  basis, at least provisionally
of the kinematics for non-
perturbative
quantum gravity\footnote{I should point out that
while I use the loop representation in this review, everything
that is done in this paper at the kinematical level could be done
in the connection representation, where the states are functions
of the Ashtekar connection $A_a^i$ (see section 3.2). In this case
the characteristic functions of loops are defined as
$\Psi_\gamma [A] \equiv Trexp(G \oint A\cdot d\gamma )$
as in ref. \cite{tedlee}. }.   We may then go on
and consider the problem of diagonalizing
the three observables  $\hat{\cal A}[{\cal S}]$,  $\hat{\cal V}[{\cal
R}]$ and
$\hat Q[\omega ]$  within this space.

We start with the areas.  A large set of eigenstates of
$\hat{\cal A}[{\cal S}]$, for all surfaces, $\cal S$, in $\Sigma$
is given by the characteristic states $\Psi_\gamma$ of simple
loops
defined by (2)\footnote{The details of the construction and
diagonalization
of the area operator is given in section 4.3.  There are
other eigenstates associated with intersecting loops, which
are given in section 4.4}.
The associated eigenvalues are
very simple:
the spectrum is, in fact, discrete!  The eigenvalue of $\hat{\cal A}[{\cal
S}]$,
associated with the eigenstate $\Psi_\gamma $
is equal to $\sqrt{6}$ times the Planck area times the number of
times\footnote{With all
intersections counted positively.}
the loop $\gamma$ pierces the surface $\cal S$.

Thus, in quantum gravity, the area of any surface can only be a
discrete
multiple of the Planck area.  This is, to my knowledge, the first
time that
a quantum theory of gravity gives a simple answer to the simple
question:
what is it that is quantized in quantum gravity?

Furthermore, the area observable allows us to give an
interpretation
to the characteristic states $\Psi_\gamma$.  They are quantum
flux
tubes, analogous   to electric flux tubes in Yang-Mills theory.
However,
they are not carrying flux of electric field-they are carrying
quantum  flux of
area.  That is, each line is carrying a unit of Planck area which
it will contribute   to the area of any surface it
crosses.

The characteristic states (2) are also all eigenstates of the operator
$\hat{Q}[\omega ]$ corresponding to the classical observable
defined
in (1)\footnote{The details of this calculation are in section 4.6}.
The eigenvalue associated with the characteristic state of a loop
$\gamma$ is simply $\sqrt{6}l_P^2  \oint_\gamma \omega $.

 Already with these two operators we can completely characterize
the classical limit of the theory.   Let us
take a moment to discuss this, as the semiclassical
limit in quantum gravity is  somewhat different from
semiclassical
limits taken in quantum field theories with only dimensionless
coupling constants.  The reason is that to define the semiclassical
limit  we must  take into account that  $\hbar$ is in the Planck
length.  This means
that a semiclassical limit in
quantum gravity must involve a limit of large distances,
so that if we make
a measurement that involves a length scale $L$, the quantum result
must agree with the classical one to order $l_P/L$,
where $l_P= \sqrt{\hbar G/c^3}$ is the Planck length.

It is easy to describe states in ${\cal
S}^{kin}_\Sigma$
which approximate a flat metric $h^0_{ab}$
in this sense.
Essentially
all we have to do is to distribute a set of loops, $\gamma_i$ so
that
any  flat surface $\cal S$ in $\Sigma$ is crossed by one of the
loops
on the average of $\sqrt{6}$ times per Planck area-where the
notions
of flatness-and the areas of the surfaces are measured according
to
$h^0_{ab}$.  Such a set of loops will be called a weave.
The weave approximates $h^0_{ab}$ in the sense that for any
surface
whose radius of curvatures are large compared to $l_P$ the state
is an eigenstate of the operator that measures the area of that
surface,
and the eigenvalue will agree with the area according to the
classical
metric $h^0_{ab}$ up to terms of the order of the ratio of the
Planck
area to the area.

The observables $Q[\omega ]$ can
also
be used to demonstrate the correspondence between the weave and
the
smooth metric $h^0_{ab}$.  This will be discussed below in section
4.7.

The semiclassical limit includes not only the classical metric,
but the quantum states of linearized gravitons that propagate
on that classical metric.  Although I will not discuss it here, it
is possible to show that the exact quantum theory can
be linearized in a neighborhood of the state space of a weave
state and that the result is exactly the Fock space\cite{gravitons} of
linearized gravitons\cite{linearization,zegwaard}.

The  weave states then provide a prototype for understanding
quantum
geometry as well as a number of issues related to both the
semiclassical
and short distance limit of quantum gravity.   In order
to discuss this, we must first
emphasize
that, at the non-perturbative level, with no background metric,
there is
no intrinsic notion of distance in the theory.  The Planck length is
a parameter of the theory, but without a metric there is nothing
to tell us what lengths, areas or volumes are Planck scale.
A notion of distance can
only be associated with a particular state of the system.

On the other hand,
most quantum states describe  quantum geometriesthat have no
classical equivalent.  If we consider, for example, the quantum
geometry of a simple unknotted loop, we see that most surfaces
have zero area.  There also do exist surfaces with area of
$n \sqrt{6} l_P^2$, for any $n$.  However, there is no smooth and
everywhere nondegenerate metric that can give this assignment of
areas to the surfaces.

Thus, classical notions of geometry can only be recovered from
special
quantum states that have the property that they approximate a
classical
metric in the sense we have just described.  Using
the
picture that each line carries one Planck unit of area we can
construct
such states by spacing the lines by Planck units according to the
classical
metric we want to recover.

This means that {\it all  states that have semiclassical limits---
that
approximate classical metrics at large scales---necessarily have
discrete
structure at the Planck scale. }  The two things are tied together
because if the
state does not have a classical limit then there is no metric which
enables
us to talk about the Planck length structure.  At the same time
from the
way the states are constructed it is impossible to pack the loops
more tightly than we have describe here because the quantum
operator
${\cal A}[{\cal S}]$
measures areas by counting intersections in Planck units.  If we
put
the loops ten times closer together, according to some
background
coordinate, then nothing really changes, because the metric that
the
state will approximate will not be the original one $h^0_{ab}$, but
one hundred times $h^0_{ab}$.  With respect to this new metric the spacing
of the loops will still be the Planck distance.

Thus, there is no reason for, and no possibility of, taking a limit in
which
the lines of the weave are taken closer and closer together.
Because
the discrete Planck scale structure is essential to the existence
of a classical limit, the states that have classical limits are those
that
involve only countable numbers of loops.  This means that the state
space we have been working in, which are states which are
normalizable
with respect to the discrete norm (3) is completely adequate for
discussing the classical limit of the theory.

We have still said nothing about the third operator mentioned
above: the one that
measures the volume of arbitrary regions.  This operator turns out
to act
nontrivially only on states that have support on intersecting loops.
Thus,
to describe its action I need to say a little more about the role of
the intersecting
loops in the state space ${\cal S}^{kin}_\Sigma$.

We must first introduce some nomenclature to describe the
intersection loops.    Given any loop $\gamma_0$ with
intersections, there
is a finite set of $M$ other loops $\gamma_i$, $i=1,...,M$ which
differ from
$\gamma_0$ by rearranging the routings through each intersection
points.  The
equivalence class of these loops under reroutings is called the
graph $\Gamma$
of the loops $\gamma_i$.

Let us begin by thinking of a particular graph $\Gamma$ with one
self-intersection
point, $p$. Let a finite number, $N$, of lines intersect at $p$.
Associated with this graph are a finite number of loops, which I
will denote $\Gamma_I$, $I=1,...,P$, which differ by changes of the
routing at
$p$.

In the loop representation the value of
the states on these different reroutings are not all independent.
This is
because there are imposed on the states on intersecting loops
a set of relations which code the fact that loops represent
holonomies
of an $SU(2)$ connection.  These relations are:
\f
\Psi [\alpha \cup \beta ] = \Psi [ \alpha \circ \beta ] +
\Psi [\alpha \circ \beta^{-1} ]
\ff
where $\alpha$ and $\beta$ are any two loops that share a common
point, so that their composition may be defined.
These are called the spin network identities because they make the
loop functionals valued on {\it spin networks}, which are discrete
models of quantum geometry introduced by
Penrose\cite{spinnetworks}.

Taking into account these identities, there are then $R <P$
independent
characteristic states associated with the graph $\Gamma$.  We
will label these also by $\Gamma_I$, where from now on the index
$I=1,...,R$ labels independent loops.  These span an $R$ dimensional
subspace of ${\cal S}^{kin}_\Sigma$ which I will denote
${\cal S}^{kin}_\Gamma$.

As I will discuss in chapter 4,
these subspaces play an important role
in the representation of a large class of operators.  The first
example
of these is the volume operator
${\cal V}[{\cal R}]$\footnote{The details concerning the
regularization
of the volume operator are in section 4.5.}.  This operator, first
of all, has a simple action when evaluated on states with support
only
on non-intersecting loops: it annihilates them.    To see what its
action
is on states with support on intersecting loops, let us consider
acting on an arbitrary state $\Phi [\alpha ]$ and then evaluating
the
result on one of the intersecting loops $\Gamma_I$.  The action is
zero unless the region $\cal R$ contains the intersection point $p$.
When it does, the action is given by
\f
{\cal V}[{\cal R}] \Phi [\Gamma_I] =
l_P^3 {\cal M}_I^{\ J} \Phi [ \Gamma_J] .
\ff
where ${\cal M}_I^{\ J}$ is an $R\times R$ dimensional real matrix
whose entries are dimensionless numbers.  That is,
the effect of the volume operator at the intersection is simply to
act with a matrix which rearranges the routings and then multiply
the result by
the Planck volume.

It is easy to generalize this to the evaluation of states on graphs
with an arbitrary number of intersection points.  The result is the
sum of the action (7) at each intersection point
in $\cal R$.    From this we can see that the eigenstates of the
volume
operator are linear combinations of the characteristic states
associated
with each graph and the eigenvalues are all equal to the Planck
volume
times arbitrary sums of the eigenvalues of the rearrangement
matrices.   Thus, just like the areas,
in nonperturbative quantum gravity the volume of
a given region has a discrete spectrum of eigenvalues proportional
to the
Planck volume.

The reader may notice an apparent contradiction between this
result and
the results I described earlier about the eigenstates of ${\cal
A}[{\cal S}]$
and $Q[\omega ]$.  This is that one of the weave states, which
correspond to the flat metric when measurements are made on
scales
larger than the Planck scale, can have zero volume.  This
will be the case when the weave has no intersections.

This is true, but the contradiction is only apparent.  A quantum
state
can indeed
approximate a smooth nondegenerate classical
metric when probed at large scales and still have  zero volume
when it
is measured with the operator ${\cal V}[{\cal R}]$.  This is because
the
latter is a completely microscopic operator, it measures something
about
the connectivity of the weave, while the correspondence to a
classical
metric only measures a long distance property of it.  A zero volume
weave is then something like a fractal geometry:
it gives us a picture of geometry
which is indistinguishable from a classical metric at large scales,
but
if one probes at Planck scales one discovers no metric structure at
all,
just some one dimensional structures associated with the loops.

If the reader still finds this disturbing, let me remind him or her
that
something like this is exactly what must happen if quantum general
relativity is to exist.   If the exact quantum states looked like
semiclassical states built around a classical metric at arbitrarily
short distances, then perturbation theory would be reliable.  But
we know that perturbation theory in fact gives only nonsense.
Conversely,   if, as we have shown, any state that has a
classical
description at large scales has discrete, completely non-classical,
structure at the Planck scale then perturbation theory , which does not
recover this feature, must be unreliable at Planck scales.  Therefore
the short distance structure of the theory cannot be studied
perturbatively and  the fact that the perturbation theory
is nonsense cannot be taken as an argument that the exact quantum
theory does not exist.

This discussion brings us to the end of our introduction to
nonperturbative
quantum geometry at the kinematical level.  We now go on to the
next
stage, which is the nonperturbative description of spatially
diffeomorphism invariant states.

\subsection{Quantum geometry at the
diffeomorphism
invariant level}

One of the early successes of the loop representation is that the
space of diffeomorphism invariant quantum states of the
gravitational
field can be exactly represented\cite{carlolee}.  To construct
this representation,
let us note that an unbroken
representation of the group of diffeomorphisms of $\Sigma $
exists
on the state space ${\cal S}^{kin}_\Sigma$, which is given by
\f
\hat U(\phi ) \Phi [\alpha ] = \Phi [ \phi^{-1} \circ \alpha ] .
\ff
where, here, $\phi \circ \alpha$ denotes the action of the
diffeomorphism $\phi$ on the loop $\alpha$.  One can check that
under the norm (3) $\hat U(\phi ) $ is a unitary
operator\footnote{This
representation is  reducible.  The irreducible representations
are classified by the knot and link classes of
$\Sigma$\cite{ls-perspectives}.}.
The diffeomorphism constraint operators are then
defined as\footnote{It is not hard to show that this
form of the operator is exactly equivalent to the one which is found
by
beginning with the classical expression for the constraint and
translating
it into quantum operators through a suitable regularization
perscription\cite{berndtjorge-onC}}
\f
\hat D (v )  \Phi [ \alpha ] = {d \over dt } U(\phi_t ) \Phi [\alpha
]|_{t=0}
\ff
where $\phi_t $ is a one parameter group of diffeomorphisms
generated
by the vector field $v$.

It is then straightforward to write down the exact solutions to the
diffeomorphism constraints,
\f
D(v) \Psi [\alpha ] =0
\ff
The solutions are that $\Psi$ must be a function of the
diffeomorphism
equivalence classes of $\alpha$.
These will
be denoted as follows:  $\{  \gamma \}$ is the diffeomorphism
equivalence
class of the loop $\gamma $.  The set of these classes are
called
the generalized
link classes of $\Sigma$.

Similarly,  $\{  \Gamma \}$ is the diffeomorphism equivalence
class of
the graph $\Gamma$. These will be called, following earlier
literature,
knotted
graphs.  Just as the case of graphs and loops, given any knotted
graph
$\{  \Gamma \}$ there are a finite set of generalized link classes
$\{  \Gamma_i \}$ which differ by rearrangements of lines through
intersection
points.

The space of diffeomorphism invariant
quantum states of the gravitational field, denoted
${\cal S}^{diff}_\Sigma$  then consists of all functionals
$\Psi [ \{ \alpha \}  ]$.  The generalized link classes are countable,
so this space has a countable basis, given by the characteristic
functions of the classes.  These will be denoted $\Psi_{\{ \gamma
\} }$;
this state is equal to one when evaluated on the class $\{ \gamma
\}$ and
zero when it is evaluated on all other classes\footnote{This is
for diffeomorphism equivalence classes of simple loops.  As at the
kinematical level, for
nonsimple loops the situation is more complicated due to the spin
network relations (6).}.

Unfortunately, although we know the exact state space of the
theory
at the diffeomorphism invariant level, we know rather less about
the
theory at this level than we would like to.   For example,  the exact
form of the inner product on this space is not presently known.   In
addition, at present we only know how to represent a few
observables on this space.

It may seem strange that there is a problem   constructing
spatially diffeomorphism invariant
observables, as an infinite number of them may immediately be
written down at the classical level, by integrating local densities
constructed from the classical fields over $\Sigma$.  Furthermore,
at the quantum level, as we have a countable basis for the state
space
${\cal S}^{diff}_\Sigma$, we may immediately write down an
infinite
number of diffeomorphism invariant quantum operators.  The
problem  is that to construct a physically meaningful quantum
observable
we must construct a relationship between a classical observable
and
a quantum operator.

Unfortunately,  to translate an expression for a
classical observable into a quantum operator, while preserving
diffeomorphism invariance, is a far from trivial operation.  As the
classical expressions for the observables involve products of
fields this
translation necessarily involves a regularization procedure.  During
the
regularization procedure extra structure is introduced which
breaks
diffeomorphism invariance.  The challenge is to remove the
dependence
on this extra structure as one takes the limit in which the
regulator
is removed.

I will mention here two diffeomorphism invariant operators that
we
know how to construct, the details of their construction are given
below in section 4.  These allow us to give a partial
physical interpretation for the states in
${\cal S}^{diff}_\Sigma$.

The first operator is based on extending the definition of surface
areas
in the kinematical theory to an operator that measures the areas of
minimal surfaces.  To describe this I need to consider the case
that
$\Sigma$ has nontrivial $\pi^2$.  Then  let
us consider a noncontractible surface $\cal S$ in $\Sigma$ and
its homotopy class
$\{ {\cal S }  \}$.  We can define, both classically and quantum
mechanically,
a diffeomorphism invariant observable, which is the minimal value
of the areas of
surfaces in $\{ {\cal S }  \}$.  This will be denoted ${\cal A}[\{
{\cal S }  \}]$.
It is clearly a diffeomorphism invariant function of the three
metric.

The operator for the area of a minimal surface is constructed by
taking
the operator for the area of a surface I described above and
minimizing
it over all surfaces within the homotopy class $\{ {\cal S}  \} $.
Given this, it is not hard to see that  the eigenstates of the
minimal surface area
operator are exactly the characteristic states,
$\Psi_{\{ \gamma \} }$
of generalized link classes.  The eigenvalues
are given by $\sqrt{6} l_P^2$ times the minimal number of
intersections (without
regard to sign so that all intersections are counted positively) of
the loops
in $\{ \gamma \}$ with the surfaces in $\{ {\cal S} \}$.

There are two interesting things about this result.  First, the
result is essentially
topological: the area of the eigenstate is gotten by counting the
intersections of
the loops associated with the eigenstate with the surfaces.  This is
an example
of a general situation which I noted above:
a property which is metrical classically is translated into an
operator which measures a topological property of the label of the
quantum mechanical state.    Second, as in the kinematical theory,
the spectrum is quantized.  The minimal area of
any class must be an integral multiple of $\sqrt{6}$ times the
Planck area.

Thus, we see that as in the kinematical case, the lines of the link
classes carry quantized area, in the sense that they contribute a
Planck area to the minimal area of any class of surfaces that they
minimally intersect.  To build up a diffeomorphism invariant
quantum
geometry that is large compared to the Planck scale, we must have
a complicated link class involving lots of lines.  For example, the
simplest  quantum geometry  we can build in the three torus
consists of unknotted
loops wrapped around the three $S^1$'s.  The areas of the minimal
surfaces wrapping each of the three two toruses are given by the
numbers of the loops in each direction in Planck units.

We have already introduced the second diffeomorphism invariant
operator,
it is simply the volume of the whole manifold $\Sigma$.  The
action of this operator, given by (7), is diffeomorphism invariant, so
it
can be extended directly to the states in ${\cal S}^{diff}_\Sigma$.
For
example, its action evaluated on graph classes with one
intersection
is given by the extension of (7):
\f
\hat {\cal V}[\Sigma ] \Phi [\{ \Gamma_I \} ] =
l_P^3 {\cal M}_I^{\ J} \Phi [ \{ \Gamma_J \} ] .
\ff
where $ {\cal M}_I^{\ J}$ are the same rearrangement matrices as
in
the kinematical case.  As in that case, this result extends trivially
to graph classes with more than one intersection, one gets
the sum of rearrangement matrices acting at each intersection.
The eigenstates then consist of
linear combinations of the characteristic states of the generalized
link classes $\{ \Gamma_I \}$ and are found by diagonalizing the
rearrangement matrices.  Further, we see that the volume of the
Universe is quantized:  its eigenvalues are sums of all the possible
eigenvalues of the rearrangement matrices taken in Planck units.
Thus,  the volume of the universe in an eigenstate is
roughly proportional to the number of intersections in the
graph that labels that state.

Four comments before we go on to physical states:  First, again as
in the kinematical case, at the quantum level the volume of the
universe
and the area of minimal surfaces are decoupled, in a
way that
indicates that the area is a more macroscopic measure of the
quantum
geometry, while the volume is more microscopic.  Second, we see
that acting on the space ${\cal S}^{diff}_\Sigma$ of
diffeomorphism
invariant states the volume operator is block diagonal: it is
represented
by simple matrices of numbers acting in each of the finite
dimensional
blocks subspaces associated with different rearrangements of
routings
within a graph class.  This is an example of a general property of
diffeomorphism invariant operators that are local in the sense that
they are constructed by a single integral over a density:  all these
operators are either block diagonal or near block diagonal, in the
sense
that they mix one graph class with graph classes that differ by the
addition or subtraction of a finite number of loops.

Third, note that the diffeomorphism  invariant states
in ${\cal S}^{diff}_\Sigma$ are not normalizable with respect to
the
kinematical inner product (3).  This is a general feature of Dirac
quantization, the states in the kernel of a constraint are not
normalizable
in the kinematical inner product.  This necessitates the choice of
new inner products at the diffeomorphism invariant and physical
levels.  This will be discussed in more detail in section 6.2.

Finally,  all of the operators I have described here are
both finite and background independent (by which I mean
independent
of the background structure  that must be used
during the regularization procedure.)  The reader will see
in the details of the constructions of the operators described
in chapter 4 how this comes about.  However, it is interesting
to mention that in general, while a finite operator need not
be background independent, it is the case that a background
independent operator must always be finite.  This is because
the regulator scale and a background metric are always
introduced together in the regularization procedure.  This is
necessary, because the scale that the regularization parameter
refers to must be described in terms of some metric and, since
none
other is available, it must be described in terms of a background
metric or coordinate chart
introduced in the construction of the regulated operator.  Because
of this the dependence of the regulated operator on the cutoff,
or regulator, parameter, is related to its dependence on the
background metric (this can be formalized into a kind of
renormalization group equation \cite{carlo-rg}.)  When one takes the
limit of the regulator parameter going to zero one isolates the
nonvanishing terms.  If these have any dependence on the
regulator parameter (which would be the case if the term
is blowing up) then it must also have a dependence on the
background metric.  Conversely, if the terms that are nonvanishing
in the limit the regulator is removed have no dependence on the
background metric, {\it they must be finite.}

This point  has   profound
implications
for the whole discussion of finiteness and renormalizability of
quantum gravity theories.  It means that any nonperturbative and
diffeomorphism invariant construction of the observables of the
theory must be finite.  A particular approach could fail in that
there
could be no way to construct the diffeomorphism invariant
observables as quantum operators.  But if it can be done, without
breaking diffeomorphism invariance, those operators will be finite.

\subsection{Physical states}

So far everything I have described is based
on only two facts about general relativity, first that the gauge
symmetry includes spatial diffeomorphism invariance and second,
that there is a representation in which the
states of the theory are functions of loops.  As I will discuss in
the next section, such a representation exists whenever the theory
has a canonical coordinate which is a connection.  But I have not
yet used anything about the dynamics of general relativity.  Thus,
all of the results I have described so far, including the existence
of the classical limit and the quantization of areas and volumes
should be true in a wider class of theories.

The real surprise of the whole development, so far, is that the
Hamiltonian
constraint has a simple action in the same representation in which
the diffeomorphism constraints can be solved.  This, as far as I
understand it, did not have to happen and exactly why it happens is
not really satisfactorily understood.

Be that as it may, the basic fact is that acting on states in
${\cal S}_\Sigma^{kin}$ or ${\cal S}_\Sigma^{diff}$
the action of the Hamiltonian constraint
is concentrated at intersections of loops.
The result is that in ${\cal S}_\Sigma^{kin}$ one can find an
infinite
dimensional space of exact solutions to the Hamiltonian
constraint\cite{tedlee,carlolee}.  These include
an infinite space of solutions which
consists of all
the states have support on loops that have no intersections.   There
are, in additional, a large number of states with support
on intersecting
loops\cite{tedlee,viqar-intersects,berndtjorge-intersects,BGP}.
Further, all of these solutions spaces
are diffeomorphism invariant, so that we can simultaneously
solve the Hamiltonian and diffeomorphism constraints.  The
result is that we have, explicitly, an infinite dimensional
space of exact physical states of the theory.

These results will be further described
in section 6.1 below.  There I will also describe
some very recent work concerning the form of the Hamiltonian
constraint in the loop representation that leads to new
(as of 1991) kinds of
solutions to all of the constraints.

 \section{Basics of the loop representation and the Ashtekar
variables}

The purpose of this chapter is to give
the  basics which are necessary to understand the
derivations
of the results just described.
The reader desiring a more detailed introduction should consult one
of
the reviews\cite{book,poona,carlo-review,gary-us} or the
original
references cited\footnote{The reader familiar with the loop
representation will find a few new things in the presentation of
the next section.}.

\subsection{The loop representation}

The loop representation is a representation, in the sense
of Dirac\cite{diracproc},
of
a gauge theory in which the Yang-Mills gauge invariance is
automatically
implemented.  In the loop representation quantum states
are represented as functions of sets of loops, of the form
$\Psi [ \gamma , \alpha , \beta  ]$, where the loops
$\alpha , \gamma ,\beta ,...$, live in the space ${\cal HL}_\Sigma$
defined in section 2.2.  It can be constructed for
any quantum
field theory which can be expressed in a form in which the
canonical
coordinate is a connection.  The loop
representation was invented independently by Gambini and
Trias\cite{gambini-loops} for Maxwell and Yang-Mills theory and by
Rovelli and the author for
quantum general relativity\cite{carlolee}.  It differs from many of
the representations
of quantum theories we are familiar with in that, as there is no
classical
variable corresponding to the position of the loops,  it is not
expressed as
a space of functions over a configuration space.   In this sense it is
analogous to using the energy eigenfunctions for the hydrogen
atom as a basis for one particle
quantum theory, in that there are no classical
variables associated with $n,l$ and $m$.

Let us consider  a theory based on a Yang-Mills
connection $A_a$ and
a conjugate electric field $\tilde{E}^a$.  We assume that these
satisfy the standard Poisson bracket relation\footnote{$i$ and $j$ denote
internal
Yang-Mills indices while $a,b,c...$ denote spatial indices},
\f
\{ A_a^i (x) ,E^b_j (y) \} = \delta^b_a \delta^i_j \delta^3 (x,y)   .
\ff

There are two ways to construct the loop representation of such
a theory.  The first
is to construct a transform from a representation in which the
states
are functions of the connection $A_a$.   The basic form for the
transform is,
\f
\Psi [\gamma ] = \int d\mu [A] \overline{T[\gamma ,A]} \psi [A]
\ff
Here, $\psi [A]$ is the state in the connection representation and
$T[\gamma ,A] $ is the trace of the holonomy of the connection $A$
around the loop
$\gamma $.  This is commonly denoted the Wilson loop variable
by particle physicists, it can be written
\f
T[\gamma ,A] = Tr U_\gamma (0,1)
\ff
where
\f
U_\gamma (s,t) = P e^{ c \int_s^t A \cdot d\gamma  }
\ff
is the parallel transport of the connection along the loop $\gamma$
from the point $s$ to $t$.   Here $c$ is the
coupling constant of the theory.  For example, in the
case of general relativity, $c$ is  Newton's  gravitational constant,
$G$\footnote{Since $G$ has dimensions, this means that the
connection, $A_a^i$, in the case of gravity does not have
the standard dimensions of inverse length.  We will
see that this is an important
fact for all of the developments of chapter 4.}.

Note that in (13) we have written the transform
for the case that we evaluate $\Psi $ on a single loop $\gamma $.
In
the general case in which the argument is a set of loops
$\{ \alpha , \beta , ...\}$ the kernel of the transform is the
product of the Wilson loops variables for the loops.

In the definition of the transform $d\mu [A]$ is an integration
measure
on the space of connections.  In order to turn this equation from a
formal statement into an actual definition of the loop
representation
it is necessary to give a precise definition of this measure.  It
turns
out that in a number of cases this can be done and the loop
representation
constructed completely from (13).

In the last several years the loop representation has been studied
in several
cases in which the quantum field theory
is already well understood.  These include
Maxwell theory\cite{gambini-loops,abhaycarlo-maxwell,selfdual},
linearized quantum gravity\cite{gravitons}
Yang-Mills theory  on a lattice\cite{latticeloops},  abelian
Chern-Simon theory\cite{maoli}
and $2+1$ gravity\cite{2+1-us}.  In each of these cases the loop
representation is
found to be
equivalent to the standard representation in which the states
are functions of the connection.  There can thus be no doubt that
the loop representation is a well defined and reliable approach to
the quantization
of field theories involving connections\footnote{Perhaps of  interest also
to particle physicists is some recent
work
which uses the loop representation as the starting point for an
alternative approach to solving lattice $QCD$
numerically\cite{latticeloops}.
The main idea is that
for a fixed lattice one can truncate the representation by cutting
the
loops off at some large, but finite size.  The Hamiltonian of $QCD$
is then represented as a sparse matrix, which is sparser and
sparser
the larger the set of loop considered.  For small lattices, and in
the $2+1$ dimensional theory, the Hamiltonian can then be
diagonalized
by sparse matrix methods\cite{latticeloops,gambini3+1fermions}.
The results are equivalent to those
found
by other methods.  Not surprisingly, it is clear that for weaker
and weaker coupling it is necessary to truncate the theory at larger
and larger loops to have a good approximation.   For the real case of
$3+1$ $QCD$,
with coupling in the scaling region, it is necessary to use
a more sophisticated algorithm to construct a good truncation of
the Hamiltonian.  Work on this direction is
proceeding and it is not
out of the question that this could lead to a method that is
competitive
with Monte Carlo methods for the solution of real problems.  This
is
especially true as the inclusion of fermions is not such a great
problem
for these methods as it is for the Monte Carlo procedure.  For
example,
calculations based on the loop representation have recently been
done for $QED$ with fermions in $3+1$
dimensions\cite{gambini3+1fermions}.}.

The case of Maxwell theory is particularly instructive. There it is
possible to explicitly compute the transform and express the
full Fock space of photons as functions $\Psi [\gamma ]$ on the
loop
space\cite{abhaycarlo-maxwell}.
I will not describe the process, but I will write down the
answer  because it is so simple.  The vacuum is given by the state
$\Psi [\gamma ]=1$.  A state with one photon of momentum $p^a$
and polarization $m^a$ is given by
\f
\Psi_{p,m} [\gamma ] = \oint ds e^{\imath p_a \gamma^a (s)}
\dot{\gamma}^a (s)  m_a
\ff
Multi photon states are given by products of these factors.
Furthermore, the inner product can be expressed simply in terms of
these factors.

The second way to construct the loop representation is to come to
it
by seeking representations of a certain algebra of classical
observables.
This algebra is called the loop algebra.  It consists of the $T$
variables
that I defined above in (14) and a complementary set of variables
that
involve both the connection and the conjugate electric field $E^a$.
These variables are parametrized by both a loop
and a point on the loop.  They are defined by
\f
T^a[ \gamma ](s) \equiv
Tr [ \tilde{E}^a (\gamma (s) U_\gamma (s,2+2\pi ) ]   .
\ff
It is then not hard to show
that,
given   (12),
the variables $T[\gamma ]$ defined by (2) and $T^a [\gamma ](s)$
defined by (17) have a closed algebra.  The basic Poisson
bracket is given
by
\f
\{ T[\gamma ]  ,T^a [\alpha ] (s)   \}   = c \int dt
\delta^3 (\alpha (s), \gamma (t)) \dot{\gamma}^a (t)
\left [ T[\gamma \circ \alpha ] - T[\gamma \circ \alpha^{-1} ]
\right ]
\ff
where $\circ$ means to combine the loops at the intersection
point.  (The
expression vanishes if there is no intersection point.)
This algebra is, furthermore, a
complete algebra, in the sense that any gauge invariant function
on the phase space of $A_a$ and $E^b$ can be expressed in terms of
them (alternatively, it can be shown that except for points of
measure
zero, they coordinatize the physical phase space\cite{GLS}).  Thus,
it is
possible
to take this algebra as the starting point for the quantization of
the
theory rather than the standard canonical algebra (12).    This can be
done for the cases already mentioned and the result is, in each
case,
equivalent to the representation constructed by means of the
transform (13).

It is easy to write down the operators that represent $T[\alpha ]$
and $T^a [\alpha ](s)$ in the loop representation.  They are defined
by
\f
\left ( \hat{T} [\alpha  ] \Psi \right ) [\gamma ]
\equiv \Psi [\alpha \cup \gamma ]
\ff
\f
\left ( \hat{T}^a [\alpha ] (s) \Psi \right ) [\gamma ] \equiv \hbar c
\int dt
\delta^3 (\alpha (s), \gamma (t)) \dot{\gamma}^a (t)
\left [ \Psi [\gamma \circ \alpha ]
- \Psi [\gamma \circ \alpha^{-1} ] \right ]
\ff
In equation (19) $\cup$ stands for union in the set of loops.  It is
because
of this definition that the loop states take values on sets of loops.
(19) tells us that the Wilson loop is represented by an operator that
is a
kind of a shift operator in the space of sets of loops.

For the calculations we will do in the next chapter, we will find it useful
to employ two elaborations of notation concerning the loop
representation.

The first elaboration concerns  a pretty geometrical expression
of the loop algebra (18).
The expressions (18) and (20) contain distributional factors.  This is
just like the local expression (12), the difference is that the
distributional
factors have support on curves rather than points.  The remedy is the
same as in ordinary local field theories, at both the classical and the
quantum mechanical level the expressions are to be integrated against
smooth test functions.  There is a very pretty way to do this, which is
the following.   As the distributional factors in (18) and (20)
are already one
dimensional, the integral over test functions must be two dimensional.
It is natural to make this an integral over a surface.  This can be done
very naturally, because the $\tilde{E}^{ai}$ is  dual to a two
form.  We may define,
\f
E^*_{bc} (x)\equiv \epsilon_{abc} \tilde{E}^{ai} (x)  .
\ff
and, from (17),
\f
T^*_{bc}[ \gamma ](s)  \equiv
Tr [ \tilde{E}^*_{bc} (\gamma (s) U_\gamma (s,s+2\pi ) ]   .
\ff

Let us have a one parameter continuous family of loops
$\gamma^a(s,u)$
where $u \in [0,1]$ and for each $u $,
$\gamma^a(s)_{u}\equiv {\bf \gamma}^a (s,u )$ is a closed loop.
We call such a family a strip, which we will denote with a hat, as in
$\hat{\gamma}$.
$s$ and $t$
then coordinatize the two dimensional surface of the strip.  We may then
define an observable
associated the strip
$\hat \gamma$,
\f
T^1[\hat{  \gamma }] \equiv \int du \oint ds
{\partial \gamma^a \over \partial s}
{\partial \gamma^b \over \partial t} T^*_{ab} [\gamma_u](s)
\ff
The $1$ refers to the fact that one factor of $\tilde{E}^{ai}$ is inserted
in the trace.

The Poisson bracket of this observable with the holonomy, $T[\gamma,A]$
is expressed in terms of
the intersection
number
of the loop $\alpha$ and the surface $\cal S$.  Denoted,
\f
I[{\cal S} , \gamma ] \equiv \int d^2 S_i^{ab} \oint d\gamma^c
\delta^3( S , \gamma ) \epsilon_{abc}
\ff
this is equal to
$\pm 1$, depending on orientation, if the loop intersects the surface,
and
zero otherwise.    It is then true that,
\f
\{ T [\gamma ]  ,T^1 [\hat{  \alpha}]   \}   = c I[ \hat{ \alpha },\gamma ]
\left [ T[\gamma \circ \alpha^* ] - T[\gamma \circ (\alpha^*)^{-1} ]
\right ]
\ff
where $\alpha^*$ is the curve in the strip that intersects $\gamma$.
This gives us a geometrical representation of the quantum operator.
\f
\left ( \hat{T}^1 [\hat{  \alpha} ]  \Psi \right ) [\gamma ] =
\hbar c
 I[ \hat{\alpha },\gamma ]
\left [ \Psi [\gamma \circ \alpha^* ]
- \Psi [\gamma \circ (\alpha^{*})^{-1} ] \right ]
\ff
This strip regularization of the loop algebra is
a prototype for the background independent regularizations that will
be introduced in chapter 4.

The second elaboration of notation involves employing a dual basis
to the representation space to write
\f
\Psi [\alpha ] = <\alpha | \Psi >
\ff
Here $<\alpha |$ is a basis in the dual space of loop functionals, which
is parametrized by the loops.   By definition, these basis states must
satisfy the identities i) to iii) introduced in section 2.2, so that
$<\alpha| $ is parametrization independent and
\f
<\alpha \circ \eta \circ \eta^{-1}| = < \alpha |
\ff
and
\f
<\alpha | = < \alpha^{-1}|.
\ff
In addition, for the application to gravity or $SU(2)$ Yang-Mills
theory we impose the $SU(2)$ spin network relations,
\f
<\alpha \cup \beta | = <\alpha \circ \beta | + <\alpha \circ \beta^{-1}|.
\ff
We may note that, by (28) $\alpha$ and $\beta$ are always equivalent to
loops with a common base point (by expressing $\beta$ as
$\beta \circ \eta \circ \eta^{-1}$, with $\eta$ an arbitrary segment
that connects to $\alpha$.)  Thus, the relation (30) is always meaningful.

The relations satisfied by the states $<\alpha|$ may be summarized
by saying that a linear combination $\sum_i c_i <\alpha_i | =0$
whenever it is the case that $\sum_i c_i T[\alpha_i,A ]=0$ for
all values of the $SU(2)$ connection $A_a^i$ \cite{abhaychris}.

We may then express the defining relations of the loop operators
in terms of the loop basis,
\f
<\gamma | \hat{T} [\alpha  ]
\equiv < \alpha \cup \gamma |
\ff
and
\f
 <\gamma | \hat{T}^1 [\hat{  \alpha} ]   =
\hbar c
 I[ \hat{\alpha },\gamma ]
\left [ <\gamma \circ \alpha^* |
- <\gamma \circ (\alpha^{*})^{-1}| \right ]
\ff

We may note that this representation on the dual space is then a cyclic
representation, as all states may be built up by repeated application
of $\hat{T}[\gamma ]$ acting on the state $<\cdot|$, where $\cdot$
stands for the trivial loop in which all of $S^1$ is mapped to an
arbitrary (by (28)) point in $\Sigma$\cite{abhaychris}.
Then the representation may be
defined by the algebra (25), together with the relation
\f
<\cdot | \hat{T}^1 [\hat{  \alpha} ] =0
\ff

 It is important to stress that the definition of  the dual basis
uses only the linear structure of the state space and  does not
involve the inner product.   Without the inner product, we do not know
what ket state $|\alpha >$ corresponds to the dual bra state
$<\alpha |$.  Thus, without the inner product, we cannot write
$<\alpha |\beta > \equiv \Psi_\beta [\alpha ]$\footnote{On the other hand,
in the connection representation, we know how
to write the bra state corresponding to a single
loop $\alpha$, it is\cite{tedlee}
\f
\Psi_\alpha [A] \equiv <A|\alpha > = T[\alpha ,A]  .
\ff
Written in terms of Dirac notation, the transform (13) is
\f
<\gamma |\Psi > = \int d\mu[A] <\gamma |A><A|\Psi> .
\ff

Thus, we see that the dual space to the loop representation, consisting
of states of the form $<\Psi | =\sum_i c_i <\alpha_i| $ corresponds to the
connection representation states of the form
$\Psi [A] =\sum_i c_i T[\alpha_i ,A]$.  Indeed, at the kinematical level,
all of the results we report in the next chapter for the dual basis of
the loop representation are true also for the corresponding states in
the connection representation.}.

The existence of a quantization based on the loop algebra (18) (or
(25)) raises
questions
about the relationship between this
quantization   and  the conventional
quantization based on the canonical algebra (12).  A rigorous study
of this question
has recently been carried out by Ashtekar and Isham for the case
of Maxwell theory and $SU(2)$ Yang-Mills theory\cite{abhaychris}.
Their
results are
not simple to state completely without using the language of the
rigorous
representation theory.  However, simply put, they find
that there exist several kinds of representations of the algebra (25)
as  function spaces on the loop space ${\cal HL}_\Sigma $.
Some of these
are equivalent to  the
standard Fock representation of Maxwell theory, but others are not
equivalent to previously known representations.
In the case of the representations
which are equivalent to Fock representations,
one can show\cite{abhaychris}
that the equivalence is expressed by an expression of the form of the
transform  (13).

Among the representations that Ashtekar and Isham show are
not equivalent to the Fock representation are those based on the
discrete norm  (3).  As the reader may check,
the operators $\hat{T}[\alpha]$ and $\hat{T}^a [\beta ](s)$ are well
defined through (19) and (20) on the space ${\cal S}^{kin}_\Sigma$ with
the norm (3).  This is the basis for my claim there that the full
kinematics of the theory is well
represented on that
space.

To summarize, in Maxwell theory, linearized general relativity,
Yang-Mills theory and $2+1$ gravity
the loop representation has been constructed and in each case it has
been found to give
results
that are equivalent to the results
found by other methods.  Further, in each of these cases
the connection
representation is also known and the loop representation can be
constructed
by means of the transform.

The reader encountering all of this for the first time will no doubt
be
wondering what Yang-Mills fields have to do with gravity.  This is
explained in the next section.

\subsection{The Ashtekar variables}

When a person first encounter the Einstein equations, they often
have two
strong impressions.  The first is how beautiful they are.  The
second,
which comes about as soon as one tries to find a solution, or
compute
something, is how complicated they are.  This seems, somehow,
unfair.
One feels that the geometrical beauty behind the equations should
serve some useful purpose when we try to do physics with the
theory.

This impression is even more strongly born out when one realizes
that,
contrary to the case with many other nonlinear partial differential
equations, a great many exact solutions to EinsteinUs equations
have
been found.  Indeed, although the theory is, presumably, not an
integrable
system, there exist several different restrictions of the theory
that
lead to integrable systems.  Two of these are the restrictions to
spacetimes with two symmetries (two killing fields) and the
restriction
to self-dual solutions.

Einstein's equations and their solutions thus have a lot of
mathematical
structure.   We may then ask, if the space of solutions has non-
trivial
structure, should this not play a role somewhere in the quantum
theory?
In order to investigate this question, it is necessary first to
express the
classical theory in a set of variables in which the structure of the
space
of solutions is more apparent.  This was the idea that led Ashtekar
to look for, and find, what are called the new variables.

Now, we know that the self-dual Einstein equations can
be solved exactly  and the space of solutions characterized in
terms of
certain free data  by using the twistor theory of
Penrose\cite{roger-twistors} and the
related heavenly methods of Newman\cite{heaven}.    Can we
use this information
somewhere in the construction of the quantum theory?  To do this
we
should express the theory in a language in which the solvability of
the
self-dual sector is more transparent.  This may be done in the
following
way.

Consider the Palatini variational principle for the Einstein
equations,
\f
S(e, \Gamma )= \int \epsilon_{ijkl} \ e^i \wedge e^j
\wedge R^{kl}(\Gamma )  .
\ff
Here $ e^i$ is the frame field, $\Gamma^i_{\ j}$ is the $O(3,1)$
connection, whose (Yang-Mills) curvature is $R^{kl}$.  I am using
form notation, the explicit indices, $i,j,k=0,1,2,3$,
are internal $O(3,1)$ indices.
The metric, $g_{\alpha \beta }$ (where now $\alpha, \beta $ are
the four dimensional spacetime indices) , is obtained from the
$e^i$ by
\f
g_{\alpha \beta} = e^i_\alpha e^j_\beta \eta_{ij}
\ff
where $\eta_{ij}$ is the Minkowski metric.

It has been known for a long time that EinsteinUs equations are
recovered if one varies both $e^i$ and $\Gamma^i_{\ j}$ in (36) .
What has not been known is that, actually we need only
half of this action in order to get the Einstein equations.  Given
any antisymmetric pair of internal Lorentz indices, such as those
on $\Gamma^{ij}$, we may split them into self-dual and
antiself-dual parts, by
\f
A^{\pm}_{ij} \equiv {1 \over 2}
\left ( \Gamma_{ij} \pm \imath \Gamma^*_{ij}   \right )
\ff
where $\Gamma^*_{ij} \equiv {1 \over 2} \epsilon_{ijkl}
\Gamma^{kl} $.
$R_{ij}$ may also be split,
\f
F^{\pm}_{ij} \equiv \left (  R_{ij} \pm \imath R^*_{ij} \right ) .
\ff
It then follows from the fact that the Lie algebra of
$SO(3,1)$ splits
into two commuting $SU(2)$ subalgebras that $F^+_{ij}$ is a
function
only of $A^+_{ij}$ (and similarly for the minus ones).  We may then
consider
taking only the self-dual half of the action (36)
\f
S^{sd} (e, A^+ )= \imath \int \epsilon_{ijkl} \, e^i \wedge e^j
\wedge F^{+ \ kl} (A^+ )  .
\ff
Now, this is a complex action.  But, as is not hard to show, it leads
to the same field equations as the Palatini action
(36) \cite{action}.
Actually this
is not quite true.  It leads to the same field equations under the
assumption that the frame fields $e^i$ have non-vanishing
determinant\footnote{The fact that new  variables are only equivalent
to the
standard formulation of general relativity when the $e^i$
are nondegenerate is important for some of the developments I will
be describing in chapter 5.}, so that a
non-degenerate metric can be formed from them
by (37).

The field equations that follow from (40) are
\f
{\delta S^{sd}  \over \delta A_{i}^{\ j} } =
\left ( \epsilon_{ijkl} {\cal D} ( e^k \wedge e^l )  \right )^+=0
\ff
\f
{\delta S^{sd}  \over \delta e^i } = 2 \epsilon_{ijkl} e^j \wedge F^{+
\ kl}  =0
\ff
In the first equation $\cal D$ is the gauge covariant exterior
derivative and the
overall $+$ means to take the self-dual part.  It is straightforward
to show that
the equation then says that $A^+$ is the self-dual part of the
Christoffel connection
of the $e^i$.  The second equation then is equivalent to the Einstein
equations.  This
can be seen because when the first equation is used, its imaginary
part vanishes by
virtue of the three index antisymmetric identity of the curvature
tensor\cite{action}.

Now, we wanted a formulation of the theory in which the
simplification of the
theory when it is restricted to the self-dual sector is manifest.  It
is easy to
restrict these equations to the self-dual sector, that can be done
by  simply
setting $F^+ = A^+ =0$ (at least locally).  But then the field
equations become
\f
\left ( \epsilon_{ijkl} d (e^k \wedge e^l ) \right )^+=0
\ff
It is not hard to show that these are equivalent to the self-dual
Einstein equations \cite{oldselfdual,robinson}.

Given the action, there is a standard procedure to construct the
Hamiltonian
formulation of  a diffeomorphism invariant theory.  This
may be done by choosing a spacelike surface, $\Sigma$, in the spacetime,
and then constructing a canonical formalism to generate the
changes
of the fields that come from evolving that surface in the
spacetime.  I will
give here only the results of applying this  procedure to the
selfdual
action (40) \cite{action}. The canonical coordinate
of the theory turns out to be the self-dual connection $A^+$,
or rather its pullback into $\Sigma$.  I will denote that simply by
$A_a^i$, where $a,b,c,..$ will from now on refer to spatial indices
in
$\Sigma$ and $i,j,k$ will be $SU(2)$ indices.  I should note that
this
$SU(2)$ is the left handed (or self-dual) part of $SL(2,C)$, not the
spatial
rotation group.   So $A_a^i$ is still the spacetime connection (or
its
spatial components) and its curvature, denoted  $F_{ab}^i$, is the
self-dual piece of the {\it spacetime} curvature tensor, pulled back
from spacetime into the three manifold $\Sigma$.

The fact that there is a canonical formalism in which all of the
components of the left handed part of the {\it spacetime}
connection
commute with each other is the basic discovery of Ashtekar that
makes
everything that follows possible\cite{abhay}.
The canonically conjugate field
to $A_a^i$ turns out to be the pullback into $\Sigma$  of
$e^j \wedge e^k \epsilon_{ijk}$.  This is easy to see from (40), it is
the coefficient of the time derivative of $A_a^i$.  It is convenient
to raise its spatial index, so that we have a variable,
\f
\tilde{E}^a_i \equiv \epsilon^{abc} \epsilon_{ijk} e^j_b e^k_c
\ff
where $e^j_b$ denotes the pullback of the one form $e^j$ into $\Sigma$
and
$\epsilon^{abc}$ denotes
the
Levi-Civita {\it density}.  Thus, $\tilde{E}^a_i $ is a triplet of
vector
densities.  (In relativity it is conventional to denote densities by a
tilde, with one tilde for each weight.)   All the components of
$\tilde{E}^a_i  $ commute with each other, moreover looking at the
action we now
see that the only occurrence of time derivatives is in
\f
S^{sd}= \imath \int d^3 x  dt  \, \tilde{E}^a_i  {d A_a^i \over dt}  + \
...
\ff
so that the canonical commutation relations are,
\f
\{  A_a^i (x) , \tilde{E}^b_j (y) \} = \imath \delta_a^b \delta^i_j
\delta^3(x,y)
\ff
The $\imath $ is important, it says that it is really the imaginary
part of $A_a^i$ that fails to commute with $\tilde{E}^a_i $.

Recall, from
the definition (38) that for real spacetimes $A_a^i$ is complex. However, it
is not an arbitrary complex connection, by its definition it is a certain
function of the frame fields $e^i$ and their derivatives.  From the
Hamiltonian
point of view this is embodied in a set of conditions which are called the
reality conditions.  They are conditions on the $\tilde{E}^{ai}$ and
$A_{Ai}$ that guarantee that the three metric $\tilde{\tilde{q}}^{ab}$ and
its time derivatives are real.  These are a set of fourth order polynomial
equations \cite{poona}.

When we are working in the classical theory we must impose the reality
conditions on the initial data.  Once imposed, they are maintained for
the whole evolution.   The question then arises as to how to deal with the
reality conditions in the quantum theory.  The key point is  that to state
the
reality conditions quantum mechanically requires the use of the inner
product, because   complex conjugation is translated into
Hermitian conjugation, which uses the inner product.    The problem of the
quantum reality conditions is then intimately connected with the inner
product.
For this reason, we put off further discussion of the reality conditions
till section 6.2, which
discusses the inner product.

Except for that $\imath$ we can interpret $A_a^i$ and
$\tilde{E}^a_i $ as
being a Yang-Mills connection and its conjugate electric fields.
Indeed,
because $\tilde{E}^a_i $ is a density, it is well
defined on our
three manifold $\Sigma$, without refering to any background
metric.  Because
of this,
general relativity can be interpreted as a special kind
of Yang-Mills field, which is constructed on a three manifold
without any
background metric or connection structure.

The dynamics of general relativity is then defined by a set of
constraints.
There are three sets, corresponding to the three gauge invariances
of the theory.  They generate, respectively,  $SU(2)$ gauge
transformations, diffeomorphisms of the three surface $\Sigma$,
and
evolution of the surface $\Sigma$ in the spacetime.  They are
called,
correspondingly, the GaussUs law, diffeomorphism and hamiltonian
constraints.  They are expressed as,
\f
{\cal G}^i \equiv {\cal D}_a \tilde{E}^{ai} =0
\ff
\f
{\cal C}_a \equiv F_{ab}^i \tilde{E}^b_i =0
\ff
\f
{\cal C}\equiv \epsilon_{ijk}F_{ab}^k \tilde{E}^{ai} \tilde{E}^{bj}  =0
\ff
We may note that the first is familiar from Yang-Mills theory  and
the
second and third  are the simplest gauge invariant combinations
of $F_{ab}^i$ and $\tilde{E}^a_i $ that could be written
down.  That is, if someone had started out trying to write down a
canonical formulation of a Yang-Mills theory on a three manifold
that
has no background structure, that someone would have reinvented
general relativity.

Indeed, the other simple terms that one could add to (47-49)
all correspond
to well known modifications of general relativity.  For example,
the cosmological constant can be added by changing (49) to
\f
{\cal C}^{\Lambda} \equiv \epsilon_{ijk}F_{ab}^k \tilde{E}^a_i
\tilde{E}^b_j
+ \Lambda \epsilon_{abc} \epsilon^{ijk} \tilde{E}^a_i \tilde{E}^b_j
\tilde{E}^c_k=0
\ff

\section{Results about quantum geometry}

We now have the background to describe the calculations that went
into
the description of quantum geometry I gave in chapter 2.
This
chapter is organized into a series of seven sections, each of which
contains the details of the calculations of a result used in that sketch.

The results described in this chapter
were found in collaboration with Abhay
Ashtekar and Carlo Rovelli.
They are  new,
and have not\footnote{Except for some comments on the weaves
in the review of Rovelli, \cite{carlo-review}.}
been reviewed previously.  They are
described
in references \cite{weaveletter,weavepaper} by Ashtekar, Rovelli and the
author.

\subsection{Why the metric at a point is not a good operator}

In the Ashtekar formulation, the metric is most simply expressed
in
terms of the doubly densitized inverse metric
$\tilde{\tilde{q}}^{ab}$,
which is given by,
\f
\tilde{\tilde{q}}^{ab}(x)= \tilde{E}^{ai}(x) \tilde{E}^b_i (x)
\ff
The problem is that, as a product of elementary fields at a
point,
there is, in the loop representation,
no elementary operator that corresponds to this quantity
or to
any other function of the metric at a point.  In order to construct
an operator corresponding to $\tilde{\tilde{q}}^{ab}$ we must build
it by means of a regularization procedure.
This means that we must find a sequence of observables
$\tilde{\tilde{q}}^{ab}_\epsilon$ such that classically
$\tilde{\tilde{q}}^{ab}= \lim_{\epsilon \rightarrow 0}
\tilde{\tilde{q}}^{ab}_\epsilon$ and such that each one of them, for
$\epsilon >0$, has a representation as a well defined operator.
The purpose of this section is to
show that  in the loop
representation  any such procedure is bound to fail when applied to
a measurement of the metric at a point.
It will fail because the resulting operator will necessarily depend on
auxiliary background structure introduced in the regularization
procedure.  In a
nonperturbative quantization, in which there are no such
background
structures, such a result is unacceptable.

We will proceed by studying a class of regularization and
renormalization
procedures that can be used to define an operator corresponding
to $\tilde{\tilde{q}}^{ab}(x)$.  From the results we will be able to
draw
some general conclusions as to why any such procedure must fail.

The regularization procedures that we will study are
 based on the old idea of point splitting.  That is, we want to pull
apart the two $\tilde{E}^{ai}$'s in (51) and use
\f
\tilde{\tilde{q}}^{ab}(x) = \lim_{y \rightarrow x} \tilde{E}^{ai}(x)
\tilde{E}^{b}_i(y).
\ff

Because in the loop representation we keep manifest the internal
$SU(2)$ gauge invariance we will consider a gauge invariant
version
of this.  To do this we make use of a useful set of observables
which
generalize the $T^a[\gamma ](s)$ defined in (17).  Given a loop
$\gamma$
and $n$ points on it $s_1,...,s_n$ we may define,
\f
T^{a_1 ....a_n} [\gamma ](s_1,...,s_n ) \equiv Tr\left [
\tilde{E}^{a_1}(\gamma (s_1))
U_\gamma (s_1, s_2 ) ...\tilde{E}^{a_n}(\gamma
(s_n)) U_\gamma (s_n, s_1 )   \right ]
\ff
These are called the $T^n$ observables, where $n$ labels the
number of
insertions of the $\tilde{E}^{ai}$'s.
The set of all the $T^n$  form a graded Poisson algebra that
generalizes (18) with  $\{ T^n , T^m \} \approx T^{n+m-1} $
(see ref. \cite{carlolee}.).   The
whole algebra can
be defined as operators on the loop representation, which
we have  already defined as
a representation of the subalgebra of the $T^0$ and $T^1$'s.

In order to write their actions in a simple form
we must take a moment and
expand our notation as to how we denote loops
formed from combinations of loops that intersect.
First of all, in the case that there is more than one intersection point,
we denote $\alpha \circ_s \beta$ to be the loop which combines
$\alpha $ and $\beta$ at the intersection point labeled by the
parameter $s$ of the first loop.  If there is no intersection
point at $\alpha (s)$ then $\alpha \circ_s \beta$ will denote the
trivial loop.   Second, if the two loops intersect at more than one
point, we have the possibility of breaking the two loops
simultaneously at more than one intersection point and
putting them together in the various reroutings.  We will denote
the loops formed by breaking and joining $\alpha$ and
$\beta$ at two
intersection points and then rerouting by
$(\alpha \circ_{s_1}  \circ_{s_2} \beta )_r$ where $r$ labels the
possible reroutings.  If there is no ambiguity, we can drop the
labels on the intersection points and write the reroutings as
$(\alpha \circ  \circ \beta )_r$ .  In general, the act of breaking and
joining at $n$ points will be denoted by $n$ $\circ$'s.

We may now write the definition of the general $\hat{T}^n$
\begin{eqnarray}
        \hat T^{a_1...a_n}[\gamma](s_1,...,s_n)\ \Psi[\alpha ]  = &
        l_P^{2n} &   \oint dt_1 \delta^3 (\gamma (t_1), \alpha (s_1) )
		\dot{\alpha}(s_1)^{a_1} ...   \\  \nonumber
		&&  \times
		\oint dt_n \delta^3 (\gamma (t_n), \alpha (s_n) )
		\dot{\alpha}(s_n)^{a_n}
 \\   \nonumber
        & &\left (  \sum_r (-1)^{q_r } \Psi[(\gamma\circ_ {t_1}...\circ_
{t_n}\alpha)_r ]  \right ).
        \label{tn}
\end{eqnarray}
The sum over $r$ is
over the $2^n$ loops obtained by rearranging the routings
at the $n$ intersections in all the possible ways. $q_r$ is the
number
of segments between intersection and intersection that must
change
orientation in order to have a consistent orientation in the loop
obtained by combining the loop $\gamma$ in the operator and the
loop $\alpha$ in the state.

An informal nomenclature has developed in conjunction with these
operators:  the points where the $\tilde{E}^{ai}$'s are represented
on the loop of the operator are called the "hands" of the operator.
We
say that a hand acts by grasping a loop in the state.  The result of
such
a grasp is to multiply the state by   a distributional factor
which is,
\f
l_P^2 \int dt \delta^3 (\gamma (s) , \alpha (t))\dot{\alpha}^a(t)
\ff
where $\gamma $ is the loop associated with the operator, $s$ is
the loop parameter at the hand and $\alpha$ is the loop in the state.
Thus, we see that the action is zero unless every hand grasps the
loop in the argument of the state.

The state is then evaluated on a loop which is gotten by combining
the two loops $\gamma$ and $\alpha$ in the following way.  One
breaks them apart at each of the hands where they meet.  Then
there
are two ways to rejoin them at each hand: so that their
orientations
agree or disagree.  The result is gotten by summing over the
evaluation
of the state at these $2^n$ new loops, with a phase factor given by
$(-1)^{q_r}$.

The appearance of $l_P^2=\hbar G_{Newton} $ in the action (54) has
profound consequences, as we will see below and so deserves some
comment here.  First of all, from dimensional analysis, the action
of a
hand must be dimensionless because it represents an insertion of a
frame field, which, being a square root of a metric,  is
dimensionless.
Thus, there must be a factor with units of area in front of (55).  Why
is
this the Planck area?  The $G_{Newton}$ comes from the fact that
there
is a $G_{Newton}$ in the definition of the parallel propagator (15).  It
is
there because Ashtekar's connection $A_a^i$ actually has
dimensions
of $(length)^{-3}$.  Thus, the action of an $\tilde{E}^{ai}$ in a
classical
poisson bracket on a holonomy brings down, by (12), a factor of
$G_{Newton}$.  The $\hbar$ is then there in the definition of the
actions (20) and
(54) on states so that the quantum commutators of these operators
are equal to $\hbar$ times the corresponding Poisson brackets.

Let us now return to the problem
at hand, which is to construct an operator to represent the
spatial metric $\tilde{\tilde{q}}^{ab}$ at a point $x$.  What we
want
to do is to represent this as the limit of a $T^2$ as the  points
where the two $\tilde{E}^{ai}$'s live are brought together.  The
regulated
observable will then depend on a loop $\gamma$ that contains
these
two points.  The choice of this loop is completely arbitrary, for
simplicity
we may introduce a uniform way of choosing it.

We may introduce an arbitrary flat background
metric
$h^0_{ab}$ in a neighborhood, $\cal U$, of $x$.
Given any two points $y$ and $z$ in $\cal U$ let us define a loop,
$\gamma_{y,z}$ which is a metric circle according to $h^0_{ab}$
and
which satisfies $\gamma_{y,z}(0)=y$ and $\gamma_{y,z}(\pi)=z$.
It
then follows that the $\lim_{z \rightarrow y} \gamma_{y,z} =
\gamma_{y,y}$ which is the degenerate loop in which the whole
circle
is mapped to the point $y$.

For short, we will define,
\f
T^{ab}[y,z] \equiv T^{ab}[\gamma_{y,z}](0,\pi )
\ff

We need one final element for our regularization procedure: a
smearing
function.We introduce a regulation of the delta
function
by means of a smooth function of compact support $f_\epsilon(x,y)$
such that
\f
        \lim_{\epsilon\to 0} f_\epsilon(x,y) = \delta^3(x,y)
\ff
and
\f
\int d^3x f_\epsilon(x,y) =1 .
\ff
For detailed calculations we will take for $f_\epsilon (x,y)$ a
gaussian,
\f
f_\epsilon (x,y) = { \sqrt{h^0 (x) } \over \pi^{3 \over 2} \epsilon^3}
e^{-{1 \over 2\epsilon^2} |x-y|^2 } ,
\ff
where the norm $| ...|$ is defined with respect to the {\it
background}
metric $h^0_{ab}$.

We now may define our regularization of the densitized inverse
metric.
It is
\f
        G_\epsilon^{ab}(x) = \int d^3y \int d^3z f_\epsilon(x,y)
        f_\epsilon(x,z)
        T^{ab}[y,z] .
\ff
For finite $\epsilon$ each of the corresponding operators is well
defined on
the kinematical representation space ${\cal S}^{kin}_\Sigma$.
Furthermore,   when evaluated on smooth geometries
\f
        \lim_{\epsilon\to 0}\ G_\epsilon^{ab}(x)
        = {\tilde{\tilde q}}^{ab}(x) .
\ff
What we need to do now is to study the same limit in the quantum
theory.
Let us first act with the operator version of (60), on a quantum state
$\Psi[\alpha ]$.  The result is
\begin{eqnarray}
 \hat{G}_\epsilon^{ab}  (x) \Psi [\alpha ]  =
        &l_P^4 & \oint ds \oint dt \dot\alpha^a(s)
        \dot\alpha^b(t) f_\epsilon(x,\alpha(s)) f_\epsilon(x,\alpha(t))
        \\ \nonumber
        &&\times
\left ( \sum_r (-1)^{q_r}\Psi [ (\gamma_{\alpha (s) \alpha (t)}
\circ \circ \alpha
)_r ]       \right )  .
\end{eqnarray}

We now want to consider what happens to this as $\epsilon
\rightarrow
0$.  There are two factors to consider: what happens in the sum
over states inside the parenthesis in (62) and what happens to the c-
number
factor multiplying it.  We consider first the factor in the loops, as
this
will be practice for similar calculations we will have to do later.

 In the sum over the reroutings in (62) there are two kinds of terms.
The first are terms of the form $\Psi [\alpha \cup \eta_\epsilon ]$
where $\eta_\epsilon$ is a loop that is shrinking to a point as
$\epsilon \rightarrow 0 $.  The second kind are terms of the form
$\Psi [\beta_\epsilon ]$ where
$\lim_{\epsilon \rightarrow 0} \beta_\epsilon =\alpha $.  The
reader
is invited to write out this sum to verify that there are
two terms of each kind and that all have, by the rules given after
equation
(54) for signs, positive signs.

We then have to evaluate the following limits
\f
\lim_{\epsilon \rightarrow 0} \Psi [\alpha \cup \eta_\epsilon ]  \
\ \ {\rm and  }
\ \ \  \lim_{\epsilon \rightarrow 0} \Psi [\beta_\epsilon ]
\ff
We will take the meaning of the limit to be such that
\f
 \lim_{\epsilon \rightarrow 0} \Psi [\beta_\epsilon ]  = \Psi[\alpha
]
\ff
whenever $\lim_{\epsilon \rightarrow 0} \beta_\epsilon =\alpha $.
This interpretation of the limit involves a certain subtlety, which
requires discussion.  This will be the subject of the next section.
For now, the reader may verify that given this last
relation, (28) and (30) together imply that
\f
\lim_{\epsilon \rightarrow 0} \Psi [\alpha \cup \eta_\epsilon ]  =
2 \Psi[\alpha ]
\ff
It then follows that the sum over the reroutings in (62) yields after
the limit a factor of $6$ times $\Psi [\alpha ]$.

We now come to the remaining c-number factor.  It is easy to
see that this is divergent as $\epsilon \rightarrow 0$.  With the
particular case that the smearing function is chosen as (59) it is not
hard to show that the leading divergent factor is
\begin{eqnarray}
& \lim_{\epsilon \rightarrow 0}  & \left [ \oint ds \oint dt
\dot\alpha^a(s)
        \dot\alpha^b(t) f_\epsilon(x,\alpha(s)) f_\epsilon(x,\alpha(t))
\right ]
\\ \nonumber
&= &
{l_P^4 \sqrt{h^0(x) } \over \epsilon^2}
\int {ds \over |\dot{\alpha}(s)|} \delta^3 (x , \alpha (s))
\dot{\alpha}^a (s) \dot{\alpha}^b(s)
\end{eqnarray}

This divergence, of course, reflects the fact that we are trying to
multiply
two distributions.  In order to define the product, we must modify
the
definition of the limit (61).  This can be done by { \it renormalizing}
the
observable.    We thus define,
\f
        ( \hat G^{ren})^{ab}(x)
        \equiv \lim_{\epsilon \rightarrow 0}
         Z {\epsilon^2 \over l_P^2 }\hat G_\epsilon^{ab}(x)
\ff
where $Z  $ is an arbitrary renormalization constant. This limit   is
finite. We find,
\f
(  \hat G^{ren})^{ab}(x) \Psi [\alpha ]
= 6 l_P^2 \sqrt{h^0(x)} Z
\int {ds \over |\dot{\alpha}(s)|} \delta^3 (x , \alpha (s))
\dot{\alpha}^a (s) \dot{\alpha}^b(s)   \Psi [\alpha ].
\ff

We have thus succeeded in defining an operator to represent the
inverse three metric on the loop states.  But at a cost:
the final form
is not really satisfactory,   because
it  depends on the background
metric
$h^0_{ab}$.  That dependence occurs in two places, in the factor of
$|\dot{\alpha}|= \sqrt{h^0_{ab}\dot{\alpha}^a \dot{\alpha}^b}$ in the
denominator and in the overall factor of $\sqrt{h^0}$ that
multiples
the result.  Since the background metric $h^0_{ab}$ is completely
arbitrary, this means that the observable $\tilde{\tilde{q}}^{ab}$
has
only been determined up to an overall conformal factor.

We must then ask whether this background dependence is a
consequence
of the particular regularization method we have used, or whether it
is a general consequence of trying to define an operator to represent
$\tilde{\tilde{q}}^{ab}(x)$ through a regularization and renormalization
procedure.  A
simple
argument makes a strong case for the latter conclusion.  Because
the state depends only on  the loop
$\alpha$, it is natural to expect that a renormalized operator
which represents
$\tilde{\tilde{q}}^{ab}(x) $ must
have have an action which depends only on the  loop.  Thus,the action
should be proportional to a distribution that has
support on $\alpha$.
Furthermore, geometrically, the only vector field in the problem is
$\dot{\alpha}^a(s)$.  Thus, the two indices in the  action of
$\hat{G}^{ab}_{ren} (x)$ must be
proportional to
$\dot{\alpha}^a (s) \dot{\alpha}^b (s)$.  Thus, making use of the
way the loops collapsed in the argument of the state, if
 $G^{ab}_{ren} (x)$
can be defined at all it must be of the form,
\f
 G^{ab}_{ren} (x) \Psi [\alpha ]= \tilde{F}(x) \int {ds \over J(s)}
\delta^3 (x,\alpha (s))
\dot{\alpha}^a(s) \dot{\alpha}^b (s)  \Psi [\alpha ].
\ff
Let us note, first, that the factor $J(s)$ must transform under
reparametrizations of the curve like a one dimensional density, so
that
the overall expression preserves the reparametrization invariance
of the
formalism.  In the regularization we have just discussed this
factor is equal to $\sqrt{|\dot{\alpha}(s)|}$.  However,
in general, we know that  some background structure will be needed
to construct such a one dimensional density, because there is none
that
can be constructed from the geometrical information at hand.

The external factor $\tilde{F}(x)$ turned out in our regularization
to
be equal to $6l_P^2 \sqrt{h^0}Z$.  This is, as we pointed out above, a
free
function.  This means that the observable has no more information
in
it other than that it lives  in the $\dot{\alpha}^a \dot{\alpha}^b$
subspace.
We may ask, however, whether it is possible that another
regularization
method will leave us with a constant $F$.  The answer is that this
is impossible.  The reason is because the delta function
$\delta^3(x,y)$, in addition to being a distribution,
is a density in its first argument\footnote{This is why it is
written
$\delta^3(x,y)$, rather than $\delta^3 (x-y)$.  The latter
expression is  only
well defined in the presence of a flat background metric.}.
The $\tilde{E}^{ai}(x)$ is a density and its product
$\tilde{\tilde{q}}^{ab}$
is thus a double density.  Now, any method that defines the product
must
give a rule to show how the product of two distributions is
proportional
to a single distribution.  However, we see that
the factor of proportionality must
be a density, in order to preserve the fact that the product is a
double
density.  This is why the $\tilde{F}(x)$ came out to be a density.
Further, as there is no natural density in the problem, if the
extension is to be well defined, the overall density factor must
come
from the background structure that is used in the regularization
procedure.

{}From this general situation there follows
an important lesson about diffeomorphism invariant quantum
field theories.    The lesson
is that renormalization will not normally be an acceptable way to
define diffeomorphism invariant observables.  A renormalization
procedure is a procedure to multiply two
well defined local operators, defined at a single point, and get another
local operator.  Since operators are distribution valued we must
also, in the absence of a background metric, take into account the
fact
that any such procedure must either change the density character
of the observable or introduce an arbitrary density as an overall
factor in the definition of the renormalized operator.   Either way,
the diffeomorphism invariance of the theory is compromised.

Thus, to conclude: we have shown in this section how to define a
regularization and renormalization procedure for the inverse
metric
at a point.  We succeeded, in that the result was a well defined
operator
on the kinematical representation space ${\cal S}^{kin}_\Sigma$.
However,
the resulting form of the operator was infected with background
dependence.  This background dependence renders the result
meaningless in a nonperturbative context.
Furthermore, we argued that this background dependence
is bound to infect any attempt to construct an operator to
represent
the metric at a point in the loop representation.

All, however, is not lost.  In the next several sections we will
show
that the problem is not with the loop representation, it is with the
idea
that the metric can be observed at a point of space.  We will do this
by
showing that several different operators that measure metric
information
on surfaces and regions may be defined through regularization
procedures
that result in operators that are finite and independent of any
unphysical
background fields.  However, first, some clarification about our
state
space.

\subsection{More about representations of
quantum theories with discrete norms}

This section is devoted to the explication of a certain subtlety associated
with the use of characteristic states and discrete measures in quantum
field theories.  It may be skipped on a first reading.

In the previous section we made an assumption about
the how the limits in
the regularization of operator products
are to be taken.
For the evaluation of (67),  and for later calculations,
we will need it to be true  that the
limits  that will appear in the
regularization
of operators in the loop representation are taken so that the
following conditions are satisfied.
If $\eta_\epsilon $ is a loop
of radius  $\epsilon,$ in some background coordinates, then it
should
be  true that
\f
\lim_{\epsilon \rightarrow 0}
< \alpha \cup \eta_\epsilon| = 2<\alpha |
\ff
and
\f
\lim_{\epsilon \rightarrow 0}
< \alpha \circ \eta_\epsilon| =<\alpha |.
\ff
As I mentioned above, the former expression actually follows from
the latter, together with the identities (28) and (30) we are imposing
on the loop space.  Now, (70) and (71) are guaranteed
if the loop functionals
satisfy some condition of continuity or differentiability on the
loop space.  For example, suppose that
\f
\lim_{\epsilon \rightarrow 0}{\left ( \Psi [\alpha \circ \eta_\epsilon
] -\Psi [\alpha ]
\right )   \over \epsilon^2  }  \equiv \nabla_\eta \Psi [\alpha ]
\ff
exists, as is often the case in applications of the loop
representation to conventional quantum field theories.  (This is the
definition of the loop derivative.  For more
information about it see \cite{berndtjorge-onC} and references
contained therein.)
Then we have
\f
\Psi [\alpha \circ \eta_\epsilon ]  = \Psi [\alpha ] + \epsilon^2  \nabla_\eta
\Psi [\epsilon ]
\ff
There are, however,  two problems with this when it comes to
the application to quantum gravity.  The first is that the
diffeomorphism
invariant states are not loop differentiable.  They are not even
continuous on the loop space.  The reason is that
$\Psi[\alpha ]$ and $\Psi [\alpha \circ \eta_\epsilon ]$ are in
different
topological classes for all $\epsilon \neq 0 $ and that, moreover,
$\Psi [\alpha \circ \eta_\epsilon ]$ does not depend on $\epsilon$
 as long as $\epsilon \neq 0$.  Thus, taken on
 diffeomorphism invariant states, the limit in (72) diverges
as $1/\epsilon^2$.

The second problem is that, even at the kinematical level, none of
the states in the space ${\cal S}^{kin}_\Sigma$ with the norm (3)
are
loop differentiable.

As the calculations discussed in this chapter are done at the
kinematical
level, we will discuss the second of these issues.  The first, having
to do with the application of the loop derivative to diffeomorphism
invariant states has not yet been resolved.

It will be helpful if we begin by talking about an analogous
problem in one dimensional quantum mechanics.  Although
it is not usually done, we can introduce characteristic
states  and discrete measures in there if we relax
a little bit what we require of the state
space\footnote{Here I am following I line of thought that I learned
from Abhay Ashtekar}.

Suppose we wanted a representation of one dimensional
quantum mechanics in which $\hat x$ and $\hat{T}(\epsilon )$
were well defined operators, where $\hat{T}(\epsilon )$ is the
translation operator defined by
$\hat{T}(\epsilon )\psi (x) =\psi (x+\epsilon )$, but one did not
require that there exists an operator $\hat{p}$ such that
$\hat{T}(\epsilon )= e^{i\hat{p}\epsilon}$.  Instead,
one wants to construct the quantum theory to be a representation
of the algebra
\f
 \hat{T}(-\epsilon )\hat x \hat{T}(\epsilon ) = \hat{x} + \epsilon
\ff
In this case, because one does not insist that $\hat{p}$ exist,
the usual uniqueness theorem does not apply and
one can
construct a representation which is inequivalent to the
usual $L^2$ hilbert space.  This representation consists of
states which are normalizable with respect to the discrete
measure, which is based on the discrete topology on $R$,
with respect to
which each point is its own open set.  Given this discrete topology,
we may construct the discrete measure
\f
\int_{\cal R} d\mu(x)_{discrete} F(x)  =  \sum_{x \in R} F(x)
\ff
The normalizable functions under this measure are those that have
support on only a countable set of points $x \in { R}$, so that
the integral in (75) converges.   A normalizable basis for this hilbert
space  is then given by the
characteristic states,
\f
\psi_{y} (x) = \delta_{xy} \equiv 1 \ \ {\rm if }  \ \ x=y , \ \ \
{\rm and} \ \ 0
\ \  {\rm otherwise}  .
\ff
It is straightforward to show that this space of functions is
a representation\footnote{This
is given by $\hat{T}[\epsilon ] \psi_y (x)
\equiv \psi_{y-\epsilon}
(x)$
and $\hat{x} \psi_y (x) \equiv y \psi_y (x)$.} for the
algebra defined by (74) and that,
with respect to the discrete measure,  $\hat{x}$
and $\hat{T}(\epsilon )$ are  hermitian and unitary operators,
respectively.  Further (although it is not important for the
one dimensional quantum mechanics), it is interesting to note
that this representation does not depend on a background metric
on $R$, so that  a unitary representation
of the one dimensional diffeomorphism group exists on this
space.

If we use the loop variables (14) and (17), with the algebra (18) as the
basis for quantization of gravity or a gauge theory, then
we are in an  analogous situation
to the case we have just considered because we want
a representation in which there exists an
operator for $\tilde{E}^{ai}$, (or for a gauge invariant
operator polynomial in it) but we do not require there to exist any
operator for  its canonical conjugate $A_a^i$.
Instead, the operators which do not commute with polynomials
in $\tilde{E}^{ai}$
are non-local operators, of the form of $\hat{T}[\gamma ]$,
which are exponential in $A_a^i$.

Now,  the algebra (74) of $\hat x$ and $\hat T(\epsilon )$ can
be represented both in terms of the standard $L^2(R)$, in which
case the operator $\hat p$ exists, or it can be represented
on the space of states normalized with respect to (75), in which
case, as can easily be shown, no operator for $\hat p$ exists.
Similarly, as shown recently by Ashtekar and Isham \cite{abhaychris},
the loop
algebra (25) has several kinds of representations.   There are
Fock representations, in which operators linear in $A_a^i$
also exist.   But there are also representations which
are defined on the space ${\cal S}^{kin}_\Sigma$ of
functions which are normalizable with respect to the
discrete measure (3).  (The reader may verify that the defining
relations (19) and (26) are well defined on this space of loop
functionals.)  These representations do not admit any operator
for $A_a^i$ and they are inequivalent to the Fock representation.

There are cases, such as Maxwell theory and linearized gravity,
where the existence of gravitons and photons requires that
there be operators linear in $A_a^i$.  In this case, even if
one begins with a loop representation defined by (19) and (26) one is
forced back to the Fock representation.
However, in the nonperturbative context there should be
no graviton operator, as these are only defined with respect
to a fixed classical background,  and we have no
reason to insist that
$A_a^i$ itself be represented by an operator.   In this case
we could choose one of these new alternative representations.

Given a choice between different inequivalent representations,
the choice must be made on physical grounds.
There are, in my opinion, two reasons for choosing
these non-Fock representations at the kinematical level for the
construction of quantum gravity.  First, as I have
already mentioned, the Fock representations
necessarily break diffeomorphism invariance, while the
discrete measure on loop space is diffeomorphism invariant.
The second reason, which is based on results that
we will derive in the rest of this chapter,
is that the classical limit of the theory is
well defined in the case that we restrict the state space to
a non-Fock representation. Thus, it is not necessary to consider
a larger class of states to recover known physics.

We can now return to the technical problem at hand,  which
is how to take the limits of shrinking small loops
involved in the regularization procedure
inside of this state space.
We will find it helpful to  use the one dimensional example to
understand the source of the problem.

The problem is to specify the topology in which the limit
$\epsilon \rightarrow 0$ is to be taken in expressions such
as (67).   Let me first show that the first, guess, which is to
use the pointwise topology on the loop space, cannot work.
For, using the pointwise
topology on the loop space to define the limit, we have
\f
\left ( \lim_{\eta \rightarrow \beta} \Psi_\eta [\alpha ]
\right )_{pointwise }= 0 \neq \Psi_\beta [\alpha]
\ff
As a result, the topology based on the discrete norm (3) is also
useless.  For if we use an inner product based on the discrete
measure on loop space, such as\footnote{We will see in the next
section that this is the form of the inner product when $\alpha$
and $\beta$ are simple loops.},
\f
< \beta | \alpha > = \delta_{\alpha \beta},
\ff
where $\delta_{\alpha \beta}$ is the kronocker delta,
we have
\f
\lim_{\eta \rightarrow \beta} \left (  < \eta | \beta > \right )
= \lim_{\eta \rightarrow \beta} \delta_{\eta \beta} =0
\ff
This is a general problem which arises when one uses a state
space  based on a discrete norm.
It occurs also in the one dimensional example. Using
the
pointwise topology we have
\f
\left ( \lim_{z\rightarrow y} \psi_z [x] \right )_{pointwise}=
0 \neq \psi_y[x]
\ff
A characteristic problem which occurs as a result has to do
with the operator $\hat{T}(\epsilon)$.  According to the
definition  $\hat{T}(\epsilon)\psi (x) = \psi (x + \epsilon )$
this operator is well defined on the space
of functions normalizable with respect to (75) for all $\epsilon \in
R$.
In particular,  $\hat{T}(0)$, is well defined and is
equal to the identity operator.  However, if we use
the pointwise topology to define the limit it is not
true that
$\lim_{\epsilon \rightarrow 0 } \hat{T}[\epsilon ] =I$
Instead, if we define the limit
in the pointwise topology we have
\f
\lim_{\epsilon \rightarrow 0} \left (  \hat{T}(\epsilon ) \psi_y (x)
\right ) = \lim_{\epsilon \rightarrow 0}  \psi_{y-\epsilon} (x ) = 0
\ff
The problem reflects the fact  that on the space of states which
are normalizable with respect to
the discrete norm, (75), the operator
$\hat{T}(\epsilon ) $ is not differentiable at $\epsilon =0$.
This is, of course, necessary, because if the derivative did exist
it would be equal to $\hat{p}$, which cannot exist in this
representation.

It is then
clear that to preserve the property that
$\lim_{\epsilon \rightarrow 0} \hat{T}(\epsilon )=1$, even when
its
action is extended to states that are normalizable under the
discrete norm (75), we have to use a different topology to define
the limit.  What we need is to induce on the states the
standard continuous topology on $R$ such that
\f
\left ( \lim_{z \rightarrow y} \psi_z (x) \right )_{continuous}
= \psi_y (x).
\ff

We can do this just as a matter of definition.  We may also
observe that the use of
such a topology is equivalent to using the topological dual of
the state space to
define the limits of operators.  Let us note, first of all, that
under the discrete norm (75) normalizable functions are dual to the
smooth bounded functions, in the sense that for any smooth
and bounded $f(x)$
\f
\int d\mu_{discrete} (x) f(x) \left (\sum_{i}  a_i \psi_{y_i} (x)
\right )
= \sum_i a_i f(y_i )  < \infty
\ff
 because $\sum_i |a_i|^2 < \infty  $.   We may then define the
limit of the translation operators such that
\f
\sum_{x \in R}  f(x) \hat{T}(0) \psi_y (x) \equiv
\lim_{\epsilon \rightarrow 0} \left  (
\sum_{x \in R}  f(x) \hat{T}(\epsilon ) \psi_y (x)       \right )
\ff
for all smooth $f(x)$.  By this definition it follows that
$\hat{T}(0)=I $.

We can then use these ideas to resolve the problem with the limits
of our
regulated operators in the loop representation. In the
remainder of this chapter will use the
 hilbert space of
states ${\cal S}^{kin}_\Sigma$ which are normalizable with
respect
to the discrete measure on the loop space ${\cal HL}[\Sigma]$.
However, when we have to define an operator on this space through
a regularization procedure, we will do this not by using the
pointwise topology on the space of states but the continuous
topology\footnote{The reader may be uncomfortable with using a topology
other than that defined by the norm on the state space to
regulate operators in a quantum field theory.  However, there
seems
to be little choice if we are to use a state space based on the
discrete
measure.
The example of taking the limit of $\hat{T}(\epsilon )$
as $\epsilon \rightarrow 0$ shows us that when dealing with
state spaces based on discrete norms, we must use such a topology
if we are not to get nonsensical results for the action of
operators.  Further, we should recall that the purpose
of a regularization procedure is only to define a certain operator on
the
state space in question.  We are free to do this in any way we like,
as long as the operator defined by the limit exists and has
reasonable
properties.}
 which is defined so that
\f
\left (  \lim_{\eta \rightarrow \beta} \Psi_\eta [\alpha ]
\right )_{continuous}= \Psi_\beta [\alpha] .
\ff

We can define this limit in terms of the topological
dual, as we did for the one dimensional case.  This
requires that we define smooth
functions on ${\cal HL}[\Sigma] $.  These will be defined to be
those
functions on ${\cal HL}[\Sigma ]$ which are infinitely
differentiable
under the loop derivatives (72) defined in \cite{berndtjorge-onC}.  We then
require  that a sequence of operators
$\hat{\cal O}(\epsilon )$ will be said to converge to an operator
$\hat{\cal O}_0$ if, for all smooth functions
$\Phi [\alpha ]$ on ${\cal HL}[M]$ and all normalizable states
$\Psi [\alpha ]$ in ${\cal S}^{kin}_\Sigma$ we have
\f
\lim_{\epsilon \rightarrow 0}
\sum_{\alpha \in {\cal HL}[M] }  \Phi [\alpha ] \hat{\cal O}(\epsilon)
\Psi [\alpha ] = \sum_{\alpha \in {\cal HL}[M] }
\Phi [\alpha ] \hat{\cal O}_0
\Psi [\alpha ]
\ff

\subsection{The construction of the area operator}

In this section, I show how to define a quantum operator that
measures
${\cal A}[{\cal S}]$,  the area of an arbitrary surface $\cal S$, in
$\Sigma$.

The key idea involved in this construction is really an idea about
regularization of the action of the $T^n$ operators defined by (54).
Recall that the action of each hand at a point $\gamma (s)$ on
a state $\Psi [\alpha ]$ is given by (55).    Recall also
from (21) that the dual of an   $\tilde{E}^{ai}$ is a two
form.   As in the construction of the ribbon operator (23) we can
integrate this two form against a  surface.   Thus,
if we let the coordinates of the surface be $S^a$
we have
\begin{eqnarray}
\int_{\cal S} d^2S^{bc}...  \hat{E}^*_{bc} (S) ... \Psi [\alpha ]&=& l_P^2
\int_{\cal S}d^2S^{bc}  \oint dt \delta^3 (S, \alpha (t) )
\dot{\alpha}^a (s) \epsilon_{abc} ...   \\  \nonumber
&=&  l_P^2
I[{\cal S},\alpha ]  ...
\end{eqnarray}
Here $I[{\cal S},\alpha ] $ is the intersection number defined by (24)
and the $...$'s indicate other factors that make the result
gauge invariant.

This makes sense of the factor of the Planck area siting in from of
the action (55).  It tells us that the hands are naturally associated
with
quanta of area; integrated over a surface, the hand wants to give a
unit of the Planck area to each line it meets in its action on a
state.

The idea is to make use of this pretty fact by finding a way to
write the area of a surface $\cal S$
in terms of the two forms $E^*_{ab}$'s integrated over the surface.

To show how to do this we
begin by writing the usual expression for ${\cal A}[{\cal S}]$
in the classical theory,
\f
 {\cal A}[{\cal S}]  = \int_{\cal S} \sqrt{h} ,
\ff
where $h$ is the determinant of the induced two metric,
$h_{ab}=q_{ab} -n_a n_b $, where $n^a $ is the unit normal to the
surface.  A
simple
calculation show that $h= \tilde{\tilde{q}}^{ab}n_a n_b$.  Now,
to avoid the problem that $\tilde{\tilde{q}}^{ab}$ cannot itself
be extended to distributional loop geometries, we must construct
the area (88) through a limit that does not need this extension.  To
do this, let us divide the surface up into   $N$ disjoint regions
${\cal S}_i$,
such that ${\cal S} =\cup_i {\cal S}_i$.  We then have
 \f
 {\cal A}[{\cal S}] = \sum_i {\cal A}[{\cal S }_i]
\ff
 We will proceed by introducing an approximation for the square of
${\cal A}[{\cal S}_i] $ which becomes exact in the limit of
infinitesimal surfaces.  This is,
\f
{\cal A}_{approx}^2[{\cal S}_i] \equiv \int d^2S^{ab}_i
\int d^2 S^{\prime \ cd}_i T^{**}(S,S^{\prime} )_{ab \ cd }
\ff
where
$T^{**}(x,y )_{ab \ cd } \equiv \epsilon_{abe} \epsilon_{cdf}
T^{ef}(x,y)$ and the latter quantity is that we defined in eq. (56).
To show that this approximates the area of the surface element
for small surfaces, we use the facts that in the limit
$T^{ab}(S,S^{\prime} )  \approx \tilde{\tilde{q}}^{ab}(S) $.  We may
invert the relation $h=\tilde{\tilde{q}}^{ab}n_a n_b$ to find that
$\tilde{\tilde{q}}^{ab}=hn^an^b-r^{ab}$ where $r^{ab}n_b=0$.
An infinitesimal element of
area is given by $d{\cal A}= d^2S^{ab}\sqrt{h}n^a \epsilon_{abc}$,
from which it follows that,
\f
d{\cal A}^2 = d^2S^{ab} d^2S^{cd} \epsilon_{abe}\epsilon_{cdf}
\tilde{\tilde{q}}^{ef}
\ff
For smooth fields  this is then equal to (90) in the limit of small
areas.
We may then consider the limit in which we divide the surface up
into smaller and smaller elements, so that $N \rightarrow \infty$.
It then follows that,
\f
{\cal A}[ {\cal S}] = \lim_{N \rightarrow \infty} \sum_{i=1}^N
\sqrt{{\cal A}_{approx}^2[{\cal S}_i]} .
\ff
For smooth, nondegenerate, metrics  this is a long way round to go
to define the area.  But, because it incorporates the approach to
regularization I described above,   this
particular definition can be carried over to the quantum theory.

I will then introduce an operator for the area of an arbitrary
surface,
$\cal S$, which I will call $\hat{\cal A}[{\cal S}]$ by following
the same procedure that we found just gave the operator
classically.  We define
the operator by equation (92), with ${\cal A}^2_{approx} [{\cal S}_i]$
defined by equation (90), but this time with the $T^{**}$ taken as the
operator.  Let us assume that in dividing up the surface
we have taken $N$ big enough that a given
curve, $\alpha $, intersects each ${\cal S}_i$ in the partition at
most once.  (For each $\alpha $ there is an $N$ such that this is the
case.)

Let us then compute
\begin{eqnarray}
\hat{\cal A}^2_{approx} [{\cal S}_i ] \Psi [\alpha ] &=&
l_P^4 \int d^2S^{bc}_i \oint dt \delta^3(S, \alpha (t)) \dot{\alpha}^a
(t)
\epsilon_{abc}  \\ \nonumber
&&\times \int d^2 S^{\prime \ ef}_i \oint dt^{\prime}
\delta^3(S^\prime, \alpha (t^\prime )) \dot{\alpha}^d (t^\prime)
\epsilon_{def}
\\ \nonumber
&&\times
\left (  \sum_r (-1)^{q_r} \Psi[ ( \gamma_{\alpha (t) \alpha
(t^\prime) }\circ \circ \alpha )_r ]        \right )  .
\end{eqnarray}
Using (24), we see that the c-number factors give the square of the
intersection number.  As $\alpha$ intersects the surface element
${\cal S}_i$ at most once the sum over loop factors is easy.
Let us for the moment restrict ourselves to the case that
the loop $\alpha$ is simple, which means it has no intersections
or retracings.   Then
both hands actually grasp on to the loop $\alpha$ at the same
point.  In this case the routings may be defined by taking
the curve joining the hands, $\gamma_{x,x}$ to be first a circle
(with respect to some background Euclidean coordinates, of radius
$\delta$ and then shrinking $\delta $ (and hence the circle)
to zero.  A factor of $6$ comes
out because the counting is then the same as in (62), in our attempt to
construct a renormalized metric operator.

We then have,
\f
\hat{\cal A}^2_{approx} [{\cal S}_i ] \Psi [\alpha ] =
\left ( l_P^2 I[{\cal S}_i, \alpha ]   \right )^2  6 \Psi [\alpha ]
\ff

It is then trivial to take the square root, to sum up the
contributions
from the different subsurfaces and then take the limit
$N \rightarrow  \infty$.  The result is,
\f
\hat{\cal A}[{\cal S}]\Psi [\alpha ] = \sqrt{6}l_P^2  I^+ [{\cal S},\alpha ]
\Psi [\alpha ]
\ff
where $I^+ [{\cal S},\alpha ]$ is the unoriented intersection number,
which is simply the number of intersections of $\alpha$ with
$\cal S$, with all intersections counted positively.

{}From this result we can deduce immediately from (27)
that for simple $\alpha$
the bra states $<\alpha| $ are eigenstates of
$\hat{\cal A}[{\cal S}]  $:
\f
<\alpha | \hat{\cal A}[{\cal S}] =
\sqrt{6}l_P^2  I^+ [{\cal S},\alpha ] <\alpha |
\ff

Note that the result is  finite, with no need of renormalization.
Further,
we see that the spectrum of the area operator in the
loop representation contains a discrete sector, in which
the eigenvalues  are quantized in integer
multiples of the Planck area.  In the next section we will argue
that, at least on the space of states normalizable
under (3),  the full spectrum is discrete.

We may also solve the eigenvector equation on the right and find
a state $\Psi_\alpha$, which has the property that for any loop
$\gamma$
\f
\hat{\cal A}[{\cal S}]\Psi_\alpha [\gamma ] = \sqrt{6}l_P^2  I^+ [{\cal
S},\alpha ]
\Psi_\alpha  [\gamma ]
\ff
If we assume the existence of an inner product with respect to which
$\hat{\cal A}[{\cal S}]$ is a self-adjoint operator, then this state
must be the hermitian conjugate of $<\alpha |$.  That is, we must have
$|\alpha > \equiv <\alpha |^\dagger = |\Psi_\alpha >$.   From this we
can find some, but not complete, information about the inner
product, as I will now show.

In particular, we may first determine $<\alpha | \beta >$ when both
loops are simple.
This follows directly from the condition that
$\hat{\cal A}[{\cal S}]$ be self-adjoint.    For we have,
for both $\alpha$ and $\beta$ simple,
\begin{eqnarray}
\left ( <\alpha | \hat{\cal A}[{\cal S}] \right ) | \beta >
&=&  \sqrt{6}l_P^2  I^+ [{\cal S},\alpha ] <\alpha |\beta >
\\ \nonumber
= \overline{ \left ( <\beta | \hat{\cal A}[{\cal S}] \right ) | \alpha  > }
&=&  \sqrt{6}l_P^2  I^+ [{\cal S},\beta ] \overline{ <\beta  |\alpha  > }
\end{eqnarray}

Thus we see that unless $I^+ [{\cal S},\alpha ] =I^+ [{\cal S},\beta ] $
for all surfaces $\cal S$ we must have $<\alpha | \beta >=0$.  For
this to be the case the support of $\alpha$ and $\beta$ must be
identical, however, since they are both assumed simple we have,
 \f
<\beta | \alpha  > = D(\alpha ) \delta_{\alpha \beta }
\ff
where $D(\alpha )$ is real.  If we normalize the eigenstates so that
$D(\alpha )=1$ we then have
 \f
\Psi_\alpha [\beta ] = <\beta | \alpha > = \delta_{\alpha \beta }
\ff
when $\beta$ is also simple.

With a little more work we can extend this result to find the
representation of the eigenstate
$ \Psi_\alpha [\beta ] = < \beta |\alpha >$ for a general $\beta$  (but
holding $\alpha $ still simple.)   Let us begin with an example.  We will
find $\Psi_\alpha [\alpha \circ \beta^2 ]$, where $\beta$ is an arbitrary
simple loop not equal to $\alpha$.  We begin with the computation
of the action of $\hat{\cal A}^2[{\cal S}_i]$.  For simplicity, we will
assume that ${\cal S}_i$ is small enough
that it intersects each segment of the loop at most once.
After a short calculation we find,
\begin{eqnarray}
\hat{\cal A}^2[{\cal S}_i]\Psi [ \alpha\circ \beta^2]&=&
6 \ l_P^4 \ I[{\cal S}_i,\alpha]^2 \Psi [\alpha \circ \beta^2 ]
\\ \nonumber
&&+ 12 \ l_P^4 \ I[{\cal S}_i,\beta ]^2 \Psi [\alpha \circ \beta^2 ]
\\ \nonumber
&&+ 4 \ l_P^4I \ [{\cal S}_i,\beta ]^2 \left (  \Psi [\alpha  ]
         		+ \Psi [ \alpha \circ \beta \cup \beta ]      \right )
\end{eqnarray}

Note that the second line comes from terms where both hands of the
$T^2$ act on the same $\beta$, while the last line comes from terms
where each hand acts on a different $\beta$.  The effect of this last
term is to induce two terms, one where the flow along the two $\beta$'s
is switched and one in which they are joined crosswise, so that they
retract via the retracing identity, leaving only $\alpha$.

If we use the spin network identity (30) we can write this result as
(restricting to the eigenstate):
\begin{eqnarray}
\hat{\cal A}^2[{\cal S}_i]\Psi_\alpha [ \alpha\circ \beta^2]&=&
6 \ l_P^4 \ I[{\cal S}_i,\alpha]^2 \Psi_\alpha  [\alpha \circ \beta^2 ]
\\ \nonumber
&&+ 8  \ l_P^4 \ I[{\cal S}_i,\beta ]^2
\left ( 2 \Psi_\alpha    [\alpha \circ \beta^2 ]
		+  \Psi_\alpha  [\alpha ]     \right )
\end{eqnarray}

We already know that the eigenvalue is $6 l_P^4I[{\cal S}_i,\alpha]^2 $,
so that the term proportional to $I[{\cal S}_i,\beta ]^2 $ must vanish.
This gives us
\f
\Psi_\alpha    [\alpha \circ \beta^2 ]
		= -{1\over 2}  \Psi_\alpha  [\alpha ]   = -{1 \over 2}
\ff

Having understood this example, it is not hard to do the general case.
Let us consider a general loop, $\gamma$, which is a particular routing
of a graph
$\Gamma$ and which may be considered
to be made out of segments $\eta_I$ such that $\gamma = \prod_I \eta_I$.
Along segments which are traced more than once,
$\hat{\cal A}^2[{\cal S}_i]$ will act to produce a term in which
two of the
tracings of that
segment are eliminated.  Let us call that loop $C_{IJ}(\gamma )$,
which indicates that the segments $\eta_I$ and $\eta_J$, which have
the same support, have been eliminated by the action of the
$\hat{\cal A}^2[{\cal S}_i]$.  Now, let us cut $\gamma$ in this
manner in all possible places.  If there are $M$ cuttings possible,
we may denote this by $C^M  ( \gamma )$.  This loop will have no multiple
segments. (If it is simple then it will also be unique, which is the case
we are interested in.)  It may then be demonstrated that if and only if
$C^M (\gamma )$ is equal to $\alpha$ then  $\Psi_\alpha [\gamma ] \neq 0$.

To show this, and to compute the explicit components, we need to show
that the value of $\Psi_\alpha [\gamma ]$, for any $\gamma$, is
determined by its value on all the loops that result from cutting
$\gamma$.  To show this, let us denote by $X_{IJ}(\gamma ) $
the other loop
that arises from the action of $\hat{\cal A}^2[{\cal S}_i]$ on
the doubled segment $\eta_I =\eta_J$.  This is the one in
which the two lines are switched, but their orientations are
not reversed, so they are not retraced.  (In the example
above $X(\alpha \circ \beta^2 ) = \alpha \circ \beta \cup \beta $.)
For notation, let us divide the segments $\eta_I$ into a set, labeled
$\eta_\rho $, which are not multiple, and cannot be cut, and the
rest, labeled by $I^\prime$ and  $J^\prime $, which are multiple.
We then have
\begin{eqnarray}
\hat{\cal A}^2[{\cal S}_i]\Psi_\alpha [ \gamma ]&=&
6 l_P^4 \sum_\rho  I[{\cal S}_i,\eta_\rho  ]^2
\Psi_\alpha  [\gamma  ]
\\ \nonumber
&+&8 l_P^4 \sum_{I^\prime }
\sum_{J^\prime {\rm s.t.} \eta_{J^\prime } = \eta_{I^\prime}}
\left (  \left [
6I[{\cal S}_i,\eta_{I^\prime } ]^2 + 6I[{\cal S}_i,\eta_{J^\prime } ]^2
\right ] \Psi_\alpha [\gamma ]  \right.
\\ \nonumber
 &+&  \left.
I[{\cal S}_i,\eta_{I^\prime } ]   I[{\cal S}_i,\eta_{J^\prime } ]
\left [
 4 \Psi_\alpha    [ C_{I^\prime J^\prime } (\gamma ) ]
		+  \Psi_\alpha  [ X_{I^\prime J^\prime } (\gamma ) ]
\right ]
\right )
\end{eqnarray}
It is clear that this process continues until one can cut no more.  The
resulting loop is either equal to $\alpha$, or equal to a loop that we
know $\Psi_\alpha $ on gives zero.  In the latter case
$\Psi_\alpha [\gamma ] =0$.  In the former case, we must solve
the equations at each level to find $\Psi_\alpha [\gamma ]$.  To
do this we should note that since we know the eigenvalue is
proportional to $I[{\cal S}_i,\alpha ]$, all the terms which are
proportional to the intersection numbers of cut segments must vanish.
This gives from (104) a linear system to solve for each cut segment.
As the reader may show, each of these can be expressed as a matrix
equation and inverted.  The result is that all the nonvanishing
coefficients of $\Psi_\alpha [\gamma ]=<\gamma |\alpha >$ can be found,
for the case of simple $\alpha$, but general $\gamma$.

In the next section we will see that the complete set of eigenstates,
in the dual representation can be found for general $\alpha$.  However,
the complete inner product, including all the values of $<\alpha |\beta >$
when neither $\alpha$ nor $\beta$ are simple is a more complicated
problem, which has not yet been solved.

\subsection{The action of operators on intersecting loop
states and the extension of  the area operator  to
the same}

We have two purposes in this section.  First, we describe some
technology that is useful when we describe operators acting
on intersecting loops.  This will be useful particularly for the
construction of the volume operator in the following section.
Then I will show, as an example, how to extend the area
operator described in the previous section to the intersecting
case and find, in particular, all the dual eigenstates of the areas.

Let us then consider a graph $\Gamma$ with $N$ intersection
points\footnote{Recall that in section 2.3 we introduced notation
for graphs.},
$p_\alpha $, $\alpha =1,..,N$.
Each intersection point $p_\alpha$ has $n_\alpha$ lines going
through it and
$M_\alpha= \{  \stackrel{2n_\alpha}{2} \}$ different
ways to route through it.  Let us have a label,
$I_\alpha =1,...M_\alpha$ which labels the different
ways to route through each of the $p_\alpha$.   We can then label
the
different loops which result from a graph $\Gamma$ by choosing
the
$N$ different routings as $\Gamma_{I_1....I_N}$.

Now, because of the spin network  relations (30),  not all of the
values
of a loop state on different members of a graph are independent.
Instead,
a general loop state, evaluated at an intersecting loop  must
satisfy a set of linear relations, one for each intersection
point, which may be written as
\f
\sum_{I_\alpha =1}^{M_\alpha} P_{n_{\alpha}}^{I_\alpha}
\Psi [\Gamma_{I_1,..,I_\alpha,..,I_N} ] =0
\ff
where the coefficients $P_{n_{\alpha}}^{I_\alpha}$ come from
repeated application  of (30).
For each $n $-fold intersection point, one
can then choose   $w(n) $ routings, which will be
a maximal set on  which the loop states are
independent.  We may then work instead with the smaller set
of loops, which are labeled by such an independent, complete
set.  We will denote these loops by
$\Gamma_{I_1^\prime ,...,I_N^\prime}$, where
$I_\alpha^\prime =1,...w(n_\alpha )$.

We now  can show that the action
of ${\cal A}[{\cal S}]$  on intersecting loops can cause
rearrangement of the routings.    We need only check the case that
the surface $\cal S$ crosses exactly one of the
intersection points of a graph $\Gamma$.  We will consider first
the simplest case,  in which
a surface $\cal S$ crosses a simple intersection point of
two curves.  The
two curves  will be labeled by $\alpha$ and $\beta$ and the
two independent routings through the intersection will
be denoted by $\Gamma_I$, where $I$ takes the two  values
$\times$
and $><$ (the symbols indicate what happens at the intersection
point \cite{tedlee}) .
Using (93) we find that,
\begin{eqnarray}
\hat{\cal A}^2_{approx} [{\cal S}_i ] \Psi [\Gamma_{\times} ]&=&
\int d^2S_1^{a_1a_2}\int d^2S_2^{b_1b_2}
\hat{T}^{**}[S_1,S_2]_{a_1 a_2 b_1 b_2} \Psi [\Gamma_{\times} ]
\\ \nonumber
&=& l_p^4 \int d^2S_1^{a_1a_2}\int d^2S_2^{b_1b_2}
\epsilon_{a a_1 a_2}  \epsilon_{b b_1 b_2}
\\ \nonumber
&&\times
\left [  \int ds \delta^3 (S_1 , \alpha (s)) \dot{\alpha}^a(s)
 \int dt \delta^3 (S_2 , \alpha (t)) \dot{\alpha}^b(t)  \right.
 \\ \nonumber
 && \times
\left (  \sum_r (-1)^{q_r}
\Psi [(\alpha \circ_s \circ_t \gamma_{\alpha (s)\alpha (t)})_r \cup
\beta]      \right )
\\  \nonumber
&& +  ( \alpha <->   \beta  )      \\ \nonumber
&&+     2 \int ds \delta^3 (S_1 , \alpha (s)) \dot{\alpha}^a(s)
 \int dt \delta^3 (S_2 , \beta(t)) \dot{\beta}^b(t)
 \\ \nonumber
 &&\left.  \times
\left (  \sum_r (-1)^{q_r}
\Psi [(\alpha \circ_s   \gamma_{\alpha (s)\beta (t)} \circ_t \beta )_r  ]
\right )   \right ]
\end{eqnarray}

At the intersection point the curves
$\gamma_{\alpha (s)\alpha (t)}$ and $\gamma_{\alpha
(s)\beta(t)}$
are trivial but their orientations give us an ordering with which to
define
the action of the splittings and reroutings. It then follows that,
\f
 \sum_r (-1)^{q_r}
\Psi [(\alpha \circ_s \circ_t \gamma_{\alpha (s)\alpha (t)})_r \cup
\beta]
= 6 \Psi [\alpha \cup \beta ] = 6 \Psi [\Gamma_{\times} ]
\ff
and
\f
\sum_r (-1)^{q_r}
\Psi [(\alpha \circ_s   \gamma_{\alpha (s)\beta (t)} \circ_t \beta )_r  ]
= 4 \Psi [\Gamma_{><} ]
-2 \Psi [\Gamma_{\times}  ]
\ff
Using the formula (24)
for
the intersection numbers we then have,
\begin{eqnarray}
\hat{\cal A}^2_{approx}[{\cal S}_i ] \Psi [\Gamma_{\times} ] &= &
6 l_P^4
\left ( I[{\cal S}_i ,\alpha ]^2 + I[{\cal S}_i ,\beta ]^2  \right )
\Psi [\Gamma_{\times} ]
\\ \nonumber
&& + l_P^4 I[{\cal S}_i ,\alpha ]I[{\cal S}_i ,\beta ]
\left ( 4  \Psi [\Gamma_{><}]  -2 \Psi [\Gamma_{\times}    \right )
\\  \nonumber
&=&   l_P^4 \left (  10 \Psi
[\Gamma_\times ] +
4 \Psi [\Gamma_{><}]     \right )
\end{eqnarray}
We have dropped the intersection numbers in the last line, because
we assume that the surface element goes through the intersection point.
Similarly, one can show that
\f
\hat{\cal A}^2_{approx}[{\cal S}_i ] \Psi [\Gamma_{><} ] =
l_P^4 \left (  10 \Psi
[\Gamma_{><} ] +
4 \Psi [\Gamma_{\times}]     \right )
\ff
These results may be summarized by introducing a {\it rerouting
matrix}, ${\cal M}_{IJ}$,  whose entries are just numbers.  We have
thus shown that
\f
<\Gamma_{I} | \hat{\cal A}^2_{approx}[{\cal S}_i ]   =
l_P^4  {\cal M}_{IJ}
<\Gamma_{J} |
\ff
It is straightforward to generalize this to the general case that a
surface
goes through an arbitrary
intersection point where $n$ lines cross.
If ${\cal S}_i$ is the section of the
surface that goes through the first intersection point of a general
graph $\Gamma$, we have
\f
\hat{\cal A}^2_{approx}[{\cal S}_i ] \Psi [\Gamma_{I_1 ,....I_N} ]=
l_P^4  {\cal M}({\cal A})_{I_1 J}^{n} \Psi [\Gamma_{J ,....I_N} ]
\ff
Here we have shown that the matrix $\cal M$ depends on the
number $n$ of lines that go through the intersection point and
we have also shown that it arises from the operator ${\cal A}$.
It is now easy to take the square root (the matrixes being finite
dimensional) and the limit (92) to find that,
\f
<\Gamma_{I_1 ,....I_N} | \hat{\cal A}[{\cal S}]  =
l_P^2 ({\cal M}({\cal A})^n)^{1 \over 2}_{I_1 J} <\Gamma_{J ,....I_N} |
\ff
We thus see that the effect of  the operator on the intersection is
to act with a finite dimensional matrix to rearrange the routings
through the point of intersection.
There are thus additional eigenstates of $\hat{\cal A}[{\cal S}]$
that come from diagonalizing the rearrangement matrices.  They will
be linear combinations of the characteristic states associated
with the routings through intersections.  As a result the
area operator has additional eigenvalues, but they are still a discrete
set.

It is not hard to see that these are all the eigenstates that can
be constructed in terms of the dual basis $<\gamma |$.  There
is one more class we have to study, which is when $\gamma$ has
retraced segments.
However, from the results of the previous section, it is clear
that the result of the action of $\hat{\cal A}[{\cal S}]$ on a dual
basis state $<\gamma |$ of a loop with retraced segments gives,
when the surface intersects a retraced segment, a mixing with other
routings through the segment  plus terms involving the dual bases
of graphs in which the segment has been cut as in (102) and (104).
However, when it acts on a state with support on simple loops
it produces a state which only has support on simple loops.  Thus,
it is not a symmetric operator.  One can then show that
because of this no eigenstate of $\hat{\cal A}[{\cal S}]$ can
have a term involving $<\gamma |$, where $<\gamma |$ has
retraced segments.  Thus, all dual eigenstates must
be linear combinations of $<\gamma |$, where $\gamma$ have no retraced
segments, but are only simple loops, or loops with intersections.  It
then follows  that in the space of states
which are normalizable under the discrete measure
the complete set of eigenstates are the
characteristic
states of simple loops together with the eigenstates of the rearrangement
matrices of intersecting graphs\footnote{It then follows that the complete
set of eigenstates of $\hat{\cal A}[{\cal S}]$ do not span the
representation space ${\cal S}^{kin}_\Sigma$ of the loop algebra.  As a
result, $\hat{\cal A}[{\cal S}]$ cannot be a Hermitian operator on this
whole space.  However,  it does have an
invariant subspace, which consists of states with support only on
simple loops and  loops with
ordinary intersections with no retracings. What we have
done in the previous section is then to choose the inner
product such that on its invariant subspace $\hat{\cal A}[{\cal S}]$
is an hermitian operator.  Similar remarks apply to the operator
$\hat{Q}[\omega ].$}.

\subsection{The construction of the volume operator}

The construction of the operator which measures the volume of a
region
quantum mechanically uses the same two ideas that were basic to
the
construction of the area operator:  first, integrate the hands
against
two dimensional surfaces and second, rather than integrate
against a density divide the region up into
pieces
and approximate the volume of the pieces.

Classically, in terms of the
new variables the volume of a region $\cal R$ is equal to,
\f
{\cal V}[{\cal R} ]= \int_{\cal R} d^3x \sqrt{det(\tilde{E}^{ai}(x))}
\ff
where
\f
det(\tilde{E}^{ai}(x))={1 \over 3!}
\epsilon_{abc} \epsilon_{ijk} \tilde{E}^{ai}(x)
\tilde{E}^{bj}(x)\tilde{E}^{ck}(x)
\ff

Before describing the right way to define an operator for
${\cal V}[{\cal R}]$, let us take a moment to explain why the
obvious
approach does not work.  The obvious approach is to define first an
operator for $det(\tilde{E}^{ai}(x))$ at a point, x,
and then integrate the
resulting operator (which must be a density) over the region
$\cal R$.  The problem with this procedure is that it is impossible
to define an operator which measures the determinant of the
metric
at a point.  This is true for the same reason we found in section 4.1
that it was impossible to define an operator to measure the
densitized inverse metric, $\tilde{\tilde{q}}^{ab}(x)$ at a point.

To see this let us consider a multi-loop $\rho$, with three
components,
$\alpha , \beta $ and $\gamma$ that intersect only at a single
intersection
point $p=\alpha (s_0) = \beta (t_0) =\gamma (u_0)$.  A naive
unregulated "calculation" gives
\begin{eqnarray}
det(\tilde{E}^{ai}(x)) \Psi [\rho  ] &=&  l_P^6
\oint ds \oint dt \oint du
 \\  \nonumber
&&\times
\delta^3 (x,\alpha (s))
\delta^3 (x, \beta (t)) \delta^3 (x,\gamma (u))
 \\  \nonumber
&&\times
 \epsilon_{abc}
\dot{\alpha}^a(s) \dot{\beta}^b (t) \dot{\gamma}^c (u)
\ \Psi [\rho ]
\end{eqnarray}

This is clearly undefined, as it involves, where it does not vanish,
the product of the three delta functions.  As in the case of the
inverse metric,
the problem is then
to define the product of the three delta functions through
a regularization procedure.  It is clear that, however this is done,
any
right answer should be expected to vanish except at the point $p$
where
there are three independent tangent vectors.   A regularization
procedure
for $det(\tilde{E}^{ai})$ can be constructed along the lines
of the one we introduced in section 4.1 for the inverse metric.

To do this we again make use of a flat background  metric
$h^0_{ab}$ in the neighborhood, $\cal U$,  of $x$ and we use the
smearing
function (59).  For each three points
$x,y,z$ of $\cal U $
we may use the euclidean coordinates based on $h^0_{ab}$  to
define a
coordinate circle
$\gamma_{xyz}^a(s)$ which passes through the points $x,y$ and $z$
at
the parameter values, $0, 2\pi/3 $ and $4\pi/3$, respectively.  We
then
define a generalization of (56)
\f
T^{abc}(x,y,z) = T^{abc}[\gamma_{xyz}](0, 2\pi/3, 4\pi /3  )
\ff
We now may define a regularized version of $det(\tilde{E}^{ai})$ by
\f
H_\epsilon (x) \equiv \int d^3y \int d^3z \int d^3w f_\epsilon (x,y)
f_\epsilon (x,z)
f_\epsilon (x,w) \epsilon_{abc} T^{abc}(y,z,w)
\ff

Interestingly enough, when we act with this operator on the state
$\Psi $ and evaluate the result on the intersecting loop $\rho$, this
operator is finite in the limit $\epsilon \rightarrow 0$.  But it
is hopelessly regularization dependent.

 The
result is of the form,
\f
 \lim_{\epsilon \rightarrow 0} \hat{H}^\epsilon (x) \Psi [\rho ]  =
 \delta^3 (x,p)
{   det(h^0)  (x) \epsilon_{abc} \dot{\alpha}^a(s_0) \dot{\beta}^b
(t_0)
\dot{\gamma}^c (u_0)
\over |\dot{\alpha}(s_0)| |\dot{\beta}(t_0)| |\dot{\gamma}(u_0)| }
\ \Psi [\rho ]
\ff
where, as in (66),  the norm
$|...|$ is taken with respect to the background metric $h^0_{ab}$.

Just as in the case of our attempt to define
$\tilde{\tilde{q}}^{ab}(x)$, the regularization
dependence comes in two places.  First, in order
to preserve reparameterization invariance  the factor
$\epsilon_{abc} \dot{\alpha}^a(s_0) \dot{\beta}^b (t_0)
\dot{\gamma}^c $
in the numerator must be balanced by the factor linear in each of
the
tangent vectors in the denominator.  In the calculation based on the
smearing function (59) this factor is
$|\dot{\alpha}(s_0)| |\dot{\beta}(t_0)| |\dot{\gamma}(u_0)|$.
However, the key point is that any such factor must depend on
additional background structure.    Second, because
$det(\tilde{E}^{ai})$ has
density weight two, there must be in the answer a factor
of density weight two because the density weight of the
$\delta^3 (x,p)$ is balanced by the $\epsilon_{abc}$.  Again, in
the particular calculation we have done
this weight two function turns out to be $det(h^0)$, but in
general some such density factor must emerge from the
calculation and, because there is no density that can
be defined only from the geometry of the problem,
this can only come from some
arbitrary, auxiliary, structure used in the regularization.

As the determinant of the metric is only a single function, it is
clear that
we get absolutely zero information from this operator.  The moral
of
the story is that we cannot measure the determinant of the metric
at a point in the loop representation.

How are we then to measure the volume of a region?  The only way
to do it
is to find a procedure for defining the volume of a region that does
not
involve ever defining the volume element at a point.  This can be
done
by mimicking the procedure by which we
defined the area of a surface.  We divide the region $\cal R$ up into
$N$
subregions ${\cal R}_i$, $i=1,...,N$ and for each of these we define
an
approximate expression for the square of the volume,
${\cal V}^2_{approx} [{\cal R}_i]$,
that is exact
in the limit of infinitesimal volume.  We then take the limit in
which we
take $N$ to infinity, making the volume of each subregion ${\cal
R}_i$
infinitesimal.
We then have,
\f
{\cal V}[{\cal R}] = \lim_{N \rightarrow \infty}
\sum_{i=1}^N \sqrt{\left |  {\cal V}^2_{approx} [{\cal R}_i]\right |}  .
\ff

We may do this in the following way.  Let us introduce a euclidean
coordinate
system, $y^{\hat{a}}$,
on ${\cal R}$ and consider a cubic lattice with $N$ cells, defined
with
a coordinate lattice spacing $d \approx N^{-1/3}$.  We will consider
the cells
to be the subregions ${\cal R}_i$.  We can define an approximation
to
the volume of the $i$'th cell in the following way.  Take three of
the
faces of the cell that share a common vertex and call them
${\cal S}_i^\alpha$, where $\alpha =1,2,3$.  We then define,
\f
{\cal V}^2_{approx} [{\cal R}_i]\equiv \int_{{\cal S}_i^1}
d^2S_1^{a_1 a_2}
\int_{{\cal S}_i^2} d^2S_2^{b_1 b_2}\int_{{\cal S}_i^3} d^2S_3^{c_1
c_2}
\epsilon_{aa_1 a_2}\epsilon_{bb_1 b_2}\epsilon_{cc_1 c_2}
T^{abc}(S_1 ,S_2, S_3 )
\ff
Now, classically, it is not hard to show that for smooth fields,
\f
T^{abc}(x,y,z)= det(\tilde{E}^{ai})(x_i) \epsilon^{abc} +
O (d \partial \tilde{E} )
\ff
where $x_i$ is a point in the (coordinate) center of the cube.  If we
now
consider infinitesimal boxes and define $s_a^\alpha =
\epsilon_{aa_1 a_2} d^2S^{a_1 a_2}_\alpha $, we see that
\f
\left | {\cal V}^2_{approx} [{\cal R}_i]\right |  =
\left |  det(q)(x_i) \epsilon^{abc}
s_a^1 s_b^2 s_c^3   \right | +O(d) = {\cal V}^2 [{\cal R}_i]  + O(d)
\ff
The relation (120) then follows, as the limit $N \rightarrow \infty$ is
to be taken in such a way that $d \rightarrow 0$.

Let us now translate  ${\cal V}^2_{approx} [{\cal R}_i] $ into
a quantum operator by writing it as (121) with the $T^3$ now
an operator.  We may now evaluate it on a quantum state.
The
calculation is similar to the evaluation of (93).   We will
evaluate the action on a multi-loop $\rho$, which will have
components that we will denote $\rho_I$.
For any multi-loop $\rho$ there will be
a lattice spacing $d$ which is small enough that each wall face,
${\cal S}^\alpha_i$ of each cell is pierced by each component of
$\rho$ at most
once.    Let us assume that $d$ is taken at least that small.   Using
the
definition of the intersection number (24) of a curve and a surface,
we
then find,
\begin{eqnarray}
\hat {\cal V}^2_{approx} [{\cal R}_i] \Psi [\rho ] &=& l_P^6 \sum_I
\sum_{J} \sum_{K}
I({\cal S}_i^1,\rho_I)I({\cal S}_i^2,\rho_J)I({\cal S}_i^3,\rho_K)
\\ \nonumber
&&\times  \left (  \sum_r (-1)^{q_r}
\Psi [ ( \rho  \circ \circ  \circ \gamma_{\rho_I (s_1^I) \rho_J (s_2^J)
\rho_K (s_3^K) }    )_r ]   \right ).
\end{eqnarray}
Here $s_\alpha^I$ is the intersection point for the $I$'th component
$\rho_I$  with
the $\alpha$'th surface.  (We can take these coordinates to vanish
when there is no intersection.)

In order to take the limit (120) we need to be able to define the square
root of this action.  However, we only need do it for an arbitrarily
small lattice spacing $d$.  Let us then consider what happens to
the factors in (124) as we take the limit $d \rightarrow 0$, making the
boxes smaller and smaller.
There are again two factors to consider: the
c-number factors which are composed of the intersection
numbers and the rearranging of the loops.

Let us first consider the c-number factors.
The main thing to see is that in the limit that we shrink the boxes
down
there will only survive one term for each  intersection point at
which there are three independent tangent vectors.  Clearly, a box
that
none of the curves pass through gives zero.   Consider next a box
that only one of the three curves, say $\alpha=\Gamma_1$,
passes through.  In this case the result is zero, because it can pass
through at most two walls of the box.

Consider then  a box that contains a double intersection point.
There are then two curves in the box, which we will call $\alpha$
and $\beta$.
The
nonvanishing terms in the intersection numbers in
(124) will come from terms in which
one
of them, say $\alpha$, passes through two walls and the second,
$\beta$, passes through the third wall.   However,
for this case, as the reader may show,
the factor coming from the rearrangement of loops
will vanish.

  This leaves only the possibility  that the
 box contains a triple intersection point $p$ at which at least
 three curves meet. To get a nonvanishing contribution,
 the tangent vectors of   three of the curves
 at the intersection point must be independent because
 only if this is the
 case will three different curves pass
 each through a different of the three walls of the box, no matter
 how small we take the box.    Each
of these meetings of a wall and a curve then
contributes a separate intersection number.  It remains
only to figure out the effect of rearranging the loops.

If we use the continuity conditions (70) and (71)
to define the limits, then
the result is just a rearrangement matrix, which reshuffles the
routings at the intersection point $p$.  To write this, we use
the notation we developed in the previous section and write
the graph containing the loops with the intersection $p$ as
$\Gamma$.  Let us assume, to begin with, that $p$ is
the only intersection point of $\Gamma$.  A particular choice of
rearrangements at $p$ gives a loop which will be labeled,
as before, $\Gamma_I$, where $I=1,...,N_p$.  Here $N_p$ is the
dimension of the finite dimensional state space $V_\Gamma$
associated with the intersection.

We then find that in the limit in which the box becomes
arbitrarily small we will have
 \f
 \hat{\cal V}^2_{approx} [{\cal R}_i] \Psi [\Gamma_I ] =l_P^6
  {\cal M}({\cal V})_{IJ} \Psi [\Gamma_J ],
 \ff
 where ${\cal M}({\cal V})_{IJ} $ is an $N_p \times N_p$ matrix
gotten by summing over the rearrangements in (124) and then shrinking
the loops down as the boxes shrink.

 Now, in the limit that $N \rightarrow \infty$ each triple
intersection
 point has a unique box surrounding it such that each of the three
lines
 pass through a separate wall.  We can then take the square
root and compute the limit in equation (120) to find,
 \f
 \hat {\cal V} [{\cal R} ] \Psi [\Gamma_I ] =l_P^3
  [{\cal M}({\cal V})^{1 \over 2}]_{IJ} \Psi [\Gamma_J ]
\ff

It is then straightforward to extend this to the general case that
the graph $\Gamma$ has $m$ such intersection points, $p_\rho$,
with $\rho=1,...,m$.   We have then a separate rearrangement matrix
${\cal M}({\cal V})_{IJ}^\rho$ at each intersection point.  We can
label the different routings with a separate index at each
intersection
point.  Thus a choice of routings will be given by
$\Gamma_{I_1,...I_m}$
with $I_\rho$ labeling the independent choices of routings
through $p_\rho$.  The result is then,
\f
 \hat {\cal V} [{\cal R} ] \Psi [\Gamma_I ] =l_P^3 \sum_{\rho=1}^m
  [{\cal M}^\rho({\cal V})]^{1 \over 2}]_{I_\rho J}
  \Psi [\Gamma_{I_1,...J,...,I_m }]
\ff

We can make some comments on this result.

1)  The eigenvalues of the volume operator are also quantized in
Planck units, being given by the Planck volume times the square roots
of the
eigenvalues
of the matrices ${\cal M}^\rho({\cal V})_{I J} $.  At present, nothing
is known about these matrixes except the simplest examples.  The
eigenstates of the volume operator are then constructed from
linear
combinations of the characteristic functions of intersecting
loops, using the eigenvectors of the matrices
${\cal M}^\rho({\cal V})_{I  J} $.  We thus see that, roughly, the
volume operator seems to count the number of triple (or greater)
intersections, in units of the Planck volume.

2)  We see that the operator is, as in the case of the area operator,
finite and background independent.  Its action in the basis
of characteristic states is {\it block diagonal}, in
that it induces a matrix in the finite dimensional
subspaces $V_\Gamma$ associated with each graph.  This is
actually a general property: a large class of operators that are
expressed classically as single integrals over densities express
themselves as rearrangement matrices in the finite dimensional
blocks associated with rearranging the routings through graphs.

3)  By taking the region $\cal R$ to be the whole three manifold
$\Sigma$ we find a diffeomorphism invariant operator: the volume
of the universe.  As discussed above in section 2.4, this operator acts
directly
on diffeomorphism invariant states, by (127).   This is the first
example
we have of an explicit construction of a spatially diffeomorphism
operator nonperturbatively.

\subsection{The construction of the operator $Q[\omega ]$}

In this section we study the construction of an operator to
represent the observable $Q[\omega ]$ introduced in section 2.3.
Let $\omega_a(x)$ be an arbitrary smooth
one
form on $M$. Classically, we defined $Q[\omega]$ to be,
\f
        Q[\omega ,\tilde{E} ] \equiv   \int d^3x
\sqrt{\tilde{E}^{ai}(x) \tilde{E}^b_i(x)
\omega_a(x)  \omega_b(x)}
\ff
For every $\omega$, the observable
$Q[\omega ,\tilde{E}]$   is a well defined observable
on the configuration space of frame fields $\tilde{E}^{ai}$.
Moreover, the collection of $Q[\omega ]$ for all smooth
$\omega$ provide a good coordinate system on the space
of these frame fields:  if we know
$Q[\omega]$ for
every smooth one form $\omega$, then we can
reconstruct  the frame fields
$\tilde{E}^{ai}$, up to local $SU(2)$ gauge transformations.

To construct a quantum operator to represent $Q[\omega , \tilde{E}]$
we must
first regulate this classical expression.  We thus write
\f
Q[\omega ] = \lim_{\epsilon\rightarrow 0} Q[\omega ]^\epsilon
\ff
where
\f
Q[\omega]^\epsilon \equiv \int d^3x
\sqrt{W_\epsilon (\omega ,x )}
\ff
and the factor inside the square root is
\f
W_\epsilon (\omega, x)
\int d^3y \int d^3z f_\epsilon (x,y) f_\epsilon (x,z)
T^{ab}(y,z) \omega_a (y) \omega_b (z) ,
\ff
where $f_\epsilon (x,y)$ is a smearing function which satisfies
(57) and (58).
We will proceed to define the action of this  operator on a state
$\Psi [\alpha ]$ through the following steps. 1)  Define
the action of the regulated factor $\hat{W}_\epsilon (\omega ,x)$.
2)  Isolate the leading term of the loop factors of
the  regulated
operator, as $\epsilon \rightarrow 0$.  3) Take the square root.
4) Integrate.
5)  Take the limit in which $\epsilon$ goes to zero.

Beginning with the first step, we
use a definition which follows the classical one, so that,
\begin{eqnarray}
\hat {W}_\epsilon (\omega ,x)  \Psi [\alpha ] &=&
\int d^3 y\int d^3 z f_\epsilon(x,y) f_\epsilon(x,z) \
        \hat T^{ab}(y,z) \omega_a(y) \omega_b(z)  \Psi [\alpha ]
        \\  \nonumber
  &=&
   l_P^4
        \int ds f_\epsilon(x,\alpha(s))
        \dot\alpha^a(s)\omega_a(\alpha(s))
    \    \int dt f_\epsilon(x, \alpha(t))
    \dot\alpha^b(t) \omega_b(\alpha(t))
        \\ \nonumber
        & & \times
\left (   \sum_r (-1)^{q_r}
	\Psi((\alpha\circ \circ \gamma_{\alpha (s)\alpha (t)})_r)
	\right )
\end{eqnarray}
We want to take the square root of this action for small $\epsilon$.
To do this we use the conditions (70) and (71), so that the operator is
expressed in terms of a leading, $\epsilon$ independent piece,
whose
square root we can take, and a correction proportional to
$\epsilon$.
We then have,
at a point where  the loop is nonintersecting,
\f
\left (  \hat{W}_\epsilon  (\omega ,x)  \right )\Psi [\alpha ]=
 6 \left (  l_P^2 \oint ds \ f_\epsilon (x, \alpha (s)) \
 \dot{\alpha}^a (s) \
\omega (\alpha (s))_a   \right )^2  \Psi [\alpha ]   + O(\epsilon )
\ff
The $6$ comes from the sum over routings in (132), as in
previous cases.

The square root is then the square root of the leading piece plus
a correction proportional to $\epsilon$ that we don't need to
compute.
It is thus,
\f
\hat{W}^{1\over 2}_\epsilon  (\omega ,x)  \Psi [\alpha ] =
\sqrt{6}  l_P^2 \left |  \oint ds f_\epsilon (x, \alpha (s))
\dot{\alpha}^a (s)
\omega (\alpha (s))_a    \right |     \Psi [\alpha ]   + O(\epsilon )
\ff
We may now integrate and take the limit to find,
\f
 (\hat Q \Psi)[\alpha] \equiv
        \lim_{\epsilon\to 0} [\hat Q_\epsilon \Psi][\alpha ]=
        6 l_P^2  \int_\alpha |d\alpha^a\ \omega_a|\  \Psi[\alpha]
\ff

Note that the resulting operator is, again, finite and independent
of the background structure that went into the definition
of the regularization.

\subsection{More about weaves and the semiclassical equivalence
of quantum states to classical geometries}

In section 2.3 I gave an introduction to the idea of a weave.  The
main idea uses the fact that the characteristic states are
eigenstates of the operators that measure metric information
and that, in particular, the lines of the loops carry a quantized
flux of area, that they contribute to the area of any surface they
cross. We can then construct a quantum state that
approximates a given classical metric $h^0_{ab}$ by building a
characteristic state on a set of loops that are distributed so that
one (or more properly $\sqrt{6}$) loops per Planck area cross
each surface, as measured by that background metric.

Here I would like to be more precise about what is meant by
approximating a classical solution, after which I will describe
the construction of one example of such a weave state.

It will be convenient to study the equivalence of a quantum state
with a classical metric by using the operator $Q[\omega]$.  Let me
first note that, classically,  if we know this observable for all
smooth one forms $\omega_a$ the classical metric $q_{ab}$ is
completely determined.  This can be shown by taking $\omega_a$
of arbitrarily small compact support; the different components
of $\tilde{\tilde{q}}^{ab}$ are then gotten by studying $Q[\omega]$
for $\omega$ that are linear combinations of some set of basis
one forms.

Now, what if we only know the value of this observable for
$\omega$
that are slowly varying on some scale $L$?   The idea is then that
we only determine the metric to within that scale $L$.  It is a
little
tricky to say this precisely, because we need a metric to describe
which one forms are slowly varying.  A good formulation turns out
to be the following:

We must first give a definition of a slowly varying one form $L$.
We
say that a one form $\omega_a$ is slowly varying on a scale $L$
with respect to a metric $q^{ab}$ if at all points of $\Sigma$
\f
{ {|  \nabla_a   \omega_b  |^2  } \over
{ |\omega_a |^2       }}
< {1 \over {L^2} } .
\ff
where $\nabla_a$ is the metric covariant derivative of
$q_{ab}$ and the norms are taken using $q_{ab}$.  We will
also use normalized one forms so that
\f
\int_\Sigma \sqrt{q} | \omega_a |^2 = 1
\ff

We then can use such slowly varying $\omega_a$ to state a criteria
for when another metric $q^\prime_{ab}$ approximates $q_{ab}$
on scales larger than $L$.

We say that $q_{ab}^\prime $
 Q-approximates $q_{ab}$ at scales larger than $L$, to an accuracy
$\epsilon$, if for all
 one forms $\omega$ which satisfy (136) and (137)
\f
      \left |Q[\omega ,q^\prime ]-Q[\omega ,q] \right |  < \epsilon
\ff

More details of this notion of one metric approximating another one
at large scales are given in reference\cite{weavepaper}.

We then can apply this notion to the quantum theory in the
following way.
Let $\Psi $ be an eigenstate of the operator $\hat{Q}[\omega ]$ so
that
\f
\hat{Q}[\omega ] \Psi = \lambda [\omega, \Psi ] \Psi.
\ff
We will then say that the state $\Psi$ $Q$-approximates a
classical
metric $q_{ab}$ on a scale $L>> l_P $ if for all slowly varying
$\omega_a$ on the scale $L$
\f
      \left |\lambda[\omega ,\Psi ]-Q[\omega ,q] \right |  < {l_P \over
L}
\ff
Note that there is no need to introduce a separate parameter
$\epsilon$ to measure the error, that will naturally come out to
be of order $l_P/L$.

I will now introduce a specific state that satisfies this criteria.
Let us fix a flat metric $h^0_{ab} $ on $\Sigma$.  Let $x$ be a
coordinate system in which the metric $h^0_{ab} $ is locally
euclidean. We
will construct a
a collection, $\Delta$,   of loops $\alpha_{\vec n}$.  Each loop
$\alpha_{\vec n}$, is a circle of radius $r$ (defined with respect to
$h^0_{ab}$).  The center  of the loop  $\alpha_{ \vec{n}}$ is placed
at the
vertex
${\vec n}$ of a square lattice with lattice spacing $d$.  The
orientation of the circles is random (each circle being oriented in a
different direction, chosen from a uniform probability distribution
on
the unit sphere, again defined with respect to $h^0_{ab}$.)     We
will call such a
collection of loops, defined with respect to a smooth metric,
$h^0_{ab}$ as an $h$-weave, or weave, for short.

It is convenient to describe
this weave   in terms of the
"average density of loops", $l$, which is  defined by
\f
        l^2 = {{\rm Volume}\over{\rm Length\ of\ the\ loops}}
        ={d^3\over 2\pi
r}
\ff
and the  dimensionless quantity
\f
        z = { r\over d}.
\ff

We may then consider the characteristic state of this collection
of loops, which will be denoted $\Psi_{weave}$.  Using (135) this is
an eigenstate of $\hat{Q}[\omega ]$ with eigenvalue,
\f
\lambda [\omega , \Psi_{weave}] = \sqrt{6}l_P^2 \sum_{\vec{n}}
\oint_{\alpha_{\vec n}} | ds^a \omega_a |
\ff

Now, we must compute what this is for slowly varying $\omega$,
assuming that the parameters $d$ and $r$ that govern the weave
are much smaller than the scale $L$ on which $\omega_a$ is
slowly varying.  We find that
\f
        \lambda [\omega ,\Psi_{weave} ]
        =  \sqrt{6} \pi l_P^2   d^{-3} r \int d^3x \sqrt{h^0} |\omega | +R
        = {\sqrt{6} l_P^2\over 2  l^2} Q[\omega , h^0 ] +R
\ff
Here $R$ is the error, which can be shown to satisfy
\f
|R| < { r+2d \over L} {\sqrt{6} l_P^2 \over 2l^2 } Q[\omega ,h^o]
\ff
We can then satisfy the condition (140) if we take the density
parameter
$l$ for the loops in the weave to satisfy
\f
l^2={\sqrt{6}\over 2} l_P^2
\ff
and the error will be of the order of the ratio $l_P/L$ if the
two parameters $r$ and $d$ are separately taken to be of the
order of $l_P$.

We close this section with some comments on these results:

1)  The definition of $Q$-approximation, given by (138) at the classical
level can be shown to be equivalent to other notions that can be
defined using the integrated norms of fields of different tensorial
character, or by directly integrating the metric against a
tensorial test density  $\tilde{f}^{ab}$, as in
$q(\tilde{f}) = \int_\sigma \tilde{f}^{ab}q_{ab}$.   We then use the
notion of $Q$-approximate quantum mechanically, because the
$Q$ operator is well defined in the loop representation, whereas
not all  observables which are functions of the metric
extend to the quantum theory.

2)  One can check that a similar notion of approximation can be
defined
using the areas of an appropriate set of surfaces and that the
resulting notion agrees with the one we have defined here both
classically and quantum mechanically.

3) There are many weave configurations for a given classical
metric $h^0_{ab}$.  By construction, the definition of
approximation
on large scales constrains nothing about the behavior of the weave
at short distances.

4)  Thus, as noted before, a state $\Psi_{weave}$ can be an
eigenstate
of the volume operator with zero volume and still approximate a
flat
metric $h^0_{ab}$ on a scale $L>>l_P$ when measured by the
operators
for $Q[\omega ]$ or areas.  There is no problem with this, as the
volume
operator measures something about the connectivity of the weave
it has
to do with the short distance structure and has, thus, nothing to do
with
the semiclassical limit.

5)  Finally, we should emphasize that a weave state
$\Psi_{weave}$ is
{\it not} going to be the vacuum state of the theory, either
perturbatively
or nonperturbatively.  It is an eigenstate of the three metric,
whereas the
vacuum  presumably must minimize the product of the uncertainty
between
the three metric and its evolution in time.  The problem of the
construction of the ground state is a key problem, which is
discussed
in some detail below, in section 6.3.

\section{Some recent developments in the classical theory}

In this chapter I would like to review some very interesting recent
developments in the classical formulation of general relativity.
These build on the Ashtekar formulation and concern four subjects.
In section 5.1 I describe an important  result  of Capovilla, Dell and
Jacobson.    They show that it is possible,
to solve the classical Hamiltonian and diffeomorphism constraints
exactly, except for a set of configurations of measure zero.  This makes it
possible to write new forms for the action for general relativity, one
of which is described in section 5.2.  This
leads to an interesting connection between general
relativity and Chern-Simon theory.

Since the beginning of work on the Ashtekar formulation, we have known
that the new forms of the constraints and equations of motion admit
solutions that are not also solutions to the Einstein's equations because
the metric is either degenerate or distributional.  Recently, a
lot has been
learned about such
solutions and it is becoming increasingly clear that
they play a role in the
quantum theory.  Essentially, the quantum geometry we uncovered in
chapter 4 can be described in terms of classical distributional geometries.
This is the subject of section
5.3.   Finally,
these classical distributional geometries make possible a new kind of
discretization of general relativity, analogous to the Regge calculus.  This
is the subject of section 5.4.

\subsection{How to solve the classical constraint equations exactly}

One of the more striking developments of the last two years is that
the Hamiltonian and diffeomorphism constraints can be solved,
{\it in closed form} for all but a measure zero of cases\cite{CDJ}.
Indeed, the
solutions are so simple that it is surprising in retrospect that they
were not discovered earlier\footnote{Partial results in the right
direction were found by Renteln and
Ashtekar\cite{rentelnthesis,book}.}  To construct these
solutions we do
something
that may seem at first strange.  We treat the gravitational
RmagneticS
fields,
\f
\tilde{B}^{ai} \equiv {1 \over 2} \epsilon^{abc} F_{bc}^i
\ff
as frame fields for the three manifold $\Sigma$.  That is, we
assume
that,
\f
det(\tilde{B}^{ai}  ) \neq 0
\ff
so that any vector field can be expanded in terms of them.  In
particular,
we can so expand the $\tilde{E}^{ai}$'s,
\f
\tilde{E}^{ai} = M^i_{\ j} \tilde{B}^{a j}
\ff
It is straightforward to show that the diffeomorphism and
hamiltonian constraints, (48) and (50) correspond, respectively,
to the conditions
that the inverse of $M^i_{\ j}$, which we will denote $\phi^i_{\ j}$,
is symmetric and that the trace is fixed by the condition
\f
\phi^i_{\ i} = -3 \Lambda
\ff

Thus, the theory can be expressed completely in terms of $A_a^i$
and
a symmetric matrix of scalar fields $\phi^i_{\ j}$ restricted by the
condition (150).  In terms of these fields the only remaining
constraint
is the GaussUs law constraint, which now, however, takes a new
form
\f
{\cal G}^i = \tilde{B}^{aj } {\cal D}_a [\phi^{-1}]^i_{\ j} = 0
\ff
Given a fixed $A_a^i$ this is a first order differential equation
for $\phi^i_{\ j}$.  Its solutions give pairs $(A_a^i , \phi^i_{\ j})$
that
are the free data for general relativity.

One should note that this form of the theory requires that (148) be
true everywhere.  Thus, there is a set of measure zero of solutions
in which $\tilde{B}^{ai}$ are degenerate that cannot be expressed
this way.  Among these solutions is, of course, flat spacetime.
Whether this is a good or a bad thing depends  on what one
wants to use the formalism for.  Whether it is a good or a bad thing
for the quantum theory is not known.

\subsection{New lagrangians and a connection with Chern-Simon theory}

Using the solutions of Capovilla, Dell and Jacobson
one can construct a Lagrangian form of the
theory that does not involve the metric or the frame field.  There
are actually several of these\cite{CDJ}, some of which just involve
$A_a^i$ and a single scalar density.  These were found by
Capovilla, Dell and Jacobson, in the same paper in which the
solutions to the constraints
were found\cite{CDJ}. One very nice form, which
involves $A_a^i$ and $\phi^i_{\ j}$ is\cite{CDJ,CS-on-edge},
\f
S(A,\phi ) = \int F^i \wedge F^j [\phi^{-1}]_{ij}
\ff
In this form of the variational principle, $\phi^i_{\ j}$ is to be
varied respecting its symmetry and the trace condition (150).   It is
then straight forward to show that if we {\it define}
$\tilde{E}^{ai}$
by
\f
\tilde{E}^{ai} = [\phi^{-1}]^i_{\ j} \tilde{B}^{a j}
\ff
the constraints (47), (48) and (50)  are satisfied.

Using the results of Capovilla, Dell and Jacobson
a variety of results can be derived about the
solution space to general relativity.
I will mention two here.  The first, due to Samuel\cite{sam1}, is
that in the presence
of a cosmological constant, every self-dual solution to the Einstein
equations comes from a self-dual $SU(2)$ Yang-Mills field that
satisfies,
\f
\tilde{E}^{ai} = -\Lambda \tilde{B}^{a i}
\ff
This means that the self-dual sector of the theory corresponds to
$\phi^i_{\ j}$ that satisfy
\f
\phi^i_{\ j} = - \Lambda \delta^i_j
\ff
The second result, due to Torre, is that for $\Lambda >0$ and
$\Sigma$
compact, the moduli space of gravitational instantons is
discrete\cite{charles-discrete}.
There are no continuous parameters as in the case of Yang-Mills
theory.

Related to the topics I have been discussing is  an interesting
connection between general relativity and Chern-Simon
theory\cite{CS-on-edge}.  Suppose we
wanted to base a quantization of the theory on the action (152).
Then we
would likely need to define the variational principle on a four
manifold $\cal M$ with
boundary $\Sigma$.
A boundary is a good thing to have if one wants to
construct observables for the theory or give a meaning to the path
integral.
We must then ask what kind of boundary conditions are
allowed and what boundary terms need to be added to the
action so that the variational principle is still well defined.  It
turns
out (the details will be presented elsewhere\cite{CS-on-edge})
that there is a very pretty
variational principle in which we demand that the connection be
self-dual
on the boundary $\Sigma$.  Making use of Samuel's result (154) it is
then easy
to see that the variational principle in this case is
\f
S(A,\phi ) = \int_{\cal M} F^i \wedge F^j [\phi^{-1}]_{ij}
+ \int_\Sigma Y_{CS} (A)
\ff
where $Y_{CS}$ is the Chern-Simon action.
Thus, if we construct a variational principle for general relativity,
with a non-vanishing cosmological constant, in
which the connection is restricted to be self-dual on the boundary,
we
induce the Chern-Simon action for the dynamics of the connection
on the
boundary\footnote{This result is
perhaps also related to a result
of Kodama\cite{kodama1}, who showed that
$exp[ \Lambda^{-1} \int Y_{CS} (A) ]$ is
an exact
solution to all the constraints of general relativity, in the
connection
representation in which states are functions of $A_a^i$, in
the presence of the cosmological constant $\Lambda$. }.

\subsection{Distributional frame fields as the classical analogue
of quantum geometry}

One of the main points that we developed in the last chapter is that
there are quantum states which approximate classical metrics, as
long
as we probe on large scales.  These states emerged from
diagaonalizing
certain operators that are functions only of the three metric.  Now,
one might ask the following question:  usually in quantum
mechanics
when we diagonalize a  complete set of commuting operators
 corresponding to the coordinates of a configuration space, the
resulting
set of eigenvalues are a subset of that classical configuration
space.
However, what we found was not quite this, because the resulting
eigenvalues  have a distributional
character.
This is natural given the fact that the eigenstates are
characteristic
functions of particular loops.  But we may ask what happened to
the
expected connection between the eigenvalues and the classical
configurations?  Is there only a connection to the classical theory
in the sense of approximation at large scales?

The answer is that the eigenvalues of the operators we studied
in chapter 4 can
be related directly to classical configurations of the field, but
these
are distributional configurations\cite{weavepaper}.
Normally, the Einstein equations
do
not allow distributional configurations, because those equations
are non-polynomial.  But, in the Ashtekar form, all equations are
polynomial and, as I will now describe, distributional solutions
to the equations exist.

Let us then consider configurations of the frame fields which have
exactly the form of the eigenvalues of the operators of the operators
we studied in chapter 4.  That is, given a loop $\alpha$, let us
consider
\f
\tilde{E}^{ai}_\alpha  (x) \equiv a^2 \int ds \ \delta^3(x,\alpha (s))
\dot{\alpha}^a (s)  e^i_\alpha (s)
\ff
We see that this is naturally a vector density.  The $e^i_\alpha (s)$
is valued in the lie algebra of $SU(2)$ and will be taken to be
dimensionless.  Then the constant $a$ has dimensions of length.  If
we compare this with the expression (55), which is the result of
applying an operator containing an insertion of $\tilde{E}^{ai}(x)$
on a loop state, evaluated at $\alpha$, we see that in any such
expression arising from the quantum theory we will have
$a \approx l_P$.

Given a collection of loops $\Delta= \cup_i \alpha_i$, we can construct
a distributional frame field by adding the contributions from the
separate loops:
\f
\tilde{E}^{ai}_\Delta (x) \equiv \sum_i \tilde{E}^{ai}_{\alpha_i }(x) .
\ff

I would now like to show that
the configuration $\tilde{E}^{ai}_\alpha  (x) $ is the point of the
classical configuration space associated with the eigenstate
$|\alpha >$ of the observables we studied in chapter 4.    To accomplish this
one can  show that the observables $Q[\omega ]$, ${\cal A}[{\cal S}]$
and ${\cal V}[{\cal R}]$ can be evaluated on $\tilde{E}^{ai}_\alpha  (x) $,
and that the values are equal to the corresponding eigenvalues
of $|\alpha >$, when $a=6^{1 \over 4} l_P$ and $|e_\alpha (s)|=1$.

I will not give the details of the calculations here, they are very
similar to the quantum calculations that were described in chapter 4.
To begin with, one may try to define the inverse metric
$\tilde{\tilde{q}}^{ab}$ by taking the square of (157); one will find that
just as in the quantum case the product cannot be defined without
introducing an arbitrary density into the problem.  However, the
other operators can be defined through the same regularization procedures
that we found worked in the quantum theory.  We find that,
\f
Q[\omega, \tilde{E}_\alpha ] =a^2 \oint ds \left | \dot{\alpha}^a (s)
\omega_a (\alpha (s))  \right | | e_\alpha (s)|
\ff
and
\f
{\cal A}[{\cal S}, \tilde{E}_\alpha]= a^2 \sum_i |e_\alpha (s^*_i )|
\ff
where the sum is over the intersection points of the curve $\alpha$
with the surface $\cal S$ and $s^*_i$ are the parameters at the
intersection points.

For the volumes there is a result analogous to (127).  The volume of a
configuration $\tilde{E}_\alpha^{ai}$ is zero unless it contains
intersection points at which there are three independent tangent
vectors.  In the simplest case, where three segments $\alpha_i$,
$i=1,2,3$ meet at a point $p$, the volume in any region containing
$p$ is,
\f
{\cal V}[{\cal R}, \tilde{E}_\alpha ] = a^3 Tr[ e_{\alpha_1}e_{\alpha_2}
e_{\alpha_3}]
\ff
If there is more than one intersection point the volume is the sum
of the contributions from the intersection points.

Thus we see that in spite of having support on a set of measure zero,
distributional frame fields of the form of (157) carry finite values of
geometrical quantities such as areas and volumes.  Furthermore,
one may repeat the arguments of section 4.7 to show that any
smooth metric $h^0_{ab}$
can be approximated by a distributional frame field, where the notion
of approximation is defined also by equation (138).  Thus, the
classical configuration (157),
where the set of loops is given by the weave $\Delta$ defined in
section 4.7  (with $|e_\alpha (s)|=1$),
is a classical distributional geometry that approximates
the flat metric $h^0_{ab}$ up to terms of order $a/L$.
The distributional frame fields  are thus the
closest we can get in the classical
theory to a description of quantum geometry.  Indeed,
with $a$ set equal to the Planck length (times $6^{1\over 4}$),
they are something like the
Bohr orbits of quantum gravity: they give a classical picture that
we can use to see to a first approximation what quantum gravitational
effects are like.

In the next section we will give a further argument for taking these
distributional field configurations seriously.  We will see that the
constraints and equations of motion of the theory can be extended
to them.

\subsection{A new classical discretization of the Einstein equations}

I would like to bring together the topics of the last three sections
to show that the dynamical equations of general relativity, in the
Ashtekar form, can be completely solved for distributional
configurations of the form (157).   To begin, let us  return
to the ansatz of Samuel, equation (154).  The reader may verify
that if we plug this into the constraint equations (47), (48) and
(50),  it
solves them with a fixed value of the cosmological constant,
as long as $det(\tilde{E}^{ai}) \neq 0$.  However, notice also
that if the determinant vanishes we still have a solution,
but now for any cosmological constant, including zero.

Since the determinant of $\tilde{E}_\alpha^{ai}$, defined by (157)
vanishes, at least for simple loops $\alpha$,  we might try to find
a solution
consisting of such a frame field and $\tilde{B}^{ai}$ given by
a distribution with the same support,
\f
\tilde{B}^{ai}_\alpha  (x) \equiv   \int ds \delta^3(x,\alpha (s))
\dot{\alpha}^a (s)  b^i_\alpha (s)   .
\ff

Such solutions exist, if the constraints are defined through a regularization
process analogous to that we introduced in section 4.5 for the
volume operator\cite{ls-CS,ls-discrete}.  I will now
describe a large class of these.

To get nontrivial solutions we should take field configurations with support
on loops with intersections.  Let us consider one familiar example
of such a set of loops, which is a cubic lattice.  Given a coordinate chart
on $\Sigma$ let us define a standard cubic lattice with coordinate
lattice spacing $a$.  The vertices will be labeled by three integers
$\vec{n}$ and the links by the pair $(\vec{n}, \hat{a})$.
Thus, $\gamma_{\vec{n} \hat{a}} (s)$, with $s \in (0,1)$ will
be taken to refer to the link leaving the vertex $\vec{n}$ in the
positive $\hat{a}$ direction.

We will then take the frame field and curvature to be of the
suggested forms
\f
\tilde{E}^{ai}  (x) \equiv   a^2 \sum_{\vec{n}, \hat{a}} \int_0^1 ds
\delta^3(x,\gamma_{\vec{n} \hat{a}}(s))
\dot{\gamma}_{\vec{n} \hat{a} }^a (s)  e^i_{\vec{n} \hat{a}}
\ff
and
 \f
F_{ab}^i   (x) \equiv    {1 \over  G}  \sum_{\vec{n}, \hat{a}} \int_0^1 ds
\delta^3(x,\gamma_{\vec{n} \hat{a}}(s))
\dot{\gamma}_{\vec{n} \hat{a}}^c (s) \epsilon_{abc}
b^i_{\vec{n} \hat{a}}.
\ff
Here, the $  G$ is put in for dimensions. We would
like the free factors $b^i_{\vec{n} \hat{a}}$ to be dimensionless.
In Ashtekar's
formalism it is $  GF_{ab}^i$ that
has the dimensions of curvature, which is inverse length squared.

For this to be a valid ansatz,
we must find a connection $A_a^i$ whose
curvature is (164).  This is a standard problem from Chern-Simon theory,
where we sometimes have to consider connections whose curvatures
are valued on curves \cite{ls-CS}.  There the usual thing is to introduce
an auxiliary background metric to define the connection  as in the
usual solenoid problems of first year electromagnetism.  However,
there is an alternative which is very natural in this context, which
is to pick $A_a^i$ that are also distributional.  A natural
choice turns out to be to pick $A_a^i$ that have support
on the faces of the lattice.  If we call
${\cal S}_{\vec{n}\hat{a}\hat{b}}$ the two dimensional face which
leaves the vertex $\hat{n}$ in the positive $\hat{a} > \hat{b}$ directions,
we can write a distributional connection as
\f
A_a^i (x) =  {1 \over G} \sum_{\vec{n}\hat{a}\hat{b}} \int d^2S^{bc}
\epsilon_{abc}
\delta^3 (x,{\cal S}_{\vec{n}\hat{a}\hat{b}} (S) )
a^i_{\vec{n}\hat{a}\hat{b}}
\ff
The curvatures may be expressed in terms of these connections in
a simple way.  Any link $\gamma_{\vec{n} \hat{a}}$ is the meeting
point of four faces.  Let us label  these for simplicity  face one through
face four (choosing one as the arbitrary starting point.)  Then, using the
nonabelian stokes theorem, one can
show that,
\f
Tr\left [ e^{b }\right ] = Tr\left [
e^{a_1  }e^{a_2  }e^{a_3  }e^{a_4 }\right ]
\ff
where $b=b^i \tau_i$ and similarly for the $a^i$'s.

A field configuration is then given by associating a Lie algebra element
$a^i$ to every face and another one, $e^i$ to every link of the lattice.

We may then extend the constraints to these
configurations\cite{ls-discrete}. The Gauss's
law constraint, (47), is a straightforward calculation.  The result is that
\begin{eqnarray}
{\cal G}^i (\vec{n}) &=& \sum_{\hat{a}}
\left (  e^i_{\vec{n} \hat{a}}  -  e^i_{(\vec{n}-\hat{a}) \hat{a}}
\right )
\\ \nonumber
& +&{1 \over 8} \epsilon_{ijk} \epsilon^{\hat{a}\hat{b} \hat{c}}
\left (  e^j_{\vec{n} \hat{a}}  +  e^j_{(\vec{n}-\hat{a}) \hat{a}} \right )
\left ( a^k_{\vec{n} \hat{b}\hat{c}}
	+ a^k_{(\vec{n}-\hat{a}) \hat{b}\hat{c}}
	+a^k_{(\vec{n}-\hat{b}) \hat{b}\hat{c}}
	+ a^k_{(\vec{n}-\hat{a}-\hat{b}) \hat{b}\hat{c}}
   \right )
=0.
\end{eqnarray}

The Hamiltonian and diffeomorphism constraints involve
products of distributions and must be regularized.  This can be
done by following the same procedure developed in section 4.5
for the volume operator.  The volume is divided up into boxes and
an approximation for the constraint integrated over the box is
developed as an expression in which the frame fields and curvatures
are integrated over the walls of the box.  The resulting integrals
are all intersection numbers and one finds in the end that,
\f
{\cal C}(\vec{n})_{\hat{a}}= \epsilon_{\hat{a}\hat{b}\hat{c}}
 e^i_{\vec{n} \hat{b}} b^i_{\vec{n} \hat{c}}
\ff
\f
{\cal C}(\vec{n}) = \epsilon_{ijk} \epsilon_{\hat{a}\hat{b}\hat{c}}
 e^i_{\vec{n} \hat{a}}  e^j_{\vec{n} \hat{b}} b^k_{\vec{n} \hat{c}}
\ff

These last two constraints may be solved directly, using the
Capovilla-Dell-Jacobson trick (149), which is just algebraic.  The
result is that we are left with three sets of difference equations
given by the discrete analogue of (151).

Thus, we see that there are distributional solutions to the initial
data constraints in the Ashtekar form.  As we have seen that
such distributional configurations can approximate any smooth
configuration, what we have is a discrete approximation to the
intitial data equations of general relativity.

Finally, the evolution equations may be developed for these
distributional solutions.  This may be done by taking the equations
of motion and writing them as evolution equations for the
$e^i_{\vec{n} \hat{a}} $'s and $a^i_{\vec{n} \hat{a}\hat{b}} $'s.
Equivalently, one can make a Hamiltonian system as follows.
Take as the phase space the collection of all
$(e^i_{\vec{n} \hat{a}} ,a^i_{\vec{n} \hat{a}\hat{b} } )$ pairs and define
the Poisson brackets,
\f
\{ a_{\vec{n}\hat{a}\hat{b}}^i , e_{\vec{m}\hat{c}\ }^j \} =
{ \imath G \over a^2 }
\delta^{ij} \delta_{\vec{n} \vec{m}} \epsilon_{\hat{a}\hat{b}\hat{c}}
\ff
One may then show that  for  test fields $f_{ai}(x)$
and $\tilde{g}^{bj} (y)$, slowly varying on a scale $L$,
\f
\{ \int \tilde{E}^{ai}f_{ai} , \int \tilde{g}^{bj}A_{bj} \}   = \imath \int
\tilde{g}^{ai}f_{ai}  + O(a/L) .
\ff

The equations of motion may then be found by taking Poisson brackets
with the Hamiltonian constraint, I will not write them down here.
Their exact form, and more details about these distributional solutions,
may be found in \cite{ls-discrete}.

\section{Three open questions}

\subsection{Solutions to the Hamiltonian constraint}

As I mentioned in section 2.5, we know how to construct
an infinite dimensional space of solutions to the Hamiltonian
constraint.  This was, indeed, one of the first results achieved
in the program of nonperturbative quantization based on the
Ashtekar
variables.  The solutions were first found in the connection
representation (in which the states are functions of the connection
$A_a^i$) \cite{tedlee} before the invention of the loop
representation.

Since the discovery of these solutions in 1986 there has been
disagreement as to whether the set of solutions found then
was a complete set, or whether they were, in some sense,
accidental
of unphysical.  Recently there has been progress in several
directions
which sheds light on this issue.  I would like to briefly describe
these developments.

First, a few words of background, as the problem of finding
solutions to the Hamiltonian
constraint was not reviewed in detail above.
The key point, which makes it possible to find exact solutions
to the Hamiltonian constraint,
is that when the Hamiltonian constraint acts on
a loop function, all of the action happens
at singularities of loops.
By a singularity of a loop I mean here a nondifferentiable point of
a loop, which may be a kink or an intersection point.

To be more specific, the Hamiltonian constraint (49) involves a
product of
operators at a point and thus needs to be regulated.  It is then
represented
by a sequence of operators, ${\cal C}^\delta (x)$, which are, for
$\delta $ strictly positive, well defined operators.  The condition
that the Hamiltonian constraint annihilate a state $\Psi $ then
becomes\cite{tedlee}
\f
\lim_{\delta \rightarrow 0}
\left (  {\cal C}^\delta (N ) \Psi [ \gamma ] \right ) = 0   .
\ff
Here ${\cal C}^\delta (N ) \equiv \int d^3 x N(x) {\cal C}^\delta (x ) $
is
the constraint smeared with a { \it smooth } lapse $N$.
The topology in which the limit (172) is taken is usually
taken to be the pointwise topology, which is to say the expression
must vanish when it is evaluated at every
loop $\gamma $, as
well as for every smooth $N$.  This is, of course, the same thing
as asking that it vanish in terms of the norm (3).

What is  meant by saying that the action is concentrated on loops
with
singular points is then that $ {\cal C}^\delta (N ) \Psi [ \gamma ] $
is of
order
$\delta$ unless the loop $\gamma $ has a singular point.  If it does
then there are terms of order $\delta^{-2}$, $\delta^{-1}$ and order
$1$,
which do not go away when we take the limit $\delta \rightarrow
0$.

Now, the first implication of this is the following\cite{tedlee,carlolee}:

{\it Any loop functional
$\Psi [\gamma ]$ with support only on nonsingular loops is a
solution
to the Hamiltonian constraint.}

If $\Psi [\gamma ]$ is also diffeomorphism invariant then it is a
function only on the link classes
of the loop (which don't intersect because they have no singular
points.)
We can then state the result another way\cite{carlolee}:

{\it Associated to every  link invariant $I[ \{ \gamma  \} ]$ is an
exact  quantum
state of the gravitational field which is defined by
$\Psi [\gamma ] = I[\{ \gamma \}] $ if $\gamma$ is simple and
$\Psi [\gamma ] =0$ if $\gamma $ is nonsimple.  }

As there are an infinite number of link classes, we  have an
infinite dimensional space of exact quantum states of general
relativity.

 The question we would like to ask is then whether this set of
solutions is complete.  A great deal of work has been done
on this question recently, mostly by Berndt Bruegman, Rodolfo
Gambini and
Jorge Pullin\cite{berndtjorge-onC,berndtjorge-intersects,BGP}.  I would
like to discuss what is
known, presently, about this question.

First of all, what do we mean by completeness of the space of solutions
of the constraints?  As far as physics is concerned, what we need is
that the space of physical states carry a nontrivial, and perhaps
faithful, representation
of the algebra of physical observables.

Unfortunately, this criteria
is very hard to evaluate at the present time, given our ignorance
about the physical observables.  Given
this situation, we can look to other criteria for guidance.  However, we
must keep in mind that any other criteria we consider must be in the
end secondary to the one just enunciated.

We can consider criteria for completeness of the solutions to
the Hamiltonian which are mathematical: do we have, in fact, the complete
set of solutions to the equation as an infinite dimensional linear
differential
equation?
In order to be able to investigate
this question  we must be specific about both
the operator in question and the space of states within which
we are asking for the complete set of solutions.    Let us first
consider the space of states in question to be the space
${\cal S}^{kin}_\Sigma$ with the discrete norm (3) that has been
the subject of most of this paper.   We must, further, be specific
about the operator ordering and regularization procedure that goes
into the construction of the operator ${\cal C}^\delta (N ) $.  Let
me take this to be the original ordering and regularization as was
given in the papers \cite{tedlee,carlolee}.  This is an ordering in
which,
in essence, the two $\tilde{E}^{ai}$'s are put to the right and hence
act on the state before the $F_{ab}^k$.

The answer in this case is that the solutions based on the
nonintersecting
link classes are not complete.  There are, in addition, a large
number
of solutions based on intersecting loops.  In the language developed
in section 2.3 the leading term, as $\delta \rightarrow 0$
of the action of the Hamiltonian constraint, evaluated on a
loop $\Gamma_I$
which is part of a graph $\Gamma$ is proportional to $1/\delta$
times
a matrix which acts in the finite dimensional subspace
$V_\Gamma$
of rearrangements of the routings through the graph.  The problem
of finding solutions has been studied for the case that
two\cite{tedlee}, three\cite{viqar-intersects},
four or five\cite{berndtjorge-intersects} lines meet at a point.  In each
case
solutions are found which are linear combinations of routings
through the points of intersection.

Unfortunately, little is known generally about these intersection solutions.
We have neither a proof of existence for an arbitrarily complicated
intersection
nor a general characterization of the solutions.
More importantly, we are still missing a proof of completeness.
However, there is an
important
consideration, which bears on this question.

This is that, as was suggested by Ted Jacobson\cite{tedpersonal} and
demonstrated
recently by Bruegman and Pullin\cite{berndtjorge-intersects},
the method of finding solutions employed
in constructing all of the intersecting solutions does not use all of the
information
in the Hamiltonian constraint operator.  All that is used is
the antisymmetry of the action of the two $\tilde{E}^{ai}$'s, which are on
the
right in the ordering used.  That is, all of the solutions described in
\cite{tedlee,carlolee,viqar-intersects,berndtjorge-intersects} are also
annihilated by any operator of the form
\f
\hat{\cal O} =  \hat{\cal W}^{ck} \epsilon_{abc}\epsilon_{ijk}
\tilde{E}^{ai}\tilde{E}^{bj}
\ff
where $\hat{\cal W}^{ck}$ is an arbitrary operator.
The fact that $\hat{\cal W}^{ck}$
is actually the curvature is not used.   This means, in particular, that
all of these states are annihilated by the operator (115) for the determinant
of the metric, which is of the form of (173) with ${\cal
W}^{ck}=\tilde{E}^{ck}$.

We should mention that this does not mean that the states in the loop
representation cannot have a good physical interpretation.  Indeed, one of the
main themes of this review is that states which are in the kernel
of the operator that represents the determinant of a metric can, and do,
have a good classical limit in terms of observables measured over scales
larger than  the Planck length.
If all states are annihilated by
the determinant of the metric at a point, this points to a breakdown of the
classical picture at short scales, rather than of the theory.  Furthermore,
another objection that has been offered-that the theory is
thereby insensitive to the value of the cosmological constant-is also not
fatal.  It is certain  that whether or not the physical state space is
insensitive to  $\Lambda$,    the physical observables,
and through them the inner product, will certainly depend on the its
value.

However,  it is still very interesting to investigate whether there
could be another class of
solutions to the Hamiltonian constraint which
are not annihilated by the determinant of the metric or other operators.
This is a problem which is currently under investigation, from several
different points of view.

To study this problem it is necessary to understand better the action
of the Hamiltonian constraint operator.
 Recently a lot of progress has been made on this problem.
As this is a rather
technical area, I will refer the reader to the
original papers\cite{carlolee,miles,berndtjorge-intersects,BGP}.
As a byproduct of this work  a new  kind of
solution to all the constraints   was constructed
by Bruegmann, Gambini and Pullin\cite{BGP}.
These solutions cannot be found by constructing the Hamiltonian
constraint operator in the kinematical representation space
${\cal S}^{kin}_\Sigma$, they are based on a representation of
the loop observables developed by Gambini and Leal\cite{GL}
that is, apparently, inequivalent at the kinematical level to the
one developed here.    Furthermore, this new method suggests
connections between quantum gravity, on the one hand, and
knot theory and Chern-Simon theory\cite{ed-jones}, on the other hand,
which are rather suggestive.  This is an extremely interesting
discovery, and we can expect further developments in this direction
in the near future.

\subsection{The problem of the choice of the inner product}

 As mentioned many times in the previous pages, the problem of the inner
product is perhaps the key unsolved problem in nonperturbative quantum
gravity.  Even if we have the complete set of physical states we
cannot compute anything in quantum mechanics
without an inner product.  Here I would like to briefly discuss three
aspects of it.  First,
I will explain why it is such a hard problem.  Second, I will summarize
what we learned about it in chapter 4.  Third, I will propose a strategy
for attacking it.

In free and perturbative quantum field theories
the inner product  is determined by the symmetries of the theory,
and in particular by Poincare invariance.   The problem is that in
nonperturbative quantum gravity there is no Poincare symmetry,  so
we have to find another criteria.  There is a criteria which could work
in this context:  this is to fix the inner product by the reality conditions
of the theory.  Indeed, this criteria is available generally to fix
the inner product of any quantum mechanical system.
The classical algebra of observables is a star algebra,
which means that the complex conjugate of any observable is
defined within the algebra.  One  can then proceed to quantize a classical
system in two steps.  First, one finds a representation of the classical
algebra of observables as an algebra of linear operators on some
representation space $\cal S$.   Second, one chooses an inner product
on $\cal S$ such that the reality conditions on the classical observables
are represented in terms of hermiticity conditions on their quantum
representatives.

This procedure has been developed formally by
Ashtekar\cite{poona} and many
examples have been worked out for finite dimensional systems by
Ashtekar and Tate\cite{poona,tate-lectures}.
It also works in the case of linear quantum
field theories, such as Maxwell\cite{abhaycarlo-maxwell,selfdual}
theory and linearized gravity\cite{gravitons}.
In these cases it yields the Poincare inner product.  It is especially
useful in nonstandard quantizations of Maxwell and linearized gravity
in which one chooses to quantize complex combinations of the field
variables, such as the self-dual variables\footnote{The representations
in which the quantum state is a function of the self-dual part of the
linearized field are unusual, in that the self-dual part of the field is
the positive frequency part of the left helicity plus the {\it negative
frequency} part of the right handed helicity.  Thus the right handed
part of the field must be quantized in an anti-Bargmann representation.
At first sight this may seem a problem, but it can be done\cite{selfdual}}.

In section 4.3 we found that the reality conditions did give us nontrivial
information about the inner product at the kinematical level.  By
requiring that the area observable, ${\cal A}[{\cal S}]$, be
hermitian\footnote{At least on a subspace of the state space.}
we found that the inner product is diagonal on simple loops.  However,
it is non-trivial when the loops involve intersections and retracings.
To completely determine it we must lift the degeneracies of the
area operator by studying the reality conditions for some operator
that does not commute with it.  This is a straightforward problem,
that is currently under attack.

However,
when we come to the problem of the physical inner product
we must face another difficulty.
For each of the three representation spaces, the kinematical,
diffeomorphism invariant and the physical, we must determine the
inner product using the reality conditions.  Further, normally
the states which solve constraints are not normalizable in the inner
product of the kinematical theory.  This means that the inner product
of the physical state space should be determined in terms of the reality
conditions on the physical observables.

The difficulty is that in general relativity we know almost nothing
at the classical level about the physical observables.   Indeed,
except for a handful in the asymptotically flat case we know
not a single  physical observable  explicitly in the full theory.
This is because
the Hamiltonian is a constraint, so that the problem of finding time
reparametrization invariant observables is a dynamical problem.  This
is, indeed, the key problem that separates the quantization of a
gravitational theory from a nongravitational theory.  The result,
and the problem, is that since the physical inner product should
be chosen to realize the reality conditions of the physical
observables its choice is a dynamical problem.

 The problem of finding and interpreting the physical observables in
quantum gravity is exactly the problem of time.    The problem of time
is currently very controversial and has been the subject of a number of
recent articles\cite{problemoftime,carlo-time}.
I will say only a few words about it here.

Because physical observables must commute with the Hamiltonian
constraint, which generates time reparametrization, they cannot
depend on a background time coordinate.  However this does not
mean they do not contain information about evolution in time.  As
first stressed by DeWitt\cite{bryce-time} and recently emphasized by
Rovelli\cite{carlo-time}, the
 physical time evolution must be specified in terms of
physical clocks, that are dynamical degrees of freedom of the theory.
One must pick some degree of freedom of the theory, perhaps arbitrarily
to call a clock.  There are then physical, that is time coordinate
invariant, observables that tell us   the value of other observables
as a function of the physical clock degree of freedom.    For
example, if one couples the gravitational field dynamically to a
point particle carrying a physical clock, one can construct a physical
observable that will tell us what the Ricci scalar is at the position
of the particle for every reading of the clock.

In practice,  the construction of such an observable as an
explicit  function
of the canonical coordinates of the theory  involves using solutions
of the equations of motion of the theory.  In a theory such as general
relativity, which is not integrable, and for which we do not expect
to be able to write the general solution to the equations of motion
in closed form, this will always be a difficult problem.  It clearly
must be approached through some systematic approximation
procedure.  But there is no objection, in principle, to the construction
of such observables at the classical level.

The problems  of time and of the interpretation of physical observables
are then essentially difficulties of the classical theory.  Once the problem
is solved there, one can proceed to the quantum theory.  As argued
in several recent papers by Rovelli\cite{carlo-time},
if the problem can be solved at
the classical level, there is no objection in principle to using the
results at the quantum level.  That is, given the classical physical
observables at the classical level, perhaps expressed in some systematic
approximation, one can seek operators which represent them acting on
the space of physical quantum states.  This may be a difficult
technical problem, involving ordering and regularization difficulties,
but there is no reason this should be a more difficult problem than
finding representatives for the spatially diffeomorphism invariant
observables.  If this can be done then the reality conditions can
be expressed in terms of these physical operators and an inner product
found on the physical state space which represents them.

This, then, is a program for solving the problem of the physical
inner product.  I should mention that this is a controversial proposal,
particularly cogent objections have been put forward by
Kuchar\cite{problemoftime}.
I suggest that the reader consult the review of Kuchar for the
statement of his objections.  One point  Kuchar makes is the,
entirely reasonable, assertion  that at some level difficulties in practice,
such as operator ordering difficulties may become difficulties in principle.
Further, as Kuchar describes, there are other proposals for solving the
problem of time in quantum gravity.  In the absence of a solution to this
key and fundamental problem, it seems to me  one should not argue too
strongly for any one point of view.  Real results on any of these proposals
would be most welcome.

As I indicated in passing above, if we are to make progress on constructing
the physical observables, the only hope seems to be to invent a systematic
approximation procedure.  If, as is strongly suggested by the results
I described in chapter 4, the problems of the divergences of the theory
are solved at the kinematical level  it is perhaps appropriate now to
return to some form of perturbation theory.  But what is required is a
purely quantum mechanical perturbation theory, based on expansion
around a nonperturbative quantum state such as the weave, rather
than around a classical background.

\subsection{The question  of the existence of a ground state}

It used to be that
the key issue to be resolved by
a successful quantum theory of gravity was renormalizability.
Certainly, if there was going to be a successful perturbative theory
this was the main problem.  Or rather, the problem was to find a theory
that is both perturbatively renormalizable and has a stable
ground state.   General relativity is perturbatively nonrenormalizable,
and by now all hopes of fixing the problem through a modification
of perturbation theory, such as a $1/N$ expansion\cite{1/N}, are dead.
It is not
hard to invent theories that are perturbatively renormalizable and
much work in the 1970's and 1980's was put into studying such
theories.  The problem is that none of them have a ground state.  This
is true for a simple reason: perturbative renormalizability requires
that the Lagrangian include all terms of dimension four consistent
with the gauge symmetry, but as the metric is dimensionless these
will necessarily include terms with four derivatives, leading to the
well known instabilities of higher derivative theories.

For this reason, many people began working on string theories in the
1980's.  The main idea of string theory is that if the theory is well
defined it will be finite perturbatively, because of modular invariance.
The main problem faced by string theory is then stability.  The bosonic
string theories are unacceptable because of tachyon instabilities.
The superstring theories seem to solve this problem, at least
perturbatively.   However,
as  emphasized by Eliezer and Woodard \cite{eliezerwoodard},
stability and the existence
of a ground state are potential problems for nonperturbative formulations
of string theory.

What I would like to argue here is that the situation is much the same
for quantum general relativity, nonperturbatively.  If the theory is
well defined at all  it will be finite.  But there are serious questions
about whether the theory has a ground state.

One of the key themes of this review has been that if a spatially
diffeomorphism invariant observable can be translated into a quantum
operator, without breaking the diffeomorphism invariance, that operator
must be finite.  I gave an argument for this at the end of section 2.4.
The key idea of this argument is that i)  any diffeomorphism invariant
operator must be constructed through a regularization procedure,
ii)  that regularization procedure introduces both a background metric,
$h_{ab}^0$ and a regulator or cutoff scale $\epsilon$, and that iii) the
later is measured in units determined by the former, so that if the
operator, in the limit the cutoff is taken away, has no dependence on
the background metric (which must be the case if diffeomorphism
invariance is to be preserved) it can have no dependence on the cutoff.

It is then a nontrivial problem to invent regularization procedures
which result in finite and background independent operators.  We
saw in chapter four examples of both unsuccessful and successful
ways to do this.  In accordance with the argument, we did find that
the resulting operators were finite whenever they were background
independent.

Let me then turn to the question of the existence of a ground state.
This problem must be approached differently than conventional
quantum field theories, because gravitational theories, in general,
do not have Hamiltonians.  Because of the inseparability of
evolution and time reparametrization, the classical evolution
is generated  by the Hamiltonian constraint.  However, there are
special cases in which Hamiltonians exist.  These cases are when
the spacetime diffeomorphism invariance is broken by the imposition
of boundary conditions.

One way that this can be done is to require that the gravitational
field be asymptotically flat.  Classically, this
means that the spatial topology
is taken noncompact and that on a region of $\Sigma$, meant to
be a neighborhood of spatial infinity\footnote{This is normally
taken to be the exterior of a compact region of $\Sigma$.  For a
full statement of the conditions for asymptotic flatness, see
\cite{book,poona}.} one can
impose a flat metric $h_{ab}$ such that, given a radial coordinate, $r$
defined with respect to it, the spatial metric $q_{ab}$ approaches
$h_{ab}$ as $r \rightarrow \infty.$   One can show that this means
that "at infinity" there is a lorentz invariance so that we can imagine
there exists a family of lorentz observers sitting at infinity watching
what is going on inside of the spacetime.  We can then define a
hamiltonian which evolves the whole spacetime according to the
time of one of these observers.  As we have broken the diffeomorphism
invariance by the imposition of $h_{ab}$ that hamiltonian is no
longer just a constraint, it has an additional term.  This additional
term is, however, a boundary term.  Thus, the Hamiltonian looks
like
\f
H(N)= \int_\Sigma N C   +  \int_{\partial \Sigma}  N B
\ff
where $N$ is the lapse function that measures the increase in
coordinate time.    In the new variables formalism, the boundary
term, $B$, is very simple, when $C$ vanishes the Hamiltonian is just
\f
H(N)= \int_{\partial \Sigma} d^2S_a N A_{b i}  \epsilon^{abi} ,
\ff
where we have assumed an asymptotic form for the frame field,
$\tilde{E}^{ai} = \delta^{ai}$.

We can now begin to address the problem of whether the theory
has a ground state.   Before attacking this problem quantum mechanically,
however, we must discuss the situation classically.

Equation (174) is not a sum of squares and it is a nontrivial problem to
show that it is positive.  Indeed, the positive energy theorem is a
justly celebrated result that was open for a long time before
being proved by Schoen and Yau\cite{positive-SY} and
Witten\cite{witten-positive}.  The proof
of Witten is, in fact, closely related to the Ashtekar formalism.
It uses the self-dual connection $A_a^i$ and was prefigured in
a paper by  Sen \cite{sen}  that was also a precursor of Ashtekar's work.
However, there is an important point, which was noticed only
recently.  This is that the proof of Witten assumes the nondegeneracy
of the frame field $\tilde{E}^{ai}$.

As I discussed in sections 5.3 and 5.4,
the space of solutions of the Ashtekar
form of the constraints includes solutions in which $\tilde{E}^{ai}$
is degenerate\cite{degenerate,ls-CS}.
Is there an extension of the positive energy theorem
to these configurations?  The answer, for better or worse, is no:
an explicit class of counterexamples was discovered
by Varadarajan\cite{degenerons}.
These are spherically symmetric solutions in which $\tilde{E}^{ai}$
is nondegenerate and asymptotically flat in a neighborhood of
infinity, nondegenerate in a neighborhood of the origin, but degenerate
in an intermediate region.    They can have any value for the Hamiltonian,
positive or negative.  Furthermore, these solutions are
everywhere
nonsingular, so that even for the positive energy ones there is no
singularity as there is in the Schwarzchild solution.   Because of
the crucial
role of the degenerate regions in these solutions, they are called
degenerons.

 The existence of the degeneron solutions raises many problems for
the theory.   For example, the transition between the lagrangian
theory and the hamiltonian theory, sketched in section 3.2,
depends on  the nondegeneracy of $\tilde{E}^{ai}$.  When this breaks
down their solution spaces are not the same.  Thus, the degenerons
are not solutions to the four dimensional field equations (41-42).
Indeed,
one can prove a positive energy theorem for the solutions to (41-42)
which includes the case that the four dimensional $e^i$'s are
degenerate\cite{sam-positive}.   The crucial difference between
the field equations (41-42) and the constraint equations
(47-49) are the density
weights.  The existence of the degeneron solutions turn out to
depend on the fact that the hamiltonian constraint (49) is a density
of weight two.

The existence of the degenerons certainly suggests that the problem
of the existence of the ground state in the
quantum theory is going to be non-trivial.
This is a problem that has been the subject of some discussion during
the last year.   There are three possibilities which have emerged
from these discussion.

The first possibility is that the degeneron solutions
destabalize any asymptotically flat quantum state, so that the
theory has no ground state.  If this is the case then we will have
to abandon the attempt to quantize general relativity nonperturbatively
using the Ashtekar variables.

The second possibility is that their effect may not be very
serious because the  degeneron solutions may be
in a region of phase space or configuration space which is
disconnected from the region of phase
space
corresponding
to non-singular solutions, for which positive energy applies.   Although
this has not been shown, it is plausible that this is the case. It  may
then be possible to restrict the quantum states to the non-
degenerate
region.  For example, in $2+1$ gravity for compact spaces the
phase space has several disconnected regions and the quantum
theory can be constructed by restricting attention to just
one of them\cite{2+1-us}.

There is, however, a third possibility.  This is that the quantum
Hamiltonian may be bounded from below in spite of the fact that
the classical Hamiltonian is not.  Certainly, there are many cases in
quantum mechanics in which
the classical hamiltonian is not bounded from below, but the
quantum hamiltonian is.    Of course, the inner product plays a crucial
role in this problem because even in cases, such as the harmonic
oscillator, where the classical Hamiltonian is bounded from below,
the quantum Hamiltonian is only bounded from below for states that
are normalized with respect to the inner product.  Thus, this problem
cannot be attacked until we know more about the inner product.

Given our present state of ignorance about this problem, perhaps I
can conclude with an intuitive argument.
If it is the case that the full quantum Hamiltonian is bounded
from below then the  degeneron solutions may play an
important role: at a semiclassical level
they may destabalize the perturbative vacuum
of the theory.  This is because in examples in which the quantum
Hamiltonian is bounded, but the classical one is not,
 the ground state  gives a lot of amplitude to
those regions of
the phase (or configuration)  space where the energy is unbounded.
But this is
exactly the region of the degenerate solutions.

This means that if the theory exists any attempt to construct
a semiclassical construction of the ground state may
fail because the vacuum will fill up with a
condensate of degenerate
configurations.  Thus,
the short
distance structure of the theory will be dominated by such
configurations-which means that there will
be no smooth metric if we look to short enough distances.

However, the ground state of the asymptotically flat theory  must also
approximate a smooth
metric at large distances: the flat metric.   Otherwise
it would not be a good state of the asymptotically flat theory.
The question to ask
is then:
is it possible that the ground state could have support on
degenerate configurations
and still be asymptotically flat when probed on scales
much larger than the Planck scale?

The key conclusion of all the work I have described here is that such
states do exist at the nonperturbative level: they are the weave
states we discussed in section 4.7.  Moreover, as I showed in
section 5.3, their classical
analogue are distributional configurations that are also, except
for sets of measure zero, completely degenerate.

We may then make the following conjecture: the degenerons are the
connection between the semiclassical and nonperturbative approaches
to the theory.  If one attempts to construct the theory semiclassically
one will find that one cannot do it because the perturbative vacuum
is unstable and  space will fill up
with a condensate of degenerate solutions.   If a stable ground state
exists this must happen at the nonperturbative level.  However, at the
nonperturbative level all states look like the result of a condensate
of degenerons: they are degenerate almost everywhere.  Furthermore,
 if such a state is to be asymptotically flat we found that its structure
is not completely free: it must
have discrete structure fixed at the Planck scale.  This is, if we are
optimistic, the signal that the theory can be stable nonperturbatively.

\section{Conclusion}

In my introductory remarks, I indicated that a great
obstacle to our being able to invent a quantum theory of gravity
is the difficulty of seeing past our Newtonian preconceptions of
the nature of physical reality and physical theory.  One of the
themes of this review has been that the classical spacetime background
is one of the features of Newtonian theories that we must transcend.  In
retrospect, it might be argued that the difficulties faced by many
perturbative and semiclassical
approaches to quantum gravity are due to the fact that
they do not sufficiently achieve
this.  A major virtue, and the key difficulty, of a purely nonperturbative
approach to quantum gravity is that one eschews the use of classical
backgrounds at the beginning.  As the development of this work has shown,
this leads to two kinds of difficulties.  First, how do we construct a
description of the geometry of spacetime that is purely quantum mechanical,
and makes no use of a classical background?  Second, how do we construct
well behaved quantum operators without the classical background sneaking
in the back door through the regularization and renormalization procedures?

The chief interest of the results I have reviewed here is that they begin
to show us the way to answering these two questions.  Attention here has
been restricted to the kinematical and diffeomorphism invariant level
and to only one set of observables: those that measure the spatial
geometry.  However, the lessons of the work are more general and there
is reason to believe that the same set of ideas and technical tricks may
help us to extend this work to observables that measure the
evolution of the three geometry and, after that, to physical observables.

It should also be emphasized that all of the new results reported in
chapter 4 make use only of the spatial diffeomorphism invariance, as well
as the basic kinematics of the new variables and loop representation.
Thus, the results such as the quantization of areas
and volumes will hold in a large  class of theories, including general
relativity with arbitrary matter couplings.

Finally, although I have not touched on it here, it is worth mentioning that
the problem of constructing a physical theory without the use of fixed
background structures is a very old problem\cite{copernicanhis}.  It was at
the heart of
Leibniz's and Mach's
criticisms of Newtonian mechanics, which were the starting point
for Einstein's construction of general relativity.   If general relativity,
to a large extent, achieves this, by making the spacetime geometry dynamical,
a large part of the problem of quantum gravity is to achieve this for the
quantum theory\cite{variety}.  The results presented here indicate that  at
least a small
part of this problem, that connected with the kinematics of the theory,
can be successfully attacked.  Whether the direction described here will
develop into a complete solution to this problem, at the dynamical level,
and by doing so give rise to a consistent and useful quantum theory of
gravity, remains an open question.

\section*{ACKNOWLEDGEMENTS}

I would like to thank, first of all, the organizers and participants in the
XXII Gift International Seminar in Theoretical Physics  for a very enjoyable
and stimulating summer school.  I have tried in writing these notes to
keep some of the informality and liveliness of that very pleasant week.
Most of the new results described here were
found in collaboration with Abhay Ashtekar and Carlo Rovelli.  I am
indebted to them for
permission to quote here from joint publications
in advance of their appearance.
I have also learned a great deal about this subject from Julian Barbour,
Berndt Bruegmann, Lois Crane, John Dell, Rodolfo Gambini, Viqar Husain,
Chris Isham, Ted Jacobson,
Lou Kauffman, Jorge Pullin, Paul Renteln, Joseph Samuel,
Ranjeet Tate and Madhavan Varadarajan.  I am also very grateful to a number
of people who have both criticized and encouraged this work  including
Karel Kuchar, Vince Moncrief, Ted Newman, John Wheeler, Edward Witten
and Richard Woodard.  Their questions have been a stimulus to
the work described here.

This work was supported by the National Science Foundation under grants
from the division of Gravitational Physics
and a US-Italy cooperative research program grant.

\end{document}